\documentclass[12pt,letterpaper]{article}

\usepackage{inputenc}
\usepackage[OT1]{fontenc}

\usepackage[nodisplayskipstretch]{setspace} 
\usepackage[margin=1in]{geometry}

\usepackage[usenames,dvipsnames,svgnames]{xcolor}
\usepackage{amsmath, amssymb, amsfonts, graphicx, pdflscape, mathtools, amsthm, upgreek, bm,pgfplots,csvsimple,multirow, multicol, booktabs, import, placeins, subcaption, microtype,etoolbox,thmtools}

\usepackage{tikz}
\usetikzlibrary{shapes.geometric, arrows, fit, positioning}
\usepackage[multiple]{footmisc}
\usepackage{verbatim}
\usepackage[nottoc,numbib]{tocbibind}

\usepackage[backend=biber,sorting=ynt,style=authoryear-comp,giveninits=true,maxbibnames=9,citetracker=true,maxcitenames=3,uniquename=false,uniquelist=true,natbib,isbn=false,doi=false,url=false,eprint=false]{biblatex}
\DeclareFieldFormat{pages}{#1} 
\renewbibmacro{in:}{} 

\DeclareCiteCommand{\citeyear}
    {}
    {\bibhyperref{\printdate}}
    {\multicitedelim}
    {}

\DeclareCiteCommand{\citeyearpar}
    {}
    {\mkbibparens{\bibhyperref{\printdate}}}
    {\multicitedelim}
    {}

\usepackage{titlesec}

\titleformat{\subsubsection}[runin]
  {\normalfont\normalsize\bfseries}
  {\thesubsubsection}{1em}{}
  []
\titlespacing{\subsubsection}
  {0pt}
  {1.5ex plus 1ex minus .2ex}
  {1em}

\titleformat{\section}
  {\bfseries\Large}{%
    {\bfseries\Large\thesection}
  }{1em}{}
  \titleformat{\subsection}
  {\bfseries\large}{%
    {\bfseries\large\thesubsection}
  }{1em}{}

\titlespacing{\subsection}
  {0pt}
  {1.75ex plus 1ex minus .2ex}
  {1ex}

\titlespacing{\section}
  {0pt}
  {1.75ex plus 1ex minus .2ex}
  {1ex}

\usepackage{etoolbox}
\newcommand{\zerodisplayskips}{%
  \setlength{\abovedisplayskip}{1.5ex}%
  \setlength{\belowdisplayskip}{1.5ex}%
  \setlength{\abovedisplayshortskip}{1.5ex}%
  \setlength{\belowdisplayshortskip}{1.5ex}}
\appto{\normalsize}{\zerodisplayskips}
\appto{\small}{\zerodisplayskips}
\appto{\footnotesize}{\zerodisplayskips}

\usepackage{verbatim}
\usepackage{accents}
\newcommand{\ubar}[1]{\underaccent{\bar}{#1}}
\usepackage{needspace}

\usepackage{comment}
\usepackage{enumitem}
    \setlist[itemize]{noitemsep,nolistsep}
    \setlist[enumerate,1]{noitemsep,nolistsep,label=(\arabic*)}
    \setlist[enumerate,2]{noitemsep,nolistsep,label=(\alph*)}
    \setlist[enumerate,3]{noitemsep,nolistsep,label=(\roman*)}
\usepackage{tocvsec2}
\usepackage{threeparttable}
\usepackage[page,title,toc]{appendix}
\usetikzlibrary{calc, cd}
\setlist{noitemsep, topsep=0pt}
\usepackage[skip = 10pt]{caption}
\allowdisplaybreaks[1]
\allowdisplaybreaks[1]

\makeatletter
\renewcommand{\paragraph}{%
	\@startsection{paragraph}{4}%
	{\z@}{1.5ex \@plus 1ex \@minus .2ex}{-0.7em}%
	{\normalfont\normalsize\bfseries}%
}
\makeatother

\let\originalleft\left
\let\originalright\right
\renewcommand{\left}{\mathopen{}\mathclose\bgroup\originalleft}
\renewcommand{\right}{\aftergroup\egroup\originalright}

\definecolor{Blue}{RGB}{86,180,233}
\definecolor{Orange}{RGB}{230,159,0}
\definecolor{Green}{RGB}{0,158,115}
\definecolor{GmailBlue}{RGB}{42, 93, 176} 
\usepackage[
	colorlinks=true,
	citecolor= GmailBlue,
	linkcolor=GmailBlue,
	urlcolor = GmailBlue,
    hyperfootnotes=false
]{hyperref}

\usepackage{pgfplots}
\usepgfplotslibrary{groupplots,colorbrewer}
\pgfplotsset{compat=newest}
\pgfplotsset{cycle list/Set1}
\usepackage{tikz}
\usetikzlibrary{matrix,calc,shapes,arrows.meta,positioning}
\tikzset{
    vertex/.style = {shape=circle,draw, minimum size = 1.8em, inner sep = 0pt},
    edge/.style = {->,> = latex}
}

\usepackage[capitalize,noabbrev,nameinlink]{cleveref}

\theoremstyle{plain}
    \newtheorem{theorem}{Theorem}
    \newtheorem{lemma}{Lemma}
    \newtheorem{proposition}{Proposition}
    \newtheorem{corollary}{Corollary}

    \newtheorem{step}{Step}

\theoremstyle{definition}
  \newtheorem{assumption}{Assumption}
  \newtheorem{definition}{Definition}

  \newtheorem{remark}{Remark}

\makeatletter
\newenvironment{pf}[1][\proofname]{\par
  \pushQED{\qed}%
  \normalfont \topsep4\p@\relax
  \trivlist
  \item[\hskip\labelsep\itshape
  #1\@addpunct{.}]\ignorespaces
}{%
  \popQED\endtrivlist\@endpefalse 
}
\makeatother

\crefname{equation}{equation}{equations}
\Crefname{equation}{Equation}{Equations}
\AtBeginEnvironment{appendices}{\crefalias{section}{appendix}}
\AtBeginEnvironment{appendices}{\crefalias{subsection}{appendix}}
\AtBeginEnvironment{appendices}{\crefalias{subsubsection}{appendix}}
\crefname{appsec}{appendix}{appendices}
\Crefname{appsec}{Appendix}{Appendices}
\Crefname{appendices}{Appendix}{Appendices}
\crefname{appendices}{appendix}{appendices}
\crefname{assumption}{assumption}{assumptions}
\Crefname{assumption}{Assumption}{Assumptions}
\Crefname{lemma}{Lemma}{Lemmata}

\def\id{\mathrm{id}} 

\DeclareMathOperator{\supp}{supp} 
\DeclareMathOperator*{\argmax}{argmax}

\DeclareMathOperator*{\argmin}{argmin}

\DeclareMathOperator{\iSP}{SuperLow}
\DeclareMathOperator{\iP}{Low}
\DeclareMathOperator{\iM}{Middle}
\DeclareMathOperator{\iR}{High}
\DeclareMathOperator{\iSR}{SuperHigh}

\newcommand{\de}{\mathop{}\!\mathrm{d}}
\DeclareMathOperator{\eff}{eff}
\DeclareMathOperator{\debtwrelief}{debt}

\newcommand{\headercref}[2]{\texorpdfstring{\Cref{#2}}{#1 \ref{#2}}}

\title{The Division of Surplus and the Burden of Proof\thanks{\protect We thank Peter Achim, Andreas Asseyer, Ian Ball, Bruno Biais, Nina Bobkova, Ben Brooks, Christoph Carnehl, Roberto Corrao, Gregorio Curello, Piotr Dworczak, Jeff Ely, Albin Erlanson, Matteo Escud{\'e}, Johannes H{\"o}rner, Jan Knoepfle, Nenad Kos, Patrick Lahr, Stephan Lauermann, Jonathan Libgober, Elliot Lipnowski, Vincent Meisner, Meg Meyer, Benny Moldovanu, Axel Niemeyer, Paula Onuchic, Franz Ostrizek, Marco Ottaviani, Eduardo Perez-Richet, Kym Pram, Daniel Rappoport, Doron Ravid, Anna Sanktjohanser, Ludvig Sinander, Roland Strausz, Stephan Waizmann, Joel Watson, Frank Yang,
and seminar audiences in Aalto, Berlin, Bocconi, Bonn, Chicago, HUJ, Sciences Po, Southampton, Tel Aviv, CETC 2025, CMID 2024, Cowles 2025, EAYE 2025, ESSET 2026, ESWC 2025, SAET 2025, SEA 2024, SETF Oxford 2025, and TTW 2024. Preusser acknowledges financial support by the European Research Council (HEUROPE 2022 ADG, GA No. 101055295 – InfoEcoScience).
}}
\author{Deniz Kattwinkel\thanks{\protect University College London, Department of Economics, \textit{\href{mailto:d.kattwinkel@ucl.ac.uk}{d.kattwinkel@ucl.ac.uk}}} \and Justus Preusser\thanks{\protect Bocconi University, Department of Economics and IGIER, \textit{\href{mailto:justus.preusser@unibocconi.it}{justus.preusser@unibocconi.it}}}}

\hypersetup{
	pdftitle={The Division of Surplus and the Burden of Proof},
	pdfauthor={Deniz Kattwinkel, Justus Preusser}
}

\begin{document}

\maketitle
 
\begin{abstract}
A principal and an agent divide a surplus whose size is known only to the agent. The agent decides how much to reveal initially. Both parties can acquire costly evidence to verify the surplus. The agent’s liability is bounded by the revealed surplus. The principal commits to their own acquisition-effort and to transfers contingent on who provides evidence.
With these instruments, the principal simultaneously motivates the agent to reveal the surplus and to acquire evidence. 
We characterize optimal mechanisms using a novel technique that is based on reforming existing mechanisms and orders the instruments. It applies to related multi-instrument problems.


\end{abstract}

\vfill
\pagenumbering{gobble}
\pagebreak
\pagenumbering{arabic}

\section{Introduction}

To resolve informational asymmetries in practice, counterparties substantiate their claims with evidence. How should they share the burden of acquiring evidence? We study this question in a canonical setting.
A principal and an agent divide a surplus, but only the agent knows its true size. 
The principal has additional funds, but the agent's liability is limited by the surplus.
Both parties can acquire conclusive evidence about the surplus, but only at a cost. 
Who bears the burden of acquiring evidence, and how does the burden shape the division of surplus?
We analyze the full solution to this problem: principal-optimal mechanisms that simultaneously divide the surplus and assign the burden of proof.

This problem arises in important applications.
The state and a citizen divide the citizen's wealth with a tax; 
an investor and an entrepreneur divide the returns from an investment;
the state and a monopolist divide the costs from providing a public good.
The size of the surplus is initially known to one party (agent) but not the other (principal): the citizen privately knows their wealth;
the entrepreneur privately observes the returns;
the monopolist privately knows the costs.
But both parties can acquire conclusive proof about the size of the surplus: the agent (citizen/entrepreneur/monopolist) can obtain certified financial statements; the principal (state/investor) can conduct a costly audit.
Absent such proof, the principal can only seize what the agent voluntarily advances as payment.
Moreover, the principal can commit more credibly and has more financial resources than the agent.
Thus, we study which mechanisms the principal optimally offers when the agent's liability is limited by the surplus.

Formally, the agent initially holds the surplus---the agent's \emph{type}.
The principal can only seize the revealed part of the surplus.
There are two ways of revealing surplus.
First, the agent can reveal an amount by advancing it as payment; for example, by transferring money to the principal or an intermediary, or by revealing ownership of other assets via legal documents. 
Second, evidence by either party conclusively reveals the difference between the advance payment and the true surplus.
Each party chooses a probability of obtaining evidence---their \emph{effort}---at an increasing cost.
This cost structure captures that proving the existence of assets by advancing them is easy, but proving that the agent conceals no further assets is comparatively costly. For example, suppose the agent holds all their funds at a bank. To prove these funds exist, the agent can obtain a statement from this bank, which is relatively cheap---this is the (costless) advance payment. But the agent may also hold funds at other banks. Thus, to prove that the first bank's statement covers the agent's entire wealth, either the principal or the agent must seek confirmation from all remaining banks, which is costly.

The agent's effort is not contractible, whereas the principal's effort---the \emph{audit probability}---is contractible.
Both parties maximize their transfer net of effort costs.
Transfers must come out of the surplus or the principal's private funds.
This model augments the canonical costly state-verification model of surplus division (\citet{townsend1979optimal}; \citet{border1987samurai}; \citet{chander1998general}) with evidence provision by the agent. This hidden action problem---motivating the agent to provide evidence---is entangled with the original hidden information problem---revealing the surplus.

The Revelation Principle applies. Without loss, we focus on mechanisms with the following timing.
First, the agent advances a payment. 
Second, the agent exerts effort to acquire evidence and decides whether to disclose.
Third, the principal exerts effort to acquire evidence, but only if the agent did not already disclose evidence. 
Finally, the principal seizes the revealed surplus and may refund the agent out of the revealed surplus and the principal's funds.
The agent is incentivized to advance the full surplus in the first step.
If there is evidence that the agent concealed surplus, the principal seizes the entire surplus as punishment.

A key feature of this set-up is that the principal has multiple instruments:
three refunds depending on whether the agent, the principal, or no one provides evidence.
Further, the principal's and the agent's efforts affect the probability with which each refund is paid.
In practice, such instruments are employed simultaneously. The existing literature has studied these instruments in isolation but has not yet provided techniques to analyze their interactions. We develop the necessary techniques and apply them to our canonical set-up.
To illustrate, we explain here how the principal's and the agent's efforts interact as complements or substitutes. 
Both instruments affect (i) \emph{the principal's audit costs} and (ii) \emph{the agent's incentives}.

(i) \emph{Principal's audit costs.}
The principal incentivizes the agent to advance the full surplus by threatening to seize everything if an audit detects concealed surplus. 
But if the agent provides evidence that they advanced everything, the principal need not audit themselves. Hence, the agent's effort saves audit costs. In this sense, the two efforts act as complements in saving audit costs.

(ii) \emph{Agent's incentives.}
The agent is willing to advance the full surplus if their payoff from doing so exceeds the highest payoff from concealing surplus, which we call the \emph{information rent}.
Both efforts contribute to providing the required information rent.
First, the principal promises a high refund after an audit proves that the agent advanced the full surplus. By increasing the probability of this high refund, the principal's effort contributes to providing the information rent.
Second, unlike the principal's effort, the agent's effort is not contractible. 
The principal induces agent-effort by promising a bonus refund if the agent provides evidence that they advanced the full surplus. The agent's limited liability prevents the principal from recouping the bonus when the agent fails to provide evidence. Thus, as in a classical hidden-action setting with limited liability, the agent's expected refund exceeds their effort costs and they receive a rent. This \emph{evidence rent} contributes to providing the information rent.
In this sense, the two efforts act as substitutes in providing the information rent.

In summary, the principal saves audit costs by motivating the agent to provide evidence themselves. For that, the principal has to pay an evidence rent which counts towards the information rent the agent requires anyways to advance their full surplus. 

This multi-instrumental problem is not amenable to standard mechanism design techniques. We use a novel three-step approach that we call \emph{incentive programming}.
Our approach characterizes a class of mechanisms that we call \emph{tight}. 
Every optimal mechanism will be tight. 
To define this class, we compute for each mechanism its loss function.
This function records for each type the lowest loss attainable from deviating by concealing surplus.
An incentive-compatible mechanism is \emph{tight} if the principal cannot simultaneously extract a higher profit from every type and threaten every type with a higher deviation loss, i.e. raise the profit and loss functions at every type simultaneously.
Tightness refines incentive compatibility and is an obvious property of optimal mechanisms. Nevertheless, we will show that tightness implies considerable structure.

First, we show that each tight mechanism solves a family of simpler problems, one for each type and for the fixed loss function induced by the mechanism. At each type, we optimize only that type's instruments subject to constraints: (i) when advancing the full surplus, the given type must lose no more than their lowest deviation loss in the old mechanism, as
specified by the loss function; (ii) for every other type the loss of deviating to the given type must exceed their lowest deviation loss in the old mechanism, as specified by the loss function.
The constraints depend on the original mechanism only via its loss function. They ensure that the re-optimized mechanism inherits incentive compatibility.
This procedure---re-optimizing a mechanism type-wise given its own loss function---defines an operator on the space of all mechanisms. We show that tight mechanisms are fixed-points of this operator. The loss function summarizes cross-type incentive constraints much as a value function summarizes continuation payoffs in dynamic programming. To evoke this analogy, we call our approach \emph{incentive programming}.

Second, we develop a solution technique for the type-wise re-optimization problems. In the problem for a given type, the principal holds the mechanism for all other types fixed and adjusts only the instruments at that type. For each instrument we introduce a \emph{virtual cost}: it measures the principal’s cost of marginally providing the type’s information rent while preserving their effort incentives and the deviation losses of other types.
Some instruments cannot be adjusted in isolation. For example, increasing the refund that is paid when the agent advances an amount $x$ and no one provides evidence makes deviations to $x$ more attractive. In the type-wise problem we can only adjust other instruments at type $x$ and, therefore, deterring these deviations requires a higher audit probability at $x$. The refund’s virtual cost accounts for these audit costs.
Thus, unlike Myersonian virtual values, which internalize adjustments to the instruments at other types, our virtual costs internalize adjustments to the instruments at the same type.
The principal substitutes among instruments to provide the information rent, optimally selecting those with the lowest virtual costs.

The virtual cost comparison reveals that the principal's and the agent's efforts play fundamentally different roles. The principal's effort has the highest virtual cost of all instruments whenever it is above a minimal level needed to deter deviations. 
Therefore, the principal's effort is optimally set to this minimal level and serves only as a stick to deter deviations, and never as a carrot to provide information rent. 
In contrast, the agent's effort's hidden feature of saving audit costs reduces its virtual cost. We show that therefore it is the principal's first choice in providing information rent. 

The virtual cost comparison not only characterizes the optimized instruments but also points towards directions for local reforms.
In practice, policymakers often reform existing mechanisms locally rather than design new ones. Our type-wise re-optimization problem captures this objective, and incentive programming provides an algorithm for such reforms.
To illustrate, consider a citizen with wealth $x$.
Fixing the mechanism at other types, the re-optimization step tests whether citizen $x$ should be induced to provide more evidence, reducing the principal's audit costs.
If citizen $x$ is sufficiently wealthy, they may be induced to exert effort above the level justified by audit-cost savings alone since the evidence rent contributes to the information rent that this type demands for truth-telling.

In the third step, we ask how the optimal instruments vary across types. 
The fixed-point property of tight mechanisms (step 1) reduces this question to the comparative statics of the type-wise optimization problems. 
This allows us to fully characterize all tight mechanisms and document the rich interplay between the instruments at different types. To illustrate, we focus again on the efforts. In every optimal tight mechanism, the principal's effort is decreasing in the advance payment. The agent's effort is non-monotonic: it initially decreases in the surplus, but then alternates its slope across exactly five intervals. To build intuition, higher types are less tempting deviations for even higher types and therefore require less auditing. Thus, the principal's effort to audit decreases in the type. In line with that, the need to save audit costs with the agent's effort decreases. But higher types also have more surplus to conceal and thus demand a higher information rent. The agent's effort is also used to provide this information rent. These opposing forces underlie the non-monotonic pattern of agent-effort.

While tightness rules out profitable perturbations of the mechanism at each fixed type, optimal mechanisms must also be immune to profitable perturbations at multiple types.
These perturbations involve inter-type comparisons, in contrast to the intra-type comparisons between the virtual costs of instruments.
To close the gap between tightness and optimality, we show it suffices to study a particular \emph{audit-the-poor-or-burden-the-rich} perturbation: it trades off the costs of auditing a given type with the costs of inducing higher types to acquire evidence that they did not deviate.
Tightness is a stepping stone in this argument: in a tight mechanism, the instrument use is optimized at each type and, therefore, the perturbation accounts for all other ways of trading off profits at the respective types.
The perturbation yields necessary first-order conditions for optimality; under a regularity assumption, these conditions are also sufficient.
This certifies that tightness and the audit-the-poor-or-burden-the-rich perturbation capture all relevant forces in optimal mechanisms.

We next (\Cref{sec:model}) set up the model. \Cref{sec:tight_mechanisms} characterizes tight mechanisms as our main result. \Cref{sec:optimal_mechanisms} fills the gap between tightness and optimality. Finally, \Cref{sec:discussion,sec:related_literature} discuss modeling assumptions and related literature.

\section{Model}\label{sec:model}
\subsection{Set-up}
There are a principal and an agent.
The principal has private funds $\tau > 0$.
The agent holds a surplus $x \in [\ubar{x}, \bar{x}]$, where $0 \leq \ubar{x} < \bar{x} < \infty$. 
The surplus $x$ is the agent's private information---the agent's \emph{type}.
The type distribution is $F$, and the minimum (respectively maximum) of the support of $F$ is $\ubar{x}$ (respectively $\bar{x}$).\footnote{In the application of the monopolist providing a public good, the model is interpreted as follows. The monopolist has no initial assets. The state provides $\bar{x}$. The monopolist then privately learns the realized costs $k\leq \bar{x}$ for providing the good. The state seeks to retrieve the unused assets $x = \bar{x} - k$.\label{footnote:publicgood}}

The agent can make an \emph{advance payment} $y \in [0, x]$ to the principal.
The advance payment is contractible and proves the existence of the advanced portion of the surplus. 

After the advance payment has been made, both the agent and the principal can acquire conclusive \emph{evidence} about the difference $x - y$ between the advance payment and the true surplus, making the true surplus contractible. 
Thus, the advance payment $y$ reveals that the surplus $x$ is at least $y$, and evidence (regardless of who provides it) reveals $x$ exactly.
To obtain evidence with probability $e_{A}\in [0, 1]$, the agent incurs a cost $c_{A}(e_{A})$.
If obtained, the agent can (but need not) present the evidence. Similarly, to obtain evidence with probability $e_{P}\in [0, 1]$, the principal incurs a cost $c_{P}(e_{P})$.\footnote{\citet{border1987samurai} consider linear evidence costs. We generalize via a flexible \citet{dye1985disclosure} model in which we identify the success probability with the effort and effort costs are non-linear.}
While the agent's effort $e_{A}$ is not contractible, the principal's effort $e_{P}$ is contractible. 
We often refer to $e_{P}$ as the audit probability.
Each party can attempt to obtain evidence only once.

This cost structure captures the idea that it is easy to prove the existence of assets by advancing them, but comparatively costly to prove that the agent is not concealing any further assets. To give an example, suppose the agent holds all their funds at a bank. To prove the existence of these funds, the agent can obtain a statement from this one bank which is relatively cheap---this is the (costless) advance payment. However, the agent may also hold funds at other banks. Thus, to conclusively prove that the first bank's statement covers the agent's entire wealth, either the principal or the agent must seek confirmation from all remaining banks, which is costly.

The agent's costs $c_{A}$ and the principal's costs $c_{P}$ are thrice differentiable, strictly increasing,\footnote{Throughout the paper, increasing, convex, etc. mean weakly increasing, weakly convex, etc.} and strictly convex, and $c_{A}(0) = c_{P}(0) = c_{A}^{\prime}(0) = c_{P}^{\prime}(0) = 0$ holds. We introduce additional regularity assumptions in \Cref{sec:model:assumptions}.

The principal can seize at most the revealed portion of surplus.\footnote{Thus, the surplus represents the upper bound on what the principal can legally seize. This bound need not be interpreted as the agent's entire wealth. For example, nothing changes in the following set-up. The agent's privately known underlying wealth is $\omega\in\mathbb{R}_{+}$. There is a legal upper bound $x(\omega)$ on any payment that the agent can make to the principal. The agent can advance an amount $y$ between $0$ and $x(\omega)$. Evidence reveals the difference between $y$ and $\omega$ and, thus, the true wealth. Without evidence, the principal can seize at most $y$. With evidence, the principal can seize up to $x(\omega)$. After relabeling, this set-up is equivalent to ours.} 
Thus, the set of feasible transfers depends on the agent's advance payment and whether evidence was provided.
Specifically, if type $x$ advances $y$ and no party provides evidence, then the transfer $t$ from the agent to the principal must be in $[-\tau, y]$; if someone provides evidence, the transfer $t$ must be in $[-\tau, x]$. 
Given a transfer $t$ and efforts $e_{A}$ and $e_{P}$, the ex-post payoffs of the agent and the principal, respectively, are $x - t - c_{A}(e_{A})$ and $t - c_{P}(e_{P})$, respectively.

\begin{remark} Evidence is acquired after the agent advances a payment, i.e. evidence reveals the surplus that the agent chose to conceal. Another valid interpretation is that evidence can be acquired at any time and always reveals the surplus, provided that the principal can ask the agent for an advance payment at a time when the agent did not yet learn whether they succeeded in acquiring evidence (see \cref{appendix:revelation_principle}). 
\end{remark} 

\paragraph*{Mechanisms.} We analyze principal-optimal mechanisms. We allow for general mechanisms, possibly featuring multiple rounds of cheap-talk communication and an arbitrary order of evidence acquisition. As it turns out, it suffices to consider the following.
\begin{definition}\label{def:mechanism}
A \emph{tax mechanism} is given by a quintuple $(e_{A}, e_{P}, r_{A}, r_{P}, r_{\emptyset})$ of functions and plays out with the following timing:
\begin{enumerate}
     \item The agent makes an advance payment $y\in [0, x]$ to the principal.
    \item The principal recommends to the agent an effort $e_A(y)$ to acquire evidence. 
    \item The agent (covertly) exerts effort. If the agent obtains evidence, the agent chooses whether to disclose it.
    \item \begin{enumerate}
    \item If the agent discloses evidence, the principal does not acquire evidence. 
    \item If the agent's advance payment $y$ is in $[0, \ubar{x})$, the principal acquires evidence with probability one.
    \item Otherwise, the principal exerts effort $e_P(y)$ to acquire evidence. 
    \end{enumerate}
    \item \begin{enumerate}
    \item If there is evidence showing that the advance payment $y$ is different from the full surplus, the principal seizes the full surplus.
    \item Otherwise, the principal seizes the advance payment $y$ and refunds one of the following amounts: 
    \begin{enumerate}
        \item $r_A(y)$ if the agent provided evidence, for a total transfer $y - r_{A}(y)$ from the agent to the principal;
        \item $r_P(y)$ if the principal provided evidence, for a total transfer $y - r_{P}(y)$;
        \item $r_\emptyset(y)$ if neither provided evidence, for a total transfer $y - r_{\emptyset}(y)$.
    \end{enumerate}
\end{enumerate}
\end{enumerate}

A tax mechanism is \emph{feasible} if for all $y\in [0, \bar{x}]$ both $e_{A}(y)$ and $e_{P}(y)$ are in $[0, 1]$, and $r_{A}(y)$, $r_{P}(y)$ and $r_{\emptyset}(y)$ are in $[0, y+\tau]$.
A tax mechanism is \emph{incentive compatible (IC)} if each type $x$ of the agent finds it optimal to advance the full surplus ($y=x$),\footnote{The agent can only advance payments $y$ that are less than the full surplus $x$. In optimal mechanisms, the agent would not advance more than the full surplus if this were possible. Thus, while our environment implies uni-directional incentive constraints, this does not affect optimal mechanisms. See \citet{strausz2024unidirectional} and references therein for uni-directional incentive constraints in screening problems without evidence.}
exert the recommended effort $e_{A}(x)$, and disclose available evidence.
\end{definition}
We refer to $r_{A}$, $r_{P}$ and $r_{\emptyset}$, respectively, as the agent-evidence refund, the principal-evidence refund and the no-evidence refund, respectively.

A version of the Revelation Principle and some basic optimality considerations imply that feasible IC tax mechanisms suffice for maximizing the principal's profit (\Cref{appendix:revelation_principle}).
Henceforth, these are simply called \emph{mechanisms}.
We identify two mechanisms if they differ only in the refund $r_{P}$ for types at which the principal's effort to acquire evidence equals $0$ since then this refund is never paid; likewise for the refund $r_{\emptyset}$ if the principal's effort equals $1$.

\subsection{Incentives and profit}\label{sec:preliminaries}

The principal faces a hidden information and a hidden action problem: the agent must be incentivized to advance the surplus and to acquire evidence.
The two problems are entangled: the refunds that the agent is promised for providing evidence will strengthen the agent's incentive to advance the surplus.

Fix a mechanism $m$. 
We describe the agent's best response via backward induction.

\paragraph*{Hidden action: evidence acquisition.}
Suppose type $x$ of the agent truthfully advanced $x$.
Since acquiring evidence is costly, the agent only acquires evidence that they plan to disclose.
If the agent does not provide evidence, the refund is either $r_{P}(x)$ if the principal acquires evidence (probability $e_{P}(x)$) and else $r_{\emptyset}(x)$. We denote the expected refund by 
\begin{equation*}
    R_{NA}(x) = e_{P}(x)r_{P}(x) + (1 - e_{P}(x))r_{\emptyset}(x)
    .
\end{equation*}
If the agent provides evidence, the refund is $r_A(x)$.
Therefore, if the agent exerts effort $\tilde{e}_{A}$, their expected utility is 
$\tilde{e}_{A}  r_{A}(x) + (1 - \tilde{e}_{A}) R_{NA}(x) - c_{A}(\tilde{e}_{A})$.
The agent exerts the recommended effort $e_{A}(x)$ if and only if
\begin{equation}\label{eq:obedience}
    \tag{Ob}
    e_{A}(x)\in
    \argmax_{\tilde{e}_{A}\in [0, 1]} \,
    \tilde{e}_{A} \left(r_{A}(x) - R_{NA}(x)\right) - c_{A}(\tilde{e}_{A}).
\end{equation}
The \emph{evidence rent} of type $x$ is the extra refund net of effort costs given $e_A(x)$:
\begin{equation*}
     \mbox{evidence rent of type $x$}
     =
     e_{A}(x)  \left(r_{A}(x) - R_{NA}(x)\right) 
     - c_{A}(e_{A}(x))
     .
\end{equation*}
In \Cref{appendix:no_excessive_evidence}, we show that optimally the evidence rent equals $u_{A}(e_{A}(x))$, where
\begin{equation*}
    \forall\tilde{e}_{A}\in [0, 1]\quad
    u_{A}(\tilde{e}_{A}) = \tilde{e}_{A} c_{A}^{\prime}(\tilde{e}_{A}) - c_{A}(\tilde{e}_{A}).
\end{equation*}
Henceforth, we focus on such mechanisms and refer to $u_{A}$ as the evidence rent.
Note $u_{A}$ is strictly increasing.

The interim utility $U(x, m(x))$ of type $x$ from advancing the full surplus is given by
\begin{equation*}
    U(x, m(x)) = R_{NA}(x) + u_{A}(e_{A}(x)).
\end{equation*}
\paragraph*{Hidden information: advancing the full surplus.}
Suppose type $x$ advances a smaller amount $y\in [0, x)$.
Then, the agent will not acquire evidence since the principal would confiscate the full surplus as punishment.
If the principal acquires evidence, the principal also confiscates the full surplus $x$. Else, the principal seizes $y$ and refunds $r_{\emptyset}(y)$.
Thus, the expected utility of type $x$ after advancing $y$ equals
\begin{equation*}
    x - \left(e_{P}(y)x + (1 - e_{P}(y))(y - r_{\emptyset}(y))\right).
\end{equation*}
We define $\lambda_{m}(x)$ as type $x$'s lowest loss when concealing surplus:\footnote{In the infimum we also consider $y = x$ only to avoid the infimum over the empty set. Note that the loss in the case $y=x$ still assumes that the principal seizes everything after acquiring evidence and, hence, this loss does not equal type $x$'s loss from being truthful.}
\begin{equation}\label{eq:loss_function_def}
    \lambda_{m}(x) = \inf_{y\in [\ubar{x}, x]} e_{P}(y) x + (1 - e_{P}(y))(y - r_{\emptyset}(y)).
\end{equation}
We call $\lambda_{m}$ the \emph{loss function} of $m$.
Type $x$ advances the full surplus if and only if\footnote{Type $x$ can also advance amounts strictly below $\ubar{x}$. For such amounts the principal acquires evidence with probability $1$ and confiscates the full surplus. Thus, these advance payments can be safely ignored. Similarly, we take the domains of $e_{A}, e_{P}, r_{A}, r_{P}, r_{\emptyset}$ to be the type space $[\ubar{x}, \bar{x}]$.}
\begin{equation}\label{eq:pre_IC}
\tag{IC}
    U(x, m(x))
    \geq x - \lambda_{m}(x).
\end{equation}
We call $x - \lambda_{m}(x)$ the \emph{information rent}.

\paragraph*{The principal's profit.}
The principal's profit $\Pi_{m}(x)$ from type $x$ is given by
\begin{equation*}
    \Pi(x, m(x)) = x - R_{NA}(x) - u_{A}(e_{A}(x))  - c_{A}(e_{A}(x)) - (1 - e_{A}(x))  c_{P}(e_{P}(x)).
\end{equation*}
The principal seizes the advance payment $x$ but reimburses the agent via $R_{NA}(x)$, via the evidence rent $u_{A}(e_{A}(x))$, and must also reimburse the agent's effort costs $c_{A}(e_{A}(x))$. 
Finally, the principal's expected effort costs are $(1 - e_{A}(x)) c_{P}(e_{P}(x))$ since the principal exerts effort only if the agent does not provide evidence.

We sometimes abbreviate $U_{m}(x) = U(x, m(x))$ and $\Pi_{m}(x) = \Pi(x, m(x))$.

\paragraph*{The principal's evidence.}
The principal's effort $e_{P}$ to acquire evidence increases the loss $\lambda_{m}$ from deviations and, thus, incentivizes the agent to advance the full surplus.
Additionally, if the principal's evidence verifies that a type $x$ advanced the full surplus, the principal pays the refund $r_{P}(x)$ which increases type $x$'s utility for advancing the full surplus and which cannot be obtained by any other type.

\paragraph*{The agent's evidence.}
On the one hand, the agent's effort $e_{A}(x)$ decreases the principal's profit (i) through the evidence rent $u_{A}(e_{A}(x))$ and (ii) through surplus destruction $c_{A}(e_{A}(x))$. 
On the other hand, (iii) $e_{A}(x)$ reduces the principal's audit costs $(1 - e_{A}(x)) c_{P}(e_{P}(x))$.
Finally, (iv) $e_{A}(x)$ increases the evidence rent $u_{A}(e_{A}(x))$ and thus helps pay the information rent that the agent demands for advancing the full surplus.

\paragraph*{Efficient agent effort.}
Consider the counterfactual situation in which the principal could induce any agent effort without paying the evidence rent. In this situation, for a fixed principal effort $e_{P}(x)$, the implemented agent effort optimally trades off the agent's effort cost (ii) and the principal's expected costs (iii), ignoring effects (i) and (iv) on the evidence rent. We call this effort level the \emph{efficient agent effort $e_{A}^{\eff}(x)$}:
\begin{equation}\label{eq:def:efficient_agent_evidence}
    e_{A}^{\eff}(x) = \argmin_{\tilde{e}_{A}\in[0, 1]} c_{A}(\tilde{e}_{A}) + (1 - \tilde{e}_{A}) c_{P}(e_{P}(x)).
\end{equation}
The minimizer is unique.
The efficient agent effort $e_{A}^{\eff}(x)$ is endogenous to the mechanism and depends co-monotonically on the principal's effort $e_{P}(x)$.

\subsection{Assumptions}\label{sec:model:assumptions}
We introduce three further assumptions. \Cref{sec:discussion} discusses dropping these assumptions.
\begin{assumption}\label{assumption:budget}
    It holds $\tau \geq c_{A}^{\prime}(1) > c_{P}(1)$.
\end{assumption}
\Cref{assumption:budget} compares the principal's funds $\tau$ to the agent's marginal costs and the principal's actual costs of effort. \Cref{assumption:budget} does not imply that the agent's actual or marginal costs are uniformly higher than those of the principal. If the cost functions are symmetric ($c_{A} = c_{P}$), then $ c_{A}^{\prime}(1) > c_{P}(1)$ follows from strict convexity.
The inequality $\tau\geq c_{A}^{\prime}(1)$ implies that in relevant mechanisms the principal's funds suffice for incentivizing the agent to acquire evidence with probability one, but $c_{A}^{\prime}(1) > c_{P}(1)$ implies that doing so is not optimal. 

\begin{assumption}\label{assumption:regularity}
    It holds $\ubar{x} + \tau > c_{P}^{\prime}(1) \max_{e_{A}\in (0, 1]} (1 - e_{A}) \frac{u_{A}^{\prime\prime}(e_{A})}{c_{A}^{\prime\prime}(e_{A})} - e_{A}$.
\end{assumption}
\Cref{assumption:regularity} ensures a single-crossing property for the \emph{virtual} costs that we introduce in \Cref{sec:how_to_tighten:virtual_costs}.
The principal can use the evidence rent $u_{A}(e_{A})$ to provide information rent. Thus, raising $u_{A}(e_{A})$ may allow to reduce other payments. As we elaborate in the context of our main characterization (\Cref{sec:tight_mechanisms:stochastic_audits}), there is one particular such substitution that entails reducing $e_{P}$ at a rate mediated by $u_{A}$ and $\ubar{x} + \tau$. \Cref{assumption:regularity} ensures that the total effect on effort costs, $c_{A}(e_{A}) + (1 - e_{A}) c_{P}(e_{P})$, is positive if and only if $e_{A}$ is sufficiently high.

\Cref{assumption:regularity} is easy to verify for some common cost functions. For example, for iso-elastic costs---$c_{A}(e_{A}) = e_{A}^{k+3}$ for some $k > -2$---or exponential costs---$c_{A}(e_{A}) = \exp((k+1) e_{A}) - (k+1)e_{A} - 1$ for some $k > -1$---one may verify $k \geq \max_{e_{A}\in (0, 1]} (1 - e_{A}) \frac{u_{A}^{\prime\prime}(e_{A})}{c_{A}^{\prime\prime}(e_{A})} - e_{A}$. In these cases, \Cref{assumption:regularity} is implied by $\ubar{x} + \tau > k c_{P}^{\prime}(1)$, so that $\tau$ should be large relative to the curvature $k$ of the agent's costs and the principal's highest marginal costs $c_{P}^{\prime}(1)$.

Both \Cref{assumption:regularity,assumption:budget} hold if the principal's funds $\tau$ are sufficiently large and the principal's cost is below the agent's \emph{marginal} cost at the top, $c_{P}(1) < c_{A}^{\prime}(1)$.
In the standard mechanism design approach, the principal can commit to unbounded transfers ($\tau = \infty$).
Hence, we view as natural to assume that $\tau$ is sufficiently large to meet \Cref{assumption:regularity,assumption:budget}.
We discuss unbounded transfers in \Cref{sec:discussion}.

Finally, we impose:
\begin{assumption}\label{assumption:rent_regularity}
    The sum $u_{A}(e_{A}) + c_{A}(e_{A}) = e_{A} c_{A}^{\prime}(e_{A})$ is strictly convex in $e_{A}$.
    In particular, for all $e_{P}\in [0, 1]$, there is a unique minimizer of evidence rent plus surplus destruction, $\argmin_{\tilde{e}_{A} \in [0, 1]} u_{A}(\tilde{e}_{A}) + c_{A}(\tilde{e}_{A}) + (1 - \tilde{e}_{A}) c_{P}(e_{P})$.  
\end{assumption}

\section{Tight mechanisms}\label{sec:tight_mechanisms}

In this section, we introduce and characterize \emph{tight} mechanisms, which entail no loss of optimality. This yields our main economic insights. Characterizing tightness is a stepping stone towards optimizing the entire mechanism.

A mechanism $m$ is tight if the principal cannot simultaneously extract a higher profit $\Pi_{m}$ from every type and threaten every type with a higher deviation loss $\lambda_{m}$.
\begin{definition}\label{def:dominance}
    A mechanism $m^{\ast}$ is \emph{tighter than} a mechanism $m$ if $(\Pi_{m}, \lambda_{m}) \leq (\Pi_{m^{\ast}}, \lambda_{m^{\ast}})$.\footnote{For real-valued functions $g$ and $g^{\ast}$ with common domain we write $g \leq g^{\ast}$ to mean that $g(x) \leq g^{\ast}(x)$ holds for all $x$ in the domain. Similarly, $(g, h)\leq (g^{\ast}, h^{\ast})$ means that both $g\leq g^{\ast}$ and $h\leq h^{\ast}$ hold.}
    A mechanism $m^{\ast}$ is \emph{tight} if $m^{\ast}$ is tighter than every mechanism $m$ that is tighter than $m^{\ast}$, i.e. $(\Pi_{m^{\ast}}, \lambda_{m^{\ast}}) \leq (\Pi_{m}, \lambda_{m})$ only if $(\Pi_{m^{\ast}}, \lambda_{m^{\ast}}) = (\Pi_{m}, \lambda_{m})$.
    (By convention, both $m^{\ast}$ and $m$ mean feasible IC mechanisms.)
\end{definition}

The next lemma shows that tight mechanisms are without loss for maximizing profits. The proof is in \Cref{appendix:tight_mechanisms:abstract_tight_existence_proof}.
\begin{lemma}\label{lemma:tight_mechanisms_wlog}
    For all mechanisms $m$ there is a tight mechanism that is tighter than $m$.
\end{lemma}

Tightness is not \emph{efficiency} as defined for the setting in which only the principal can acquire evidence (\citet{border1987samurai}; \citet{chander1998general}).
Efficiency asks whether it is possible to extract a type-by-type higher revenue at a type-by-type lower principal-effort.
Efficiency is silent on the agent's effort, motivating a novel notion for our setting.
A companion note shows that tightness refines efficiency when only the principal can acquire evidence (\citet{tightsamurai}).

\subsection{The class structure of tight mechanisms}
\label{sec:tight_mechanisms:stochastic_audits}

\begin{figure}[t!]
\centering
\includegraphics[width=\textwidth]{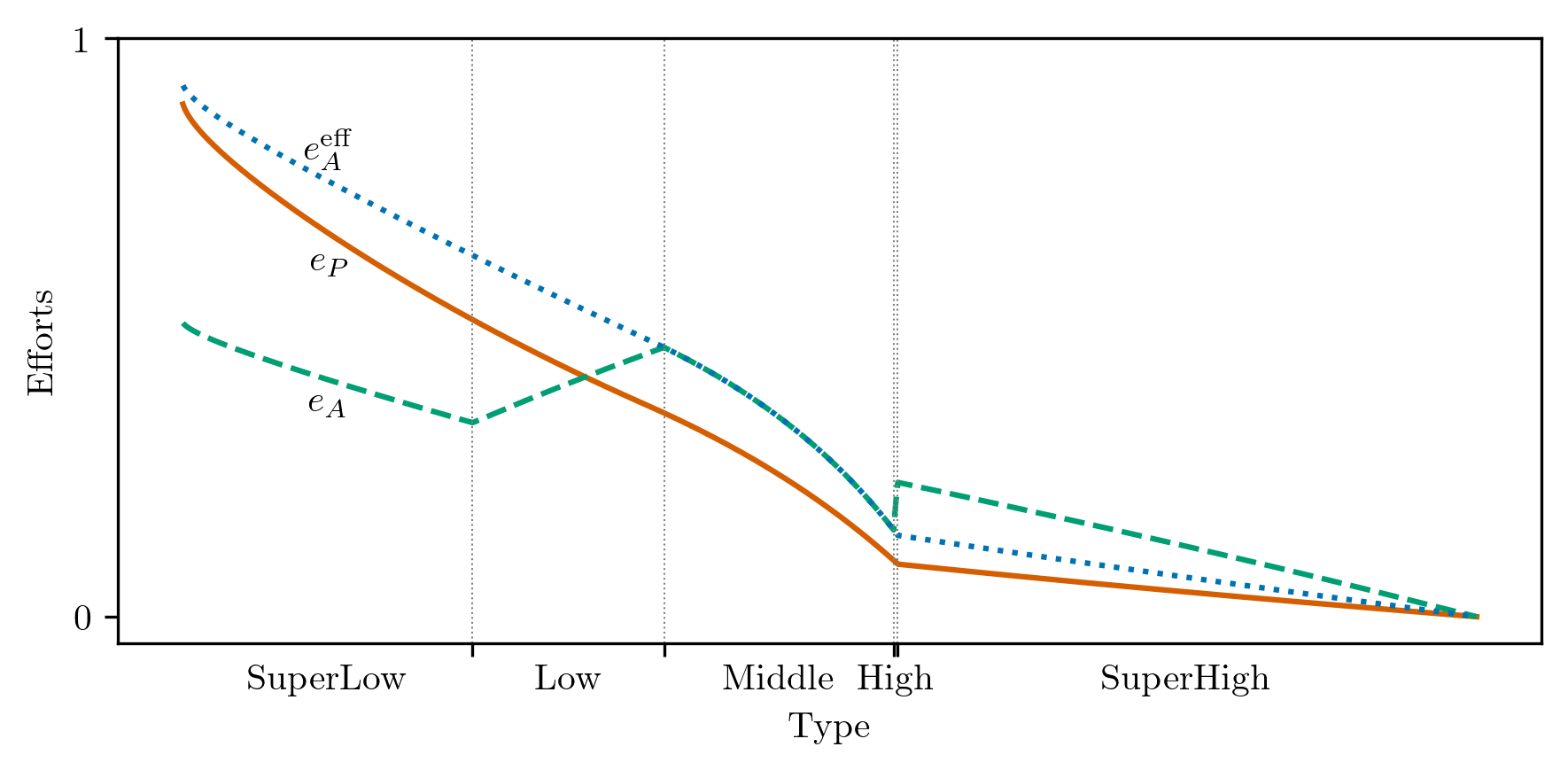}
\caption{The efforts of a tight mechanism in which $e_{P}$ is interior except at $\bar{x}$. \Cref{footnote:computation} (page \pageref{footnote:computation}) explains the computation.}
\label{fig:random_tight_mechanism_example}
\end{figure}

A mechanism has \emph{random audits} if the principal's effort $e_{P}(x)$ is interior for at least one type $x$.
In this section, we characterize tight mechanisms with random audits, later showing that this characterization applies to all optimal mechanisms.

Each tight mechanism with random audits admits a non-monotonic relationship between the principal's and agent's evidence acquisition efforts.
The mechanism divides types into seven endogenous intervals. 
In optimal tight mechanisms, we later find that the first interval is empty, the last interval contains only the highest type $\bar{x}$, but the five intermediate intervals all have non-empty interiors.
We denote the five intermediate intervals in the order of their surplus levels by \emph{SuperLow, Low, Middle, High, SuperHigh}.
The intervals differ in the offered refunds and the efforts.
\Cref{fig:random_tight_mechanism_example} illustrates the general pattern: the principal's effort $e_{P}$ is decreasing, and the agent's effort $e_{A}$ is non-monotonic.
We explain the economic intuitions of each interval after stating the theorem.

The definition and optimality of tight mechanisms are independent of the type distribution: all comparisons are type-by-type.
Our characterization is also robust to the type distribution.

\begin{theorem}\label{thm:stochastic_tight_characterization}
    Let $m$ be a tight mechanism with random audits.
    There exist five consecutive\footnote{By ``consecutive'' we mean $\sup\iSP = \inf\iP$ etc.} intervals of types---$\iSP,\iP,\iM,\iR,\iSR$---that each have a non-empty interior, that partition the set of types where $e_{P}$ is interior, and that satisfy all of the following:
    \begin{enumerate}
        \item The \emph{principal's effort} $e_{P}$ is
        \begin{enumerate}
            \item constantly $1$ on $[\ubar x, \inf\iSP)$;
            \item decreasing on $[\ubar{x}, \bar{x}]$, strictly so on $\iSP\cup \iP \cup \iM \cup \iR$;
            \item constantly $0$ on $[\sup\iSR,\bar x]$;
            \item continuous, except possibly at $\inf\iSP$, where $e_{P}$ is right-continuous.
        \end{enumerate}

        \item The \emph{agent's effort} $e_A$ is
        \begin{enumerate}
            \item constant on $[\ubar x, \inf\iSP)$;
            \item non-monotonic; specifically, strictly decreasing on each of $\iSP$, $\iM$ and $\iSR$, but strictly increasing on each of $\iP$ and $\iR$.
            \item constantly $0$ on $[\sup\iSR,\bar x]$.
            \item strictly below $e_{A}^{\eff}$ on $[\ubar{x}, \inf\iM)$, equal to $e_{A}^{\eff}$ on $\iM$, strictly above $e_{A}^{\eff}$ on $(\sup\iM, \sup\iSR)$, and equal to $e_{A}^{\eff}$ on $[\sup\iSR, \bar{x}]$;
            \item continuous, except possibly at $\inf\iSP$, where $e_{A}$ is right-continuous.
        \end{enumerate}
\item The \emph{no-evidence refund} $r_{\emptyset}$ is increasing, and such that $r_{\emptyset}(y) > 0$ if and only if $y > \inf\iSR$. 
\item The \emph{principal-evidence refund} $r_{P}$ is increasing, and such that $r_{P}(y) > 0$ if and only if $y > \inf\iM$, and $r_{P}(y) = y + \tau$ if and only if $y \geq \sup\iM$.

\item The \emph{agent's utility} $U_{m}$ is $v$-shaped: constant on $[\ubar{x}, \inf\iSP]$, strictly decreasing on $\iSP$, and strictly increasing on $[\sup\iSP, \bar{x}]$. Moreover, $U_{m}$ is bounded away from $0$, i.e. $\inf_{y\in[\ubar{x}, \bar{x}]}U_{m}(y) > 0$.
\item The \emph{agent's incentive} to advance the full surplus is
\begin{enumerate}
    \item strict for $y\in [\ubar{x}, \sup\iSP)$; specifically, $U_{m}(y) > y - \lambda_{m}(y)$;
    \item binding for $y\in [\sup\iSP, \bar{x}]$; specifically, $U_{m}(y) = y - \lambda_{m}(y)$.
\end{enumerate}
\item The \emph{principal's profit} $\Pi_{m}$ is increasing.
\end{enumerate}
\end{theorem}
The proof is in \Cref{appendix:tight_mechanisms:preparations}. We sketch the proof strategy in \Cref{sec:tight_mechanisms:how_to_tighten}.

To explain this structure, fix a tight mechanism $m$ with random audits. Assume $e_{P}$ is interior on $[\ubar{x}, \bar{x})$, i.e. the intervals $\iSP$ to $\iSR$ cover $[\ubar{x}, \bar{x})$.
This assumption simplifies the exposition and holds for all optimal tight mechanisms, as we show later.

\paragraph*{How to make information rent?} To incentivize the agent to advance their surplus truthfully, the agent's interim utility $U_{m}(x)$ must exceed the utility from the best deviation, $x - \lambda_{m}(x)$. 
The principal has multiple instruments for providing the information rent $x - \lambda_{m}(x)$: the evidence rent $u_{A}$ (via $r_{A}$ when the agent acquires evidence), and the refunds $r_{P}$ and $r_{\emptyset}$ (when the agent does not provide evidence).
Additionally, the principal's effort $e_{P}(x)$ affects the probability with which each refund is paid.
The information rent $x - \lambda_{m}(x)$ is increasing in $x$ since the principal cannot seize more than the surplus $x$ when detecting a deviation.
We derive \Cref{thm:stochastic_tight_characterization} by ordering across types with which instruments the principal provides the information rent $x - \lambda_{m}(x)$.

To describe this order, let us momentarily take as given that the principal's effort $e_{P}$ is decreasing; we explain this important property later.
The agent's efficient effort $e_{A}^{\eff}$ acts as a complement to the principal's effort $e_{P}$. 
Thus, $e_{A}^{\eff}$ is also decreasing.  

\paragraph*{Super low types.}
Making the information rent is easy for these types. First, super low types $x$ do not have much surplus to hide and, hence, demand a small information rent. 
Second, to deter higher types from deviating to super low types, the principal must audit super low types with substantial probability $e_{P}(x)$. Thus, to save auditing costs $(1 - e_{A}(x))c_{P}(e_{P}(x))$, the principal incentivizes super low types to provide evidence with substantial probability $e_{A}(x)$, yielding them an evidence rent $u_{A}(e_{A}(x))$. In particular, this rent exceeds the small information rent of all super low types. Thus, the principal, unable to offset the evidence rent against the information rent, implements an inefficiently low agent effort which minimizes surplus destruction plus evidence rent (the minimizer is unique; \Cref{assumption:rent_regularity}):
\begin{equation*}
    e_{A}(x) = \argmin_{\tilde{e}_{A}\in [0, 1]}
	u_{A}(\tilde{e}_{A}) + c_{A}(\tilde{e}_{A}) + (1 - \tilde{e}_{A}) c_{P}(e_{P}(x)).
\end{equation*}
This optimal effort is co-monotone with $e_{P}(x)$.
Thus, $e_{A}(x)$ and the corresponding evidence rent $u_{A}(e_{A}(x))$ decrease on $\iSP$. The information rent $ x - \lambda_{m}(x)$ increases. At the top of the interval, the two rents match.

\paragraph*{Low types.} 
These and all higher types have more surplus to hide; the incentive to advance the full surplus binds.
For such a type $x$, the principal can offset the evidence rent against the information rent; any evidence rent up to $x - \lambda_{m}(x)$ is effectively costless for the principal since the information rent must be paid.
For $x\in \iP$, the implemented effort $e_{A}(x)$ is still inefficiently low and, hence, the principal provides the information rent fully via the evidence rent: $u_{A}(e_{A}(x)) = x- \lambda_{m}(x)$.
Since the information rent $x - \lambda_{m}(x)$ increases, also the implemented effort $e_{A}(x)$ increases.
At the highest type in $\iP$, the implemented effort $e_{A}$ reaches the efficient level $e_{A}^{\eff}$. 

\paragraph*{Middle types.}
To provide the increasing information rent to these types exclusively via the evidence rent, the implemented effort would have to be inefficiently high. 
Instead, for all $x\in\iM$ the principal implements the efficient effort---$e_{A}(x) = e_{A}^{\eff}(x)$---and provides the missing rent $x - \lambda_{m}(x) - u_{A}(e_{A}(x))$ via a refund $r_{P}(x)$ after an audit confirms that the agent advanced the full surplus: 
\begin{equation*}
    e_{P}(x)r_{P}(x) = x - \lambda_{m}(x) - u_{A}(e_{A}(x)).
\end{equation*}
Thus, $e_{A}$ decreases co-monotonically with $e_{P}$.
The refund $r_{P}(x)$ is increasing since $e_{P}(x)$ is decreasing in $x$ while the missing rent $x - \lambda_{m}(x) - u_{A}(e_{A}(x))$ is increasing in $x$.
The highest type in $\iM$ receives the maximal refund after an audit, $r_{P}(x) = x + \tau$, a full refund of the advance payment $x$ and principal's private funds $\tau$.

\paragraph*{High types.} The principal already exhausts the maximal refund after a successful audit: $r_{P}(x) = x + \tau$ for all $x\in\iR$.
To provide the missing information rent, the principal must either raise the evidence rent $u_{A}(e_{A}(x))$ or offer a refund $r_{\emptyset}(x)$ if no one proves that the agent advanced the full surplus.
Since $r_{\emptyset}(x)$ is not conditioned on evidence and attracts other types to deviate to $x$, the principal rather raises $e_{A}(x)$ above the efficient level to provide the missing information rent: $u_{A}(e_{A}(x)) = x - \lambda_{m}(x) - e_{P}(x)(x + \tau)$.
At the highest type in $\iR$, the agent's effort $e_{A}(x)$ is so inefficiently high that it is too expensive to raise $e_{A}(x)$ further.

\paragraph*{Super high types.} To provide the information rent for super high types $x$, the principal starts paying the no-evidence refund $r_{\emptyset}(x)$ when neither the agent nor the principal prove that the agent advanced the full surplus. The refund $r_{\emptyset}(x)$ is the principal's last resort since, in contrast to all other instruments, this refund also benefits higher types that deviate to $x$. That is, while increasing $r_{\emptyset}$ pays towards the information rent for type $x$, it also increases the required information rent for all higher types. This drawback of $r_{\emptyset}(x)$ is less severe at higher types $x$ because there are fewer even higher types. The no-evidence refund $r_{\emptyset}$ increasingly replaces the payments via the evidence rent $u_{A}$ and the refund $r_P$. The efforts $e_A$ and $e_P$ decrease to $0$ at the highest type $\bar{x}$, who cannot be mimicked.
At the highest type $\bar{x}$, the entire information rent is paid via the no-evidence refund: $r_{\emptyset}(\bar{x}) = U_{m}(\bar{x}) = \bar{x} - \lambda_{m}(\bar{x})$.

\paragraph*{Monotonicity of the principal's effort.}
The principal's effort is always set as low as possible to deter deviations, as we explain in more detail in \Cref{sec:how_to_tighten:virtual_cost_comparison}.
For types $x$ in $\iSP$, $\iP$, $\iM$, or $\iR$, the no-evidence refund $r_{\emptyset}(x)$ equals $0$. 
Thus, other types deviating to $x$ expect to lose everything when the principal audits, and lose $x$ otherwise.
Consequently, the deviation to a higher type $x$ is less attractive and a decreasing audit probability $e_{P}(x)$ suffices to deter other types from misreporting to $x$.

On $\iSR$, the principal's effort $e_{P}$ is also decreasing, but the argument is more subtle.
For a fixed super high type $x$, we first explain the interactions between the three instruments $e_{P}(x)$, $e_{A}(x)$ and $r_{\emptyset}(x)$, all of which affect type $x$'s interim utility.
Consider the following perturbation. First, increase $e_{P}(x)$; this increases the probability that type $x$ gets the maximal reward $r_{P}(x) = x + \tau$, and the probability that higher types deviating to $x$ are punished. Consequently, this change increases $x$'s interim utility and slackens higher types' incentive constraints. Second, also increase $r_\emptyset(x)$ to exactly offset this slack in the others' incentive constraints. Type $x$'s interim utility increases again. Finally, reduce $e_A(x)$ (reducing the evidence rent $u_{A}(e_{A}(x))$) to exactly offset type $x$'s gain in interim utility from increasing $e_P(x)$ and $r_\emptyset(x)$.

This perturbation effectively substitutes $r_{\emptyset}(x)$ and $e_P(x)$ for $u_{A}(e_{A}(x))$ to pay type $x$'s information rent. 
The effect on the principal's profit decomposes as follows.
On the one hand, the substitution increases the principal's costs $(1-e_{A}(x))c_{P}(e_{P}(x))$: the principal acquires evidence with a higher probability $e_{P}(x)$, and this happens more often since the agent's effort $e_{A}(x)$ is reduced. 
On the other hand, the substitution reduces the costs $c_{A}(e_{A}(x))$ of the inefficiently high effort of the agent.
The agent's effort $e_{A}(x)$ and the principal's effort $e_{P}(x)$ are complementary in reducing the principal's costs $(1-e_{A}(x))c_{P}(e_{P}(x))$.

In summary, there is both a substitution and a complementarity between $e_{A}(x)$ and $e_{P}(x)$.
\Cref{assumption:regularity} implies that the complementarity dominates.
Precisely, there is a unique optimal level of $e_{A}(x)$ for the above perturbation. We can interpret this as a single-crossing property of the virtual costs that we introduce in \Cref{sec:how_to_tighten:virtual_costs}.
\Cref{assumption:regularity} is a joint condition on the evidence rent $u_{A}$, the principal's funds $\tau$, and the principal's costs $c_{P}$ since $u_{A}$ and $\tau$ mediate the substitution between the efforts, while $c_{P}$ determines the complementarity.

Thus, we can show that $e_{A}$ and $e_{P}$ are co-monotonic on $\iSR$.
We still have to explain why both efforts decrease.
For higher types the principal increasingly substitutes $r_{\emptyset}(x)$ for the evidence rent $u_{A}(e_{A}(x))$ to pay $x$'s information rent, because, as intuited earlier, there are fewer even higher types who can deviate to $x$ to claim $r_{\emptyset}(x)$.
Thus, $e_{A}$ strictly decreases.
By co-monotonicity, we can show that $e_{P}$ also decreases weakly; weakly because the principal may substitute $r_{\emptyset}$ for $e_{A}$ exactly at a rate that holds $e_{P}$ constant.

\subsection{Tight mechanisms with non-random audits}\label{sec:tight_mechanisms:debt}

In \Cref{sec:optimal_mechanisms:random_optimal}, we will show that random audits are optimal.
To provide a contrasting perspective, we now consider tight mechanisms with non-random audits---$e_{P}(x) \in\lbrace 0, 1\rbrace$ for all $x$---and characterize these mechanisms as follows: 
\begin{definition}
    A mechanism $m$ is a \emph{debt-with-relief mechanism} if there is a \emph{face value} $y_{0} \in [\ubar{x}, \bar{x}]$ and a \emph{relief} $r_{A}^{\debtwrelief}\in [0, \ubar{x} + \tau]$ such that for all $x\in [\ubar{x}, \bar{x}]$, 
    \begin{equation*}
       \underbrace{e_{P}(x)=
        \begin{cases}
        1,& x < y_{0},\\
        0,& x\ge y_0;
        \end{cases}\;
        r_P(x)=0,\;
         r_{\emptyset}(x) = \max\{x - y_{0},0\}}_{\mbox{debt contract}},\;
        \underbrace{r_{A}(x) = \begin{cases}
        r_{A}^{\debtwrelief},& x < y_{0},\\
        0,& x\ge y_0.
        \end{cases} }_{\mbox{relief clause}}
    \end{equation*}
\end{definition}
The first part resembles a debt contract {\`a} la \citet{townsend1979optimal}. When the agent advances less than the face value $y_0$ (i.e. defaults) and does not provide evidence, the principal audits with certainty and seizes the full surplus. When the agent advances more than the face value, the principal only seizes the face value. The second part is a relief clause: in the case of a default, the agent gets a relief $r_{A}$ for providing evidence. 

Debt-with-relief mechanisms are special cases of the tight mechanisms characterized by \Cref{thm:stochastic_tight_characterization}, but the five interior intervals are all empty.

We show in \Cref{appendix:tight_mechanisms:nonrandom_audits} that tight mechanisms with non-random audits are debt-with-relief mechanisms. Further, these mechanisms differ only in terms of the face value $y_{0}$; they all feature the same relief and induce the same agent effort.
Namely, let $e_{A}^{\debtwrelief}=\argmin_{\tilde{e}_{A}\in [0, 1]} u_{A}(\tilde{e}_{A}) + c_{A}(\tilde{e}_{A}) + (1 - \tilde{e}_{A}) c_{P}(1)$ be the effort that minimizes total costs plus the evidence rent when the principal audits with probability one.
For types below the face value, the principal induces $e_{A}^{\debtwrelief}$ via a relief $r_{A}^{\debtwrelief} = c_{A}^{\prime}(e_{A}^{\debtwrelief})$, reducing the principal's audit costs to $(1 - e_{A}^{\debtwrelief})c_{P}(1)$, compared to $c_{P}(1)$ in a debt contract of \citet{townsend1979optimal}.
\begin{theorem}\label{thm:deterministic_tight_characterization}
    All tight mechanisms with non-random audits are debt-with-relief mechanisms with relief $r_{A}^{\debtwrelief} = c_{A}^{\prime}(e_{A}^{\debtwrelief})$.
\end{theorem}

\subsection{How to tighten a mechanism? An incentive programming approach}\label{sec:tight_mechanisms:how_to_tighten}

We sketch the proof strategy for \Cref{thm:deterministic_tight_characterization,thm:stochastic_tight_characterization} and how to operationalize tightness. This provides a template for other mechanism design problems with many instruments.

Every tight mechanism solves a particular fixed-point problem. 
Fix a tight mechanism $m$. We first show that $m$ induces a loss function $\lambda_{m}$ that is concave, increasing, and leaves no information rent to the lowest type. 
Then, we show that, given its induced loss function, $m$ cannot be improved by perturbations that only change the mechanism at a \emph{single} type. 

Formally, let $\Lambda$ be the set of concave, increasing functions $\lambda\colon [\ubar{x}, \bar{x}]\to [\ubar{x}, \bar{x}]$ such that $\lambda(x) \leq x$ for all $x$ and $\lambda(\ubar{x}) = \ubar{x}$.
For arbitrary $\lambda\in\Lambda$, we construct a mechanism $m$ as follows: separately for all $x$, choose the instruments $m(x) = (e_{A}(x), e_{P}(x), r_{A}(x), r_{P}(x), r_{\emptyset}(x)) \in [0, 1]^{2} \times [0, x + \tau]^{3}$ to maximize the profit $\Pi(x, m(x))$ at $x$ subject to:
\begin{subequations}\label{eq:Talgorithm:constraint}
\begin{align}
    \label{eq:Talgorithm:constraint:onpath}
    x - \lambda(x) &\leq e_{P}(x) r_{P}(x) + (1 - e_{P}(x))r_{\emptyset}(x) + u_{A}(e_{A}(x));
    \\
    \label{eq:Talgorithm:constraint:deviation}
    \forall z\in [x, \bar{x}],\quad \lambda(z) &\leq e_{P}(x) z +  (1 - e_{P}(x))(x - r_{\emptyset}(x));
    \\
    \label{eq:Talgorithm:constraint:obedience}
    c_{A}^{\prime}(e_{A}(x)) &= r_{A}(x) - \left(e_{P}(x) r_{P}(x) + (1 - e_{P}(x)) r_{\emptyset}(x)\right)
    .
\end{align}    
\end{subequations}
Let $T(x, \lambda)$ be the set of maximizers of this problem. 
Here, \eqref{eq:Talgorithm:constraint:onpath} says that type $x$ obtains the information rent $x - \lambda(x)$ if they are truthful;
\eqref{eq:Talgorithm:constraint:deviation} says that higher types $z$ lose at least $\lambda(z)$ from deviating to $x$; and \eqref{eq:Talgorithm:constraint:obedience} says that type $x$ obeys the effort recommendation $e_{A}(x)$. 

\begin{theorem}\label{thm:fixed_point}
    If $m$ is a tight mechanism, then $\lambda_{m}\in \Lambda$ and $m(x) \in T(x, \lambda_{m})$ for all $x \in [\ubar{x}, \bar{x}]$.
\end{theorem}
The proof is in \Cref{appendix:tight_mechanisms:preparations}.

\Cref{thm:fixed_point} shows that every tight mechanism $m$ solves a fixed-point problem analogous to a Bellman equation. The loss function $\lambda_{m}$ plays the role of a value function: it summarizes cross-type incentive constraints; analogously, a value function summarizes future payoff consequences of current choices. 
\Cref{thm:fixed_point} characterizes $m$ as solving a family of ``static'' problems $T(x, \lambda_{m})$, one per type $x$ given its own loss function $\lambda_{m}$; analogously, Bellman's principle characterizes optimal policies via a family of static problems given the value function.
Hence, we call our approach \emph{incentive programming}.
In contrast to dynamic programming, there is not a unique loss function and fixed-point to consider: there is a gap between tightness and optimality which we tackle in \Cref{sec:optimal_mechanisms}.

The regularity property $\lambda_{m} \in \Lambda$ is a consequence of tightness and characterizes the agent's deviation loss: $\lambda_{m}$ is increasing since higher types have more to lose; it is concave since the agent deviates to minimize their loss; the lowest type $\ubar{x}$ optimally requires zero information rents, $\lambda(\ubar{x}) = \ubar{x}$. 
Our proof shows how to profitably modify any mechanism to have a regular loss function, which is the non-trivial part of \Cref{thm:fixed_point}.\footnote{Regularity is not immediate from the definition \eqref{eq:loss_function_def} of the loss function since the agent cannot advance more than they own, i.e. deviate upwards (note the range of the infimum in \eqref{eq:loss_function_def}). In essence, our proof argues that tight mechanisms are immune to upward deviations.}

Our incentive programming approach breaks down the infinite-dimensional problem into a family of five-dimensional problems, one for each type $x$ with solution $T(x, \lambda_m)$. Still, this chain of separate subproblems yields structure on the entire mechanism: the proofs of \Cref{thm:stochastic_tight_characterization,thm:deterministic_tight_characterization} reduce to comparative statics of $T(x, \lambda_{m})$ as $x$ varies.
The regularity property $\lambda_{m}\in\Lambda$ yields single-crossing properties that lead to a monotone comparative statics analysis (\citet{milgrom1994monotone}).\footnote{We also obtain \Cref{fig:random_tight_mechanism_example} by solving the problem $T$. In this example, $[\ubar{x}, \bar{x}] = [0, 1]$, $\tau = 1.01$, and $c_{A}(e_{A}) = (\exp(e_{A}) - e_{A} - 1) / (\exp(1) - 1)$ and $c_{P}(e_{P}) = e_{P}^{1.01}/1.01$ for all $e_{A}, e_{P}\in [0, 1]$. We solved $T$ for $\lambda$ given by $\lambda(x) = 1 - \exp(-0.9x)$ for all $x\in[0, 1]$. The value $1.01$ in $c_{P}$ ensures that the interval $\iR$ is pronounced in \Cref{fig:random_tight_mechanism_example}. The value $0.9$ in $\lambda$ ensures that $e_{P}$ is interior at the lowest type. The value $\tau = 1.01$ (instead of $\tau = 1$) ensures that all assumptions from \Cref{sec:model} hold strictly.\label{footnote:computation}}

To derive the structure of tight mechanisms, note that in the type-wise maximization problem each type $x$ demands a certain information rent, as per \eqref{eq:Talgorithm:constraint:onpath}.
To understand how to optimally provide this rent using the various instruments, we adapt a \emph{virtual cost} approach.

\subsubsection{Intra-type virtual costs.}\label{sec:how_to_tighten:virtual_costs}
Fix $\lambda\in\Lambda$.
For each instrument, the virtual cost captures the associated marginal profit change in units of agent-utility when perturbing the instrument:
\begin{equation*}
    \mbox{virtual cost of instrument} = - \frac{\mbox{marginal profit wrt. instrument}}{\mbox{marginal utility wrt. instrument}}.
\end{equation*}
The direct effects are on the agent's utility (the right side of \eqref{eq:Talgorithm:constraint:onpath}) and the principal's profit.
Additionally, to keep \eqref{eq:Talgorithm:constraint:deviation} intact, $e_{P}(x)$ must be above the following minimal level\footnote{If $x = \bar{x}$, we define $e_{P}^{\min}(x, r_{\emptyset}(x)) = 0$ since no other type can deviate to $x$. For expositional simplicity, suppose that the supremum in $e_{P}^{\min}(x, r_{\emptyset}(x))$ is attained and that $e_{P}^{\min}(x, \cdot)$ is differentiable at $r_{\emptyset}(x)$. Our proofs do not make these assumptions. We show that for tight mechanisms the supremum is attained and that $e_{P}^{\min}$ is absolutely continuous; see \Cref{def:alpha_beta} (\Cref{appendix:tight_mechanisms:minimal_ep}) and the subsequent discussion.}
\begin{equation}\label{eq:minimal_eP} e_{P}(x)\geq 
    e_{P}^{\min}(x, r_{\emptyset}(x)) = \left[\sup_{z\in (x, \bar{x}]} \frac{\lambda(z) - x + r_{\emptyset}(x)}{z - x + r_{\emptyset}(x)}\right]_{+}
    .
\end{equation}
Finally, to keep obedience \eqref{eq:Talgorithm:constraint:obedience} intact, we will adjust $r_{A}(x)$ in all perturbations that we consider here. Since $r_{A}(x)$ is not directly relevant for utility or profit,\footnote{
We mean adapting $r_A(x)$ while using other refunds to hold $e_{A}(x)$ and therefore $u_{A}(e_{A}(x))$ fixed.} we effectively eliminate $r_{A}(x)$ and obedience \eqref{eq:Talgorithm:constraint:obedience}.
These additional perturbations, that serve to maintain \eqref{eq:Talgorithm:constraint:deviation} and \eqref{eq:Talgorithm:constraint:obedience}, also affect profit and utility.
The virtual costs will reflect the total effects.

For $e_{A}(x)$, $e_{P}(x)$, and $r_{P}(x)$ we obtain the virtual costs by simply evaluating the partial derivatives of profit and utility. Raising $e_{P}(x)$ slackens the constraint $e_{P}(x) \geq e_{P}^{\min}(x, r_{\emptyset}(x))$, and $e_{A}(x)$ and $r_{P}(x)$ are irrelevant for this constraint.
We obtain the following virtual costs:
\begin{align*}
    &\kappa_{e_A}(x)= - \frac{\partial_{\tilde e_A} \Pi(x, m(x))}{\partial_{\tilde e_A} U(x, m(x))}
        =
        1 + \frac{c_{A}^{\prime}(e_{A}(x)) - c_{P}(e_{P}(x))}{u_{A}^{\prime}(e_{A}(x))} ,
        \\
    &\kappa_{e_P}(x)= - \frac{\partial_{\tilde e_P} \Pi(x, m(x))}{\partial_{\tilde e_P} U(x, m(x))} = 1 + \frac{(1 - e_{A}(x)) c_{P}^{\prime}(e_{P}(x))}{r_{P}(x) - r_{\emptyset}(x)} ,
        \\
    &\kappa_{r_P}(x)= - \frac{\partial_{\tilde r_P} \Pi(x, m(x))}{\partial_{\tilde r_P} U(x, m(x))} = 1 .
\end{align*}
     Since $r_{\emptyset}(x)$ enters the minimal level $e_{P}^{\min}(x, r_{\emptyset}(x))$, we distinguish two cases. First, if $e_{P}(x)=e_{P}^{\min}(x, r_{\emptyset}(x))$, then $e_P(x)$ has to be adapted and we obtain as the virtual cost:
     \begin{align*}
         \kappa_{r_\emptyset}(x) = - \frac{\partial_{\tilde{r}_{\emptyset}} \Pi(x, m(x)) + \partial_{\tilde{e}_{P}}\Pi(x, m(x)) \partial_{\tilde{r}_{\emptyset}} e_{P}^{\min}(x, r_{\emptyset}(x))}{\partial_{\tilde{r}_{\emptyset}} U(x, m(x)) + \partial_{\tilde{e}_{P}}U(x, m(x)) \partial_{\tilde{r}_{\emptyset}} e_{P}^{\min}(x, r_{\emptyset}(x))}
         \\
         =
         1 +  \frac{(1 - e_{A}(x))c_{P}^{\prime}(e_{P}(x)) \partial_{\tilde{r}_{\emptyset}} e_{P}^{\min}(x, r_{\emptyset}(x))}{  1 - e_{P}(x) + (r_{P}(x) - r_{\emptyset}(x)) \partial_{\tilde{r}_{\emptyset}} e_{P}^{\min}(x, r_{\emptyset}(x))} 
         .
     \end{align*}
     One can show that the fraction in this expression is positive: the agent's utility increases in $r_{\emptyset}(x)$ even when the principal's effort $e_{P}(x)$ is adjusted.\footnote{Since $e_{P}^{\min}(x, \cdot)$ is differentiable at $r_{\emptyset}(x)$, Theorem 1 of \citet{milgrom2002envelope} yields $\partial_{\tilde{r}_{\emptyset}} e_{P}^{\min}(x, r_{\emptyset}(x)) = \frac{1 - e_{P}^{\min}(x, r_{\emptyset}(x))}{z - x + r_{\emptyset}(x)}$, where $z$ attains the supremum in \eqref{eq:minimal_eP}. Thus, $\partial_{\tilde{r}_{\emptyset}} U(x, m(x)) + \partial_{\tilde{e}_{P}}U(x, m(x)) \cdot \partial_{\tilde{r}_{\emptyset}} e_{P}^{\min}(x, r_{\emptyset}(x)) = (1 - e_{P}^{\min}(x, r_{\emptyset}(x)))\frac{z - x + r_{P}(x)}{z - x + r_{\emptyset}(x)} > 0$ since $e_{P}^{\min}(x, r_{\emptyset}(x)) = e_{P}(x) < 1$, since $z > x$, and since refunds are positive.}
     Second, if $e_{P}(x)>e_{P}^{\min}(x, r_{\emptyset}(x))$ we do not have to adapt $e_P(x)$ and we obtain:
     \begin{align*}
 \kappa_{r_\emptyset}(x) = - \frac{\partial_{\tilde{r}_{\emptyset}} \Pi(x, m(x))}{\partial_{\tilde{r}_{\emptyset}} U(x, m(x))}=1
     \end{align*}

Our notion of virtual costs differs from Myerson's \citeyearpar{myerson1981optimal} virtual value in two ways. Myerson expresses the virtual value in units of revenue. We instead compare the virtual costs of providing information rent and therefore express the virtual costs in units of utility. 
Second, Myerson's virtual value accounts for the costs of adjusting the mechanism at other types to keep their incentive constraints intact. Our virtual cost accounts for adjusting different instruments at the same type to keep both incentive and obedience constraints intact.

\subsubsection{Virtual cost comparison.}\label{sec:how_to_tighten:virtual_cost_comparison}
The virtual cost of each instrument specifies the cost of providing the agent with one util. The principal can substitute one instrument for another to provide type $x$'s information rent, i.e. to satisfy \eqref{eq:Talgorithm:constraint:onpath}. 
Next, we explain with the virtual costs why the principal's and the agent's efforts play fundamentally different roles.

\paragraph*{Who acquires evidence?}
We first observe that the principal optimally sets $r_{P}(x) \geq r_{\emptyset}(x)$.
The marginal virtual cost of $r_{\emptyset}(x)$ is always weakly higher than the marginal virtual cost of $r_{P}(x)$, i.e. $\kappa_{r_\emptyset}(x) \geq \kappa_{r_P}(x)$.
Therefore, the principal optimally uses the no-evidence refund, $r_\emptyset(x) > 0$, only if the principal-evidence refund is at its upper bound, $r_P(x) = x + \tau$.
Intuitively, while $r_{P}(x)$ is conditioned on evidence, $r_{\emptyset}(x)$ is an unconditional refund and attracts other types to deviate. 

We next argue that $e_{P}$ is used as a stick to deter deviations, but not as a carrot to provide the information rent.
Formally, $e_{P}(x)$ is set minimally: $e_P(x)=e_P^{\min}(x, r_{\emptyset}(x))$. 
Indeed, we just noted that the principal optimally sets $r_{P}(x) \geq r_{\emptyset}(x)$.
Thus, $e_{P}(x)$ could be used to provide information rents by increasing the probability of paying $r_{P}(x)$ instead of $r_{\emptyset}(x)$.
However, if $e_{P}(x) > e_{P}^{\min}(x, r_{\emptyset}(x))$, we have $\kappa_{r_{P}}(x) = \kappa_{r_{\emptyset}}(x) = 1$, whereas $\kappa_{e_{P}}(x) \geq 1$.
Therefore, the principal could reduce $e_{P}(x)$ (without upsetting $e_{P}(x) > e_{P}^{\min}(x, r_{\emptyset}(x))$) while increasing $r_{P}(x)$ or $r_{\emptyset}(x)$ to hold type $x$'s utility fixed but saving on costs.

In contrast, $e_{A}$ is used both to provide the information rent and to reduce audit costs.
The evidence rent $u_{A}(e_{A}(x))$ contributes to the information rent, the cost of which is reflected in the term $1$ in the expression for $\kappa_{e_{A}}(x)$.
Further, $e_{A}(x)$ may reduce total effort costs since the principal need not audit after the agent provides evidence that they were truthful. This effect is reflected in the term $c_{A}^{\prime}(e_{A}(x)) - c_{P}(e_{P}(x))$ and is positive if and only if $e_{A}(x)$ is above the efficient level $e_{A}^{\eff}(x)$.
Due to this effect, $\kappa_{e_{A}}(x)$ may be above or below $0$, $\kappa_{r_{P}}(x)$, or $\kappa_{r_{\emptyset}}(x)$.
Thus, $e_{A}(x)$ may be the preferred instrument for providing the information rent.

\begin{figure}[t!]
\centering
\includegraphics[width=\textwidth]{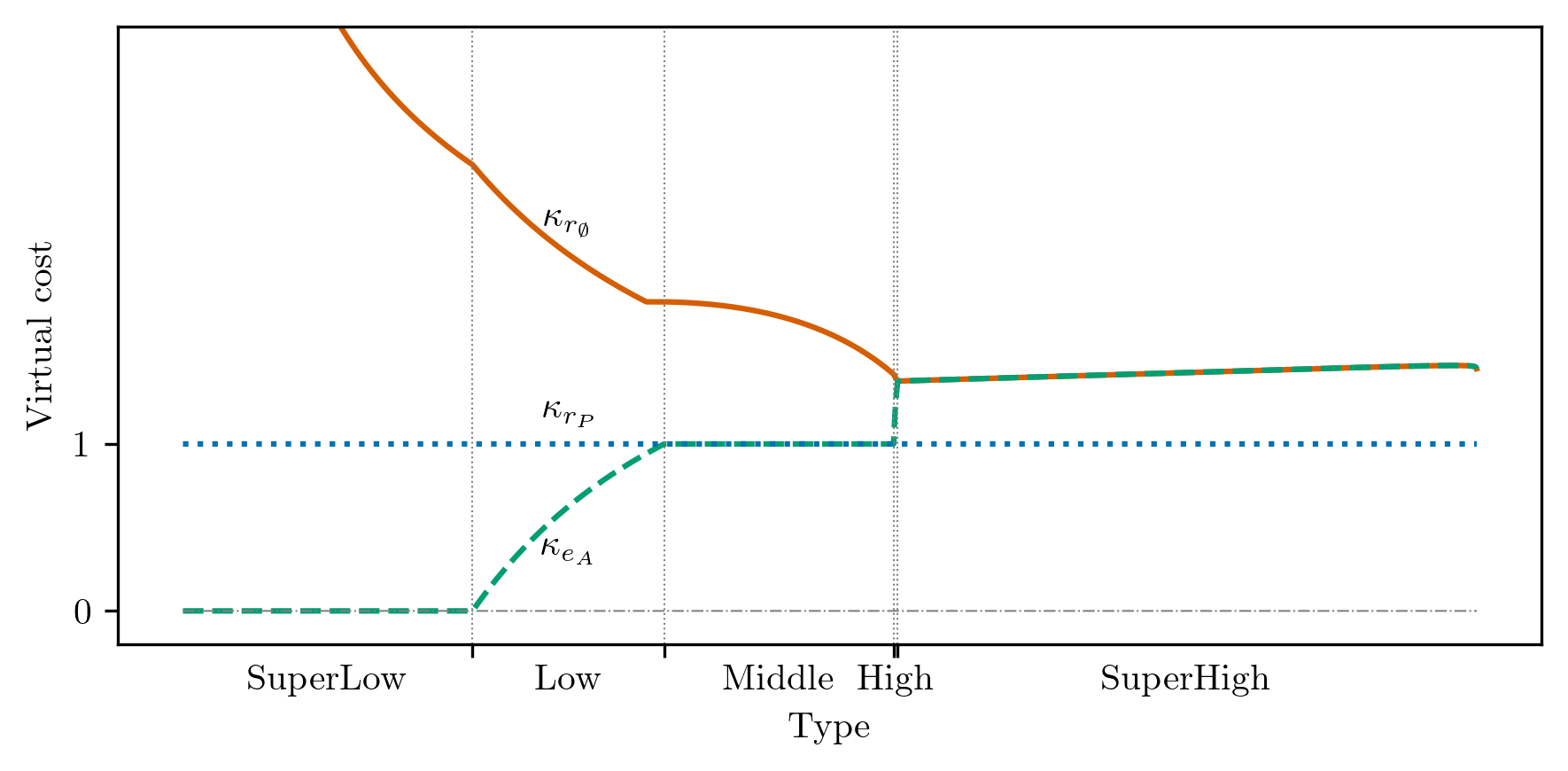}
\caption{The virtual costs in the mechanism from \Cref{fig:random_tight_mechanism_example}.}
\label{fig:virtual_costs}
\end{figure}

\paragraph{Virtual costs in action.}
Generally, comparing the virtual costs of $e_{A}$, $r_{P}$, and $r_{\emptyset}$ exactly yields the instrument use in tight mechanisms, as described by \Cref{thm:deterministic_tight_characterization,thm:stochastic_tight_characterization}.
\Cref{fig:virtual_costs} shows the virtual costs in the tight mechanism from \Cref{fig:random_tight_mechanism_example}. We have $\kappa_{e_{A}}(x) \leq \kappa_{r_{P}}(x) \leq \kappa_{r_{\emptyset}}(x)$ for all types $x$ in $\iSP$ and $\iP$; thus, the principal only uses $e_{A}(x)$ to provide the information rent to these types. When $e_{A}(x)$ reaches the efficient level for $x\in \iM$, we have $\kappa_{e_{A}}(x) = \kappa_{r_{P}}(x)$ and, hence, the principal starts using $r_{P}(x)$. For $x\in \iR$, the principal raises $e_{A}(x)$ since $\kappa_{e_{A}}(x)$ is still below $\kappa_{r_{\emptyset}}(x)$ while $r_{P}(x)$ cannot be raised further. For $x\in \iSR$, agent-effort $e_{A}(x)$ is so inefficiently high that $\kappa_{e_{A}}(x)$ and $\kappa_{r_{\emptyset}}(x)$ coincide and, hence, the principal also pays $r_{\emptyset}(x)$.
\Cref{assumption:regularity,assumption:rent_regularity}, respectively, ensure that $\kappa_{e_{A}}(x) - \kappa_{r_{\emptyset}}(x)$ and $\kappa_{e_{A}}(x)$, respectively, single-cross $0$ from below as functions of $e_{A}(x)$ and when $r_{\emptyset}(x)$ is used only if $r_{P}(x)$ is exhausted.\footnote{Precisely, for all types $x$ the auxiliary objectives $\pi_{1}(x, \tilde{e}_{A})$ and $\pi_{3}(x, \tilde{e}_{A})$ defined in \Cref{appendix:tight_mechanisms:minimal_ep} are strictly quasiconcave in agent-effort $\tilde{e}_{A}$; see \Cref{prop:pifunctions_characterization} in \Cref{appendix:tight_mechanisms:SCPs}.}

\subsubsection{Incentive programming in practice.}\label{sec:how_to_tighten:reform}
In practice, policymakers have to reform existing tax mechanisms rather than design mechanisms entirely anew. Also, they might find it difficult to solve a fixed point problem. 
Strikingly, in this reform problem, applying our re-optimization operator $T$ only once to any existing mechanism yields a tight mechanism with a higher profit.
Thus, the economic forces shaping tight mechanisms also guide the optimal direction of local reforms. 
\begin{theorem}\label{thm:reform}
    Let $m$ be a mechanism with loss function $\lambda_{m}\in \Lambda$.
    If $m^{\ast}$ satisfies $m^{\ast}(x) \in T(x, \lambda_{m})$ for all $x$, then $m^{\ast}$ is a (feasible IC) mechanism that is tight and $\Pi_{m^{\ast}} \geq \Pi_{m}$ holds.
\end{theorem}
\Cref{thm:reform} is the converse to \Cref{thm:fixed_point} and shows how to construct tight mechanisms.
The proof (\Cref{OA:how_to_tighten}) connects the type-wise maximization problem, which treats types in isolation, to tightness, which concerns the mechanism at all types.
We show that $m^{\ast}$ adheres to the structure derived for tight mechanisms (\Cref{thm:deterministic_tight_characterization,thm:stochastic_tight_characterization}). This structure yields the binding incentive-constraints, which we use to connect the type-wise problems across types.

\Cref{thm:reform} applies if the initial mechanism's loss function is regular ($\lambda_{m}\in \Lambda$). Every mechanism can be profitably modified so that its loss function is regular.
\Cref{OA:how_to_tighten} spells out a full algorithm transforming an arbitrary mechanism to a tight mechanism.

\section{Optimal mechanisms}\label{sec:optimal_mechanisms}

The expected profit of a mechanism $m$ is $\int \Pi_{m}(x) \de F(x)$.\footnote{Restricting attention to mechanisms with measurable profit has no substance since tight mechanisms are type-wise optimal (\Cref{lemma:tight_mechanisms_wlog}), and since all tight mechanisms have a measurable profit (\Cref{thm:deterministic_tight_characterization,thm:stochastic_tight_characterization}).}
A mechanism is \emph{optimal} if it maximizes the expected profit across all mechanisms.

There is an optimal mechanism that is tight, and all optimal mechanisms are essentially tight (\Cref{OA:trade-offs}).
But there is a gap between tightness and optimality. Optimal mechanisms trade off the audit costs at a type $y$ with the surplus that the principal extracts from types who contemplate deviating to $y$.
A mechanism can be tight without balancing out these trade-offs optimally since ``tighter than'' insists that profits increase at all types simultaneously. 

\subsection{Positive rents and random audits are optimal}\label{sec:optimal_mechanisms:random_optimal}

A first consequence of optimality is that the principal leaves a rent to all types and audits all types except the highest type randomly.
Thus, every tight optimal mechanism is characterized by \Cref{thm:stochastic_tight_characterization}, and the five intervals $\iSP, \ldots, \iSR$ of the mechanism cover $[\ubar{x}, \bar{x})$.
No debt-with-relief mechanism is optimal.

\begin{theorem}\label{thm:optimal_mech_fully_stochastic}
    Let $F$ have no mass point at $\bar x$.
    If $m$ is tight and optimal, then $\inf_{x\in[\ubar{x}, \bar{x}]} U_{m}(x) > 0$ and $\sup_{x\in[\ubar{x}, \bar{x}]} e_{P}(x) < 1$ and $e_{P}(x) > 0$ for all $x\in [\ubar{x}, \bar{x})$.
\end{theorem}

The proof (\Cref{appendix:thm:fully_stochastic}) first shows that the highest type $\bar{x}$ is optimally left with an information rent. Else, lower types have to be audited with certainty to deter $\bar{x}$. 
But since there is no mass point at $\bar{x}$, it is suboptimal to separate $\bar{x}$ from types just below $\bar{x}$ in this way.\footnote{For this step, it is important that $F$ has no mass point at $\bar{x}$. Suppose instead that types are binary. If the high type $\bar{x}$ is much larger and likelier than the low type, then the principal conceivably seizes everything from $\bar{x}$, even at the cost of auditing the low type with certainty.}
We then show that every type $x$ except $\bar{x}$ must optimally enjoy an evidence rent $u_{A}(e_{A}(x))$: indeed, since $c_{P}^{\prime}(0) = 0$, the principal optimally audits with non-zero probability $e_{P}(x)$ to deter and extract more from higher types who contemplate deviating to $x$;
since $c_{A}^{\prime}(0) = 0$, the principal also incentivizes non-zero agent effort $e_{A}(x)$ to reduce the costs $(1 - e_{A}(x)) c_{P}(e_{P}(x))$.
In summary, all types enjoy a rent when advancing the full surplus.
Hence, it is excessive for the principal to audit with certainty and threaten to seize everything.

\subsection{The trade-off across types}\label{sec:optimality:trade_off}

While tightness rules out profitable perturbations of the mechanism at each fixed type, optimal mechanisms must also be immune to profitable perturbations at multiple types. The trade-off across types in these perturbations depends on the binding incentive constraints: lowering the principal's effort at a type $y$ requires leaving more surplus to each higher type with a binding IC constraint to $y$.
Fix a mechanism $m$. For types $y$ and $x$ such that $y\leq x$, type $x$ is a \emph{binding IC type of $y$} if $x$'s best deviation is to $y$,
i.e. 
\begin{equation*}
\lambda_{m}(x) = e_{P}(y) x + (1 - e_{P}(y))(y - r_{\emptyset}(y)). 
\end{equation*}

For a moment, assume that the binding ICs are \emph{doubly unique}, meaning for each type $y \in (\ubar{x}, \bar{x})$ there is a unique binding IC type $\hat{x}(y)$, and type $y$ is the unique type who has $\hat{x}(y)$ as a binding IC type (unless $\hat{x}(y) = \bar{x}$). 
We will argue that then $y < \hat{x}(y) < \bar{x}$ holds.
For these mechanisms, we can use an envelope theorem to evaluate perturbations.

There are many perturbations of a mechanism that involve many types, such as changing the mechanism at $y$, $\hat x(y)$, $\hat x (\hat x(y)),$ etc. simultaneously to maintain incentives.
We next describe a particular perturbation that only involves two types $y$ and $\hat x(y)$ and explain in which sense this is exhaustive to establish optimality.\footnote{In the setting studied by \citet{border1987samurai}, where only the principal can acquire evidence, simultaneous perturbations at infinitely many types are necessary to characterize optimality.}

\paragraph*{The relevant perturbation: audit-the-poor-or-burden-the-rich.}

Fixing $y$, the principal increases agent effort $e_{A}(\hat{x}(y))$ (by increasing the refund $r_{A}(\hat{x}(y))$), thereby raising type $\hat{x}(y)$'s interim utility $U_{m}(\hat{x}(y))$ and reducing the incentive of $\hat{x}(y)$ to deviate to $y$.
Thus, the principal can decrease $e_{P}(y)$, perturbing some of $e_{A}(y)$, $r_{A}(y)$, $r_{P}(y)$, and $r_{\emptyset}(y)$ to hold type $y$'s interim utility $U_{m}(y)$ constant.
This perturbation trades off the costs of auditing the poor type $y$ with the costs of inducing the rich type $\hat x(y)$ to carry some of the burden of proof.
Under the assumption of double-uniqueness, the perturbation does not affect other types:
first, only $\hat{x}(y)$ can enjoy the refund $r_{A}(\hat{x}(y))$ for presenting evidence;
second, $e_{P}(y)$ has no first-order effect on types other than $\hat{x}(y)$ since $\hat{x}(y)$ is $y$'s unique binding IC type;
third, perturbing $\hat{x}(y)$'s interim utility has no first-order effect on $\hat{x}(y)$'s incentive to deviate to a type other than $y$ since $y$ is the unique type with $\hat{x}(y)$ as a binding IC type.

\paragraph*{Closing the gap between tightness and optimality.} We use the audit-the-poor-or-burden-the-rich perturbation to derive necessary first-order conditions for a tight mechanism with doubly unique binding ICs to be optimal (\Cref{OA:trade-offs}).
To argue that this perturbation is exhaustive, we provide a regularity condition on the environment under which the necessary condition is also sufficient for optimality, even without double uniqueness.
Intuitively, the perturbation trades off the profits at $y$ and $\hat{x}(y)$ using specific instruments ($e_{P}(y)$ and $r_{A}(\hat{x}(y))$, respectively), subject to IC.
Since the mechanism is tight, at each type the shadow costs of all instruments are equalized and the perturbation accounts for other ways of trading off profits at the respective types. 
Perturbations that affect more than two types can be decomposed into several perturbations of the above kind.
Therefore, the audit-the-poor-or-burden-the-rich perturbation suffices to close the gap between tightness and optimality.  
 
\paragraph*{Closing the gap approximately.}
There are tight mechanisms without doubly unique binding ICs.
However, we prove an approximation result.
For every optimal tight mechanism $m$, we construct a sequence of tight mechanisms $(m_{n})_{n\in\mathbb{N}}$ with doubly unique binding ICs such that the mechanism $m_{n}$ and the profit $\Pi_{m_{n}}$, respectively, converge pointwise on the interior $(\ubar{x}, \bar{x})$ of the type space to $m$ and $\Pi_{m}$, respectively, as $n\to\infty$.\footnote{Double uniqueness is related to the shape of the loss function $\lambda_{m}$, which is concave in a tight mechanism $m$. If $\lambda_{m}$ is not differentiable at a point $x$, then $x$ may be a binding IC type of an interval of types. If $\lambda_{m}$ is affine on a subinterval, then there may be a type having all types in the subinterval as binding IC types. We construct the described sequence of tight mechanisms by approximating $\lambda_{m}$ via differentiable, strictly concave functions, and applying the algorithm from \Cref{sec:tight_mechanisms:how_to_tighten}.}
Since every optimal mechanism is a limit of doubly unique IC mechanisms, all economically interpretable perturbations that shape the optimal mechanism can be understood via audit-the-poor-or-burden-the-rich perturbations and tightening perturbations.
\paragraph*{Optimal tax evasion.}
In each optimal tight mechanism with doubly unique binding ICs $\hat x$, for each type $x$ with a binding incentive to deviate, the optimal deviation is to advance only $\hat x^{-1}(x)$.\footnote{Types below $\sup\iSP$ have no binding incentive to deviate. For types $x$ in $(\sup\iSP, \bar{x})$, the inverse $\hat x^{-1}(x)$ exists, by double uniqueness. The best deviation for type $\bar{x}$ is infinitesimal, $\bar{x} = \hat{x}^{-1}(\bar{x})$, if the mechanism is optimal. This also implies that no interior type has $\bar{x}$ as a binding IC type.} We show that this deviation is continuous and strictly increasing.
Intuitively, since the principal seizes everything when detecting a deviation, the principal's effort $e_{P}$ more effectively deters high types $x$ who have more to lose. Since $e_{P}$ is decreasing, $\hat{x}^{-1}(x)$ is increasing in $x$. 
Further, the binding ICs are non-local: $\hat{x}^{-1}(x) < x$ if $x < \bar{x}$. Intuitively, since the agent risks everything by evading any amount of tax, worthwhile deviations must be to distant types where the agent saves a lot of surplus if undetected.

\section{Discussion}\label{sec:discussion}

\paragraph*{Funds for incentivizing effort.}
\Cref{assumption:budget} implies that the principal could but does not wish to incentivize the agent to acquire evidence with probability one.
Without this assumption, the virtual cost comparison is unchanged, but we would have to track additional boundary constraints, namely for when the principal exhausts their funds when incentivizing effort and for when incentivizing the agent to acquire evidence with probability one.

\paragraph*{Regularity assumptions on costs.}
\Cref{assumption:regularity} implies that the virtual costs of $e_{A}$ single-cross the virtual costs of $r_{\emptyset}$ on a relevant domain. 
This assumption is not needed to show that the instruments are optimally ordered as described by our tightness characterization, and that the principal substitutes $r_{\emptyset}$ for $e_{A}$ if the agent's effort is too high.
However, with multiple crossing-points, the characterization of $e_{A}$ and $r_{\emptyset}$ on $\iSR$ is more delicate.
For example, $e_{A}$ may conceivably decrease discontinuously at some types, causing $e_{P}$ to jump up. 

\Cref{assumption:rent_regularity} implies that the virtual costs of $e_{A}$ single-cross $0$ from below. As with \Cref{assumption:regularity}, we can thus rule out discontinuities in the agent-effort $e_{A}$ in the interval $\iSP$ of each tight mechanism.

\paragraph{Bounded funds $\tau$.}
In practice, all feasible transfers are bounded, $\tau < \infty$.
Moreover, by studying bounded transfers our insights are more easily portable to other mechanism design problems, such as problems without transfers that naturally admit a bound on the utility that the principal can provide to the agent.

If the principal has enormous private funds ($\tau\to\infty$), the interval $\iR$ vanishes. Intuitively, the benefit of using the evidence rent to provide rewards for truth-telling is small if the principal can pay a high reward $\tau$ after auditing the agent.
The interval $\iSR$ need not vanish, and the principal's effort is strictly positive but close to $0$ throughout this interval.

The analysis of the limit $\tau = \infty$ is delicate due to existence issues.
For certain types, the principal wishes to make a large transfer with vanishing probability after auditing.
An optimal mechanism fails to exist. (This does not mean that the principal can approximately extract the full surplus.)
Such existence issues are known in the literature. \citet{border1987samurai} also impose bounds on transfers.

\section{Related literature}\label{sec:related_literature}

We study the canonical surplus division model (\citet{townsend1979optimal}; \citet{border1987samurai}; \citet{chander1998general}) with voluntary evidence provision by the agent.\footnote{See also \citet{gale1985incentive,mookherjee1989optimal,monnet2005optimal,wang2005dynamic,ravikumar2012optimal,popov2016stochastic}.}
This hidden action changes how the principal approaches the original hidden information problem: the agent's evidence reduces the principal's audit costs, and the resulting hidden action rent substitutes for other instruments in providing the information rent.
As a concrete implication, consider that the corporate finance literature discusses Townsend's model as a rationale for debt but notices that debt can be suboptimal (e.g. \citet{tirole2010theory}).
\citet[p. 176]{chander1998general} also argue that deterministic mechanisms (debt) are not precluded from being optimal among possibly stochastic ones.
In contrast, we show that debt is never optimal.

The model of \citet{allingham1972income}, which is the basis of much empirical work on tax compliance, does not model evidence acquisition by the agent and tends to underpredict compliance (\citet{andreoni1998tax}).
Though our mechanism design approach departs from Allingham-Sandmo, we make a conceptual contribution by showing that the agent's rent from providing evidence strengthens the incentive to reveal the surplus. 
\citet{palonen2022mechanism} also study taxation with agent-effort, but in a setting where this effort reduces the chance of being detected by the principal's audit.

A growing literature studies entangled hidden information and hidden action problems with productive effort {\`a} la \citet{holmstrom1979moral}.\footnote{See \citet{castro2024disentangling,castropires2025howadverseselectionflexible,liu2025flexible,krahmer2025optimal,martimort2025optimal} and references therein.}
In our model, the agent's hidden action has no exogenous benefit; its benefit from revealing the surplus is endogenous to the mechanism.

The question of who should acquire costly evidence arises in many problems other than surplus division. One strand of the literature focuses on evidence acquisition by a principal with commitment power (verification).\footnote{E.g. \citet{hu2024screening,pham2024adverse,malenko2019optimal,erlanson2024optimalallocationscapacityconstrained,benporath2014,epitropou2019optimal,erlanson2019note,halac2020commitment,patel2022costly,ball2024optimalauctiondesigncontingent,kattwinkel2023costless,li2020mechanism,li2021mechanism,li2024dynamics,khalfan2023optimal,erlanson2020costly,kaplow2011optimal,kaplow2011burdens,beshkar2017cap,siegel2023judicial,chen2022informationdesignallocationcostly,khalfan2024sequential,brzustowski2024evidence,ball2023should,vravosinos2024bidimensional,ahmadzadeh2024mechanism}.}
Another strand focuses on mechanism design problems and games in which players without commitment can acquire or present evidence at a cost.\footnote{E.g. \citet{kartik2012implementation,bull2008costly,bull2008mechanism,verrecchia1983discretionary,jovanovic1982truthful,perezrichet2024scorebasedmechanisms}. In \citet{ben2023sequential,pram2023learningevidence,preusser2022evidence,asseyer2024certification,jiang2024sequentialevidence,madarasz2025cost,rappoport2017incentivizing,whitmeyer2022costlyevidencediscretionarydisclosure, siegelstrulovici2019pbs}, the agents also learn about a state.}
We combine the two strands to ask how to share the burden of proof.\footnote{\citet{ben2025evidence} survey work on evidence in games and mechanisms.} 
Only a few papers study counterparties that provide verifiable information.
\citet{ben2023sequential} study evidence acquisition by multiple agents; in contrast to us, evidence acquisition is deterministic and about distinct objects, and the principal cannot acquire their own evidence.  
\citet{bester2021signaling} study signaling when the firms without commitment can acquire evidence about the worker's type.
\citet{stahl2017certification} compare lemons markets with evidence acquisition either by buyers or by sellers, but not simultaneously.
\citet{lichtig2024optimal} allow the receiver in a disclosure game to publicly design a private signal.

\citet{ben2019mechanisms} relate a class of mechanism design problems with costly principal evidence to Dye-disclosure games. In contrast, we show that in our set-up costly agent evidence and principal evidence are not substitutable.

We also make a methodological contribution. Key to our approach of incentive programming is to define and fix a loss function and then re-optimize the mechanism type-wise.
A strength of this method is that the loss function can be defined in any principal-agent problem.
In the existing literature, the problem is often split into simpler subproblems (implicitly) by fixing one part of the mechanism while optimizing the rest type-wise.
For example, in settings with simple type dependence {\`a} la \citet{benporath2014}, the strategy is to optimize given fixed allocations of the worst-off types.\footnote{See \citet{erlanson2019note,erlanson2020costly,li2020mechanism,ben2019mechanisms}. In the single-agent version of \citet{benporath2014}, the loss function is a constant $\lambda$ since the agent's preferences are type-independent. When the principal does not wish to allocate to all agent-types, it is easy to characterize optimal mechanisms by first maximizing type-wise given $\lambda$, and then optimizing expected profits with respect to $\lambda$.}
\citet{patel2022costly} fix an increasing interim-expected allocation rule.
\citet{border1987samurai} optimize transfers given a fixed profile of audit probabilities.\footnote{In \citet{tightsamurai}, we show how our approach can reproduce \citet{border1987samurai}.} 
Our method offers a unified perspective on such type-wise optimization. Our virtual costs then allow for an intuitive solution method in the resulting type-wise subproblems even if they are complex.

In information design, \citet{candogan2023optimal} consider the problem of persuading multiple privately informed receivers. They provide a decomposition into multiple subproblems that is reminiscent of the fixed-point property of our tight mechanisms.

\section{Concluding remarks}

We study the interplay between evidence acquisition by the principal and the agent in the canonical model of surplus division. 
We approach this multi-instrumental mechanism design problem by introducing tightness and intra-type virtual costs.
Essentially, our approach asks how the principal's cost of providing the agent with utility compares across instruments.
This is a fundamental question in any mechanism design problem with many instruments. We therefore believe that tightness and intra-type virtual costs are useful more broadly.

\appendix 
\crefalias{section}{appendix}
\crefalias{subsection}{appendix}
\crefalias{subsubsection}{appendix}

\section{Proofs}\label{appendix:proofs}

\subsection{Revelation Principle}\label{appendix:revelation_principle}

Following \citet{myerson1982optimal}, a general mechanism has the following timing.  
\begin{enumerate}
    \item The agent reports the type (surplus) via a cheap talk message.
    \item The principal recommends an advance payment and an agent effort.
    \item The agent advances a payment, and then covertly exerts effort to acquire evidence (recall that evidence can only be acquired after an advance payment has been made). If the agent obtains evidence, the agent chooses whether to disclose it.
    \item The principal exerts effort and then implements a transfer that is constrained by the advance payment and the evidence (as described in \Cref{sec:model}).
\end{enumerate}
Additionally, the agent finds it optimal to report the type truthfully and exert the recommended effort.
Myerson's Revelation Principle captures any grand mechanism featuring multiple rounds of cheap-talk. Our only assumptions on the evidence technology are that the agent can only acquire evidence after advancing a payment,\footnote{If the agent can acquire evidence about the full surplus and the principal can ask the agent for the advance payment at a time when the agent did not yet learn whether they succeeded in acquiring evidence, then the principal optimally first asks for the advance payment since waiting for the agent's evidence acquisition only adds incentive constraints.} and that both agent and principal can try to acquire evidence only once. 
In such a grand mechanism, the agent may exert effort randomly following some randomized cheap talk.
The recommended agent effort following a report $x$ then replicates the ex-ante probability that type $x$ acquires evidence in the grand mechanism in equilibrium, and the principal's effort following report $x$ is the ex-ante probability that the principal acquires evidence when the agent is type $x$. 

We now argue that it is without loss for the principal to use a feasible incentive compatible tax mechanism.
First, whenever the agent discloses evidence, the principal optimally does not exert effort to acquire evidence; indeed, the principal can already condition on the true surplus; thus, by possibly re-defining transfers, the principal can reduce their own effort costs.
Second, rather than sending a cheap-talk message about the type, the principal demands the agent advance their type as payment; the principal commits to treating this payment as if the agent had reported the payment as their type; since the agent found it optimal to report truthfully, the agent now finds it optimal to advance the type.
Third, if there is evidence that the agent did not advance the full surplus, the principal optimally seizes everything since doing so only strengthens the incentive to advance the full surplus.
Similarly, if the agent advances strictly less than the lowest type $\ubar{x}$, the agent could not have advanced the full surplus, and hence the principal optimally acquires evidence with certainty.
We obtain a feasible incentive compatible tax mechanism.

\subsection{Evidence rent}\label{appendix:no_excessive_evidence}
To justify the expression for the evidence rent from \Cref{sec:preliminaries}, we show that the principal does not give the agent a strict incentive to acquire evidence with certainty.
\begin{lemma}\label{lemma:no_excessive_evidence}
    Let $\tilde{m}$ be a mechanism.
    There is a mechanism $m$ such that $\Pi_{\tilde{m}} \leq \Pi_{m}$ and such that $e_{A}(x)(r_{A}(x) -  R_{NA}(x)) - c_{A}(e_{A}(x)) = u_{A}(e_{A}(x))$ for all $x\in [\ubar{x}, \bar{x}]$.
\end{lemma}

\begin{pf}[Proof of \Cref{lemma:no_excessive_evidence}]
    It suffices to find a mechanism $m$ such that $\Pi_{\tilde{m}}= \Pi_{m}$ and $ r_{A}(x) - R_{NA}(x) \leq c_{A}^{\prime}(1)$ for all $x$; the inequality implies that $e_{A}(x)$ satisfies the first-order condition $c_{A}^{\prime}(e_{A}(x)) = r_{A}(x) - R_{NA}(x)$, yielding $e_{A}(x) (r_{A}(x) - R_{NA}(x)) - c_{A}(e_{A}(x)) = u_{A}(e_{A}(x))$.
    If $x$ is a type such that $\tilde{r}_{A}(x) - \tilde{R}_{NA}(x) > c_{A}^{\prime}(1)$, then type $x$ acquires evidence with certainty: $\tilde{e}_{A}(x) = 1$.
    Given $\tilde{e}_{A}(x) = 1$, the principal never expects themselves to acquire evidence on-path if facing type $x$.
    Thus, the principal may as well set $e_{P}(x) = 1$ and set $r_{P}(x) = \tilde{r}_{A}(x) - c_{A}^{\prime}(1)$. 
    This choice of $r_{P}(x)$ is feasible since $x+\tau\geq \tilde{r}_{A}(x) - c_{A}^{\prime}(1) > 0$ holds.
    Modify the mechanism in this way at all types $x$ such that $\tilde{r}_{A}(x) - \tilde{R}_{NA}(x) > c_{A}^{\prime}(1)$, leaving all other parts of the mechanism unchanged.
    The modification raises for all types the loss from deviating, but weakly raises the principal's profits and the agent's payoff from advancing the full surplus.
\end{pf}

\subsection{Preparations to characterize tight mechanisms}\label{appendix:tight_mechanisms:preparations}

We begin with the maximization problem from \Cref{sec:tight_mechanisms:how_to_tighten} (restated in \Cref{appendix:tight_mechanisms:pointwise_maximization} below).
To characterize tightness, we shall study the comparative statics of this problem with respect to the type.
The next subsections simplify the problem.
Along the way, we characterize the curvature of loss functions induced by tight mechanisms and prove \Cref{thm:fixed_point}.
Next, in \Cref{appendix:tight_mechanisms:SCPs}, we develop crucial single-crossing properties on the effect of agent-effort on profit.
Finally, the proof of \Cref{thm:stochastic_tight_characterization} follows in \Cref{appendix:tight_mechanisms:random_audits}.

\subsubsection{Pointwise maximization.}\label{appendix:tight_mechanisms:pointwise_maximization}
Let $\Lambda_{0}$ denote the set of functions $\lambda\colon [\ubar{x}, \bar{x}] \to [-\tau, \bar{x}]$ such that $\lambda\leq\id$.
For every mechanism $m$, the loss function $\lambda_{m}$ is in $\Lambda_{0}$.

    Given $y\in [\ubar{x}, \bar{x}]$ and $\lambda\in\Lambda_{0}$, let $M(y, \lambda)$ be the set of tuples 
    $(\tilde{e}_{A}, \tilde{e}_{P}, \tilde{r}_{A}, \tilde{r}_{P}, \tilde{r}_{\emptyset}) \in [0, 1]^{2}\times [0, y + \tau]^{3}$
    such that all of the following hold:
    \begin{subequations}
    \begin{align}
        \label{eq:def:virtual_problem:on_path_utility}
        \lambda(y) &\geq y - \left(\tilde{e}_{P} \tilde{r}_{P} + (1 - \tilde{e}_{P}) \tilde{r}_{\emptyset}\right) - u_{A}(\tilde{e}_{A});
        \\
        \label{eq:def:virtual_problem:relaxing_lambda}
        \forall x\in [y, \bar{x}],\quad \lambda(x) &\leq \tilde{e}_{P} x + (1 - \tilde{e}_{P}) (y - \tilde{r}_{\emptyset});
        \\
        \label{eq:def:virtual_problem:feasible_eA}
        c_{A}^{\prime}(\tilde{e}_{A}) &= \tilde{r}_{A} - \left(\tilde{e}_{P} \tilde{r}_{P} + (1 - \tilde{e}_{P}) \tilde{r}_{\emptyset}\right)
        .
    \end{align}
    \end{subequations}
    Denote the profit from such a tuple by
    \begin{equation*}
        \Pi(y, \tilde{e}_{A}, \tilde{e}_{P}, \tilde{r}_{A}, \tilde{r}_{P}, \tilde{r}_{\emptyset}) =  y - \tilde{e}_{P} \tilde{r}_{P} - (1 - \tilde{e}_{P})\tilde{r}_{\emptyset} 
        - u_{A}(\tilde{e}_{A}) - c_{A}(\tilde{e}_{A}) - (1 - \tilde{e}_{A})c_{P}(\tilde{e}_{P}).
    \end{equation*}
    Finally, let $T(y, \lambda) = \argmax_{m(y) \in M(y, \lambda)} \Pi(y, m(y))$.\footnote{For all $y$ and $\lambda\in\Lambda_{0}$, the set $M(y, \lambda)$ is non-empty; e.g., let $\tilde{e}_{P} = 1$, $\tilde{e}_{A} = 0$ and $\tilde{r}_{P} = \tilde{r}_{A} = y + \tau$ (and $\tilde{r}_{\emptyset}$ arbitrary). By compactness and continuity, $T(y, \lambda)$ is non-empty.}
    We write $m\in T(\lambda)$ (resp. $m\in M(\lambda)$) to mean $m(y) \in T(y, \lambda)$ (resp. $m(y)\in M(y, \lambda)$) for all $y$.

We first verify that applying $T$ tightens a mechanism.
\begin{lemma}\label{lemma:tight_operator_improvement}
    If $m$ is a mechanism and $m^{\ast}\in T(\lambda_{m})$, then $m^{\ast}$ is a mechanism and tighter than $m$, i.e. $(\Pi_{m}, \lambda_{m}) \leq (\Pi_{m^{\ast}}, \lambda_{m^{\ast}})$.
    If $\lambda\in\Lambda_{0}$ and $m^{\ast} \in T(\lambda)$, then $\lambda \leq \lambda_{m^{\ast}}$.
\end{lemma}

\begin{pf}[Proof of \Cref{lemma:tight_operator_improvement}]
    We observe $m(y)\in M(y, \lambda_{m})$ for all $y$.
    Indeed, \eqref{eq:def:virtual_problem:on_path_utility} says that $m$ is IC for type $y$, \eqref{eq:def:virtual_problem:relaxing_lambda} follows from the definition of $\lambda_{m}$, and \eqref{eq:def:virtual_problem:feasible_eA} says that type $y$ finds the agent effort optimal.
    If $m^{\ast}\in T(\lambda_{m})$, then under $m^{\ast}$ all types $y$ lose at most $\lambda_{m}(y)$ from being truthful, but at least $\lambda_{m}(y)$ from deviating.
    Hence, $m^{\ast}$ is feasible, IC, and induces a loss function $\lambda_{m^{\ast}}$ which is pointwise above $\lambda_{m}$.
    The profit under $m^{\ast}$ is also pointwise higher than the profit under $m$ since $m\in M(\lambda_{m})$.
    Thus, $m^{\ast}$ is tighter than $m$.
    The second claim is analogous.
\end{pf}

\subsubsection{Optimal refunds.}\label{appendix:tight_mechanisms:optimal_refunds}
We relate $T$ to another maximization problem by optimizing the refunds $\tilde{r}_{P}$ and $\tilde{r}_{\emptyset}$, following the virtual-cost heuristic developed in \Cref{sec:how_to_tighten:virtual_cost_comparison}.
Fix a type $y$. Since the no-evidence refund $\tilde{r}_{\emptyset}$ attracts other types to deviate to $y$, the refund $\tilde{r}$ should be non-zero only if $\tilde{r}_{P}$ equals its upper bound $y + \tau$.
The refunds should be minimal subject to $y$'s receiving their information rent.
We obtain a maximization problem involving only the efforts.

    For all $y\in [\ubar{x}, \bar{x}]$ and $\lambda \in \Lambda_{0}$, let $E(y, \lambda)$ be the set of $(\tilde{e}_{A}, \tilde{e}_{P}) \in [0, 1]^{2}$ such that
    \begin{equation}
        \forall x\in [y, \bar{x}],\quad
        \lambda(x) \leq \tilde{e}_{P} x + \min\left\lbrace (1 - \tilde{e}_{P}) y, \lambda(y) + u_{A}(\tilde{e}_{A}) + \tilde{e}_{P}\tau\right\rbrace.
        \label{eq:relaxed_operator:IC}
    \end{equation}
    For $(\tilde{e}_{A}, \tilde{e}_{P}) \in E(y, \lambda)$, let 
    \begin{equation}\label{eq:relaxed_operator:objective}
        \hat{\Pi}(y, \lambda(y), \tilde{e}_{A}, \tilde{e}_{P})
        =\min\left\lbrace y - u_{A}(\tilde{e}_{A}), \lambda(y)\right\rbrace - c_{A}(\tilde{e}_{A}) - (1 - \tilde{e}_{A}) c_{P}(\tilde{e}_{P}).
    \end{equation}
    Finally, let $\hat{T}(y, \lambda)= \argmax_{(\tilde{e}_{A}, \tilde{e}_{P}) \in E(y, \lambda)} \hat{\Pi}(y, \lambda(y), \tilde{e}_{A}, \tilde{e}_{P})$.    

The minimum in the profit \eqref{eq:relaxed_operator:objective} will be given by $\lambda(y)$ when type $y$ has a binding incentive to advance the full surplus.
The minimum in \eqref{eq:relaxed_operator:IC} will be given by $\lambda(y) + u_{A}(\tilde{e}_{A}) + \tilde{e}_{P}\tau$ if $r_{P}(y) = y+\tau$.
These facts are made precise in the proof of the next lemma, which establishes an equivalence between $\hat{T}$ and $T$.

\begin{lemma}\label{lemma:relaxed_operator_wlog}
    Let $y\in [\ubar{x}, \bar{x}]$ and $\lambda\in\Lambda_{0}$.
    If $(\tilde{e}_{A}, \tilde{e}_{P}, \tilde{r}_{A}, \tilde{r}_{P}, \tilde{r}_{\emptyset})\in M(y, \lambda)$, then $(\tilde{e}_{A}, \tilde{e}_{P})\in E(y, \lambda)$ and $\Pi(y, \tilde{e}_{A}, \tilde{e}_{P}, \tilde{r}_{A}, \tilde{r}_{P}, \tilde{r}_{\emptyset}) \leq \hat{\Pi}(y, \lambda(y), \tilde{e}_{A}, \tilde{e}_{P})$.
    Conversely, if $\lambda \geq 0$ and $(\tilde{e}_{A}, \tilde{e}_{P})\in E(y, \lambda)$, then there are $\tilde{r}_{A}, \tilde{r}_{P}, \tilde{r}_{\emptyset}$ such that $(\tilde{e}_{A}, \tilde{e}_{P}, \tilde{r}_{A}, \tilde{r}_{P}, \tilde{r}_{\emptyset})\in M(y, \lambda)$ and $\Pi(y, \tilde{e}_{A}, \tilde{e}_{P}, \tilde{r}_{A}, \tilde{r}_{P}, \tilde{r}_{\emptyset}) = \hat{\Pi}(y, \lambda(y), \tilde{e}_{A}, \tilde{e}_{P})$.
    Finally, if $\lambda \geq 0$ and $m\in T(\lambda)$, then for all $y\in [\ubar{x}, \bar{x}]$,
    \begin{subequations}\label{lemma:relaxed_operator_wlog:formulae}
    \begin{align}
        \label{lemma:relaxed_operator_wlog:formulae:profit}
        \Pi(y, m(y)) &= \hat{\Pi}(y, \lambda(y), e_{A}(y), e_{P}(y));
        \\
        \label{lemma:relaxed_operator_wlog:formulae:utility}
        U_{m}(y) &= \max\lbrace u_{A}(e_{A}(y)), y - \lambda(y)\rbrace.
    \end{align}
    \end{subequations}
\end{lemma}
This lemma also implies that if $\lambda \geq 0$, then for all $y$ and $(\tilde{e}_{A}, \tilde{e}_{P})\in [0, 1]^{2}$, there are $\tilde{r}_{A}, \tilde{r}_{P}, \tilde{r}_{\emptyset}$ such that $(\tilde{e}_{A}, \tilde{e}_{P}, \tilde{r}_{A}, \tilde{r}_{P}, \tilde{r}_{\emptyset})\in T(y, \lambda)$ if and only if $(\tilde{e}_{A}, \tilde{e}_{P})\in\hat{T}(y, \lambda)$.
In this sense, $T$ and $\hat{T}$ are equivalent once we know it suffices to consider $\lambda \geq 0$, as we shall argue later.
\begin{pf}[Proof of \Cref{lemma:relaxed_operator_wlog}]
    First, for all $(\tilde{e}_{A}, \tilde{e}_{P}, \tilde{r}_{A}, \tilde{r}_{P}, \tilde{r}_{\emptyset}) \in M(y, \lambda)$, we verify $(\tilde{e}_{A}, \tilde{e}_{P})\in E(y, \lambda)$ and $\Pi(y, \tilde{e}_{A}, \tilde{e}_{P}, \tilde{r}_{A}, \tilde{r}_{P}, \tilde{r}_{\emptyset}) \leq \hat{\Pi}(y, \lambda(y), \tilde{e}_{A}, \tilde{e}_{P})$.
    Specifically,  
    \eqref{eq:def:virtual_problem:on_path_utility} and $\tilde{r}_{A}, \tilde{r}_{P} \geq 0$ imply $\Pi(y, \tilde{e}_{A}, \tilde{e}_{P}, \tilde{r}_{A}, \tilde{r}_{P}, \tilde{r}_{\emptyset}) \leq \hat{\Pi}(y, \lambda(y), \tilde{e}_{A}, \tilde{e}_{P})$.
    Next, \eqref{eq:def:virtual_problem:on_path_utility} and $\tilde{r}_{P} \leq y + \tau$ imply $(1 - \tilde{e}_{P})(y - \check{r}_{\emptyset}) \leq \lambda(y) + u_{A}(\tilde{e}_{A}) + \tilde{e}_{P}\tau$.
    This inequality, $\tilde{r}_{\emptyset} \geq 0$, and \eqref{eq:def:virtual_problem:relaxing_lambda} imply \eqref{eq:relaxed_operator:IC}; that is, $(\tilde{e}_{A}, \tilde{e}_{P}) \in E(y, \lambda)$.

    Conversely, if $(\tilde{e}_{A}, \tilde{e}_{P}) \in E(y, \lambda)$ and $\lambda \geq 0$, let $\tilde{r}_{\emptyset}$ solve 
    $(1 - \tilde{e}_{P})(y - \tilde{r}_{\emptyset}) = \min\lbrace (1 - \tilde{e}_{P})y, \lambda(y) + u_{A}(\tilde{e}_{A}) + \tilde{e}_{P}\tau\rbrace $,
    let $\tilde{r}_{P}$ solve $\tilde{e}_{P}(y - \tilde{r}_{P}) + (1 - \tilde{e}_{P})(y - \tilde{r}_{\emptyset}) = \min\lbrace y, \lambda(y) + u_{A}(\tilde{e}_{A})\rbrace$, and let $\tilde{r}_{A}$ solve $c_{A}^{\prime}(\tilde{e}_{A}) = \tilde{r}_{A} - \left(\tilde{e}_{P} \tilde{r}_{P} + (1 - \tilde{e}_{P})\tilde{r}_{\emptyset}\right)$.
    By inspection, $\tilde{r}_{P}$ and $\tilde{r}_{\emptyset}$ are in $[0, y + \tau]$.
    \Cref{assumption:budget} and $\lambda \geq 0$ imply $\tilde{r}_{A} \in [0, y + \tau]$.
    Finally, $\Pi(y, \tilde{e}_{A}, \tilde{e}_{P}, \tilde{r}_{A}, \tilde{r}_{P}, \tilde{r}_{\emptyset}) = \hat{\Pi}(y, \lambda(y), \tilde{e}_{A}, \tilde{e}_{P})$ by construction.

    The formulae for the profit and the agent's utility when $m\in T(\lambda)$ and $\lambda \geq 0$ follow from the choice of the refunds in the previous paragraph; this argument uses that since the efforts are held fixed the value of the profit pins down the value of the agent's utility.
\end{pf}

\subsubsection{Shape of the loss function.}\label{appendix:tight_mechanisms:shape_of_loss}
Let $\Lambda$ be the set of increasing concave $\lambda\colon[\ubar{x}, \bar{x}]\to\mathbb{R}$ such that $\lambda(\ubar{x}) = \ubar{x}$ and $\lambda\leq\id$.

By inspection, all $\lambda\in\Lambda$ are Lipschitz-continuous with constant $1$.
The difference $y - \lambda(y)$ is increasing but the ratio $\frac{\lambda(y)}{y}$ is decreasing in $y$ for $y > 0$.

The next lemma will be used to show $\lambda_{m}\in \Lambda$ for all tight mechanisms $m$.
\begin{lemma}\label{lemma:lambda_regularization}
    Let $m$ be a mechanism, and let $\lambda_{m}$ be its loss function. 
    For all $x\in[\ubar{x}, \bar{x}]$, let $\lambda^{+}(x) = \max\lbrace \ubar{x}, \sup_{x^{\prime}\in [\ubar{x}, x]} \lambda_{m}(x^{\prime})\rbrace$, and
    \begin{equation}\label{lemma:virtual_loss_definition}
        \tilde{\lambda}(x) = \inf_{y\in[\ubar{x}, \bar{x}]} e_{P}(y) x + \min\left\lbrace(1 - e_{P}(y))y, \lambda^{+}(y) + u_{A}(e_{A}(y)) + e_{P}(y)\tau\right\rbrace.
    \end{equation}
    Then $\tilde{\lambda}\in\Lambda$ and all $x\in [\ubar{x}, \bar{x}]$ satisfy
    $(e_{A}(x), e_{P}(x))\in E(x, \tilde{\lambda})$, $\lambda_{m}(x)\leq\tilde{\lambda}(x)$, and $\Pi_{m}(x) \leq \hat{\Pi}(x, \lambda^{+}(x), e_{A}(x), e_{P}(x)) \leq \hat{\Pi}(x, \tilde{\lambda}(x), e_{A}(x), e_{P}(x))$.
\end{lemma}

\begin{pf}[Proof of \Cref{lemma:lambda_regularization}]
    To verify $\tilde{\lambda}\in\Lambda$, note that $\tilde{\lambda}$ is increasing and concave as $\tilde{\lambda}$ is the pointwise infimum of increasing affine functions.
    Finally, $\tilde{\lambda}(\ubar{x}) = \ubar{x}$ and $\tilde{\lambda}\leq\id$ follow from inspecting \eqref{lemma:virtual_loss_definition} and using $\lambda^{+}\leq\id$ and $\lambda^{+}(\ubar{x}) = \ubar{x}$.
    
    In an intermediate step, we show $\lambda^{+}(x) \leq \tilde{\lambda}(x)$ for all $x$.
    By inspection, $\ubar{x}\leq\tilde{\lambda}(x)$.
    Thus, we show $\sup_{x^{\prime}\in [\ubar{x}, x]} \lambda_{m}(x^{\prime}) \leq \tilde{\lambda}(x)$.
    Let $x^{\prime}\in[\ubar{x}, x]$ and $y\in [\ubar{x}, \bar{x}]$ be arbitrary.
    First, if $x^{\prime} \leq y$, then since $\lambda^{+}$ is increasing and $\lambda_{m}(x^{\prime}) \leq \lambda^{+}(x^{\prime}) \leq x^{\prime}$ holds, we find 
    \begin{align*}
        \lambda_{m}(x^{\prime}) \leq \lambda^{+}(x^{\prime}) &\leq e_{P}(y) x^{\prime} + \min\left\lbrace (1  - e_{P}(y))x^{\prime}, \lambda^{+}(x^{\prime})\right\rbrace
        \\
        &\leq e_{P}(y) x + \min\left\lbrace (1  - e_{P}(y))y, \lambda^{+}(y) + u_{A}(e_{A}(y)) + e_{P}(y)\tau\right\rbrace.
    \end{align*}    
    Second, let $y \leq x^{\prime}$. Since $m$ is a (feasible IC) mechanism, it holds $m(y)\in M(y, \lambda_{m})$. By \Cref{lemma:relaxed_operator_wlog}, thus $(e_{A}(y), e_{P}(y)) \in E(y, \lambda_{m})$. Using also $\lambda_{m}(y) \leq \lambda^{+}(y)$, we find 
    \begin{align*}
        \lambda_{m}(x^{\prime}) &\leq e_{P}(y) x^{\prime} + \min\left\lbrace (1 - e_{P}(y))y, \lambda_{m}(y) + u_{A}(e_{A}(y)) + e_{P}(y)\tau\right\rbrace 
        \\ & \leq e_{P}(y) x + \min\left\lbrace (1 - e_{P}(y))y, \lambda^{+}(y) + u_{A}(e_{A}(y)) + e_{P}(y)\tau\right\rbrace,
    \end{align*}
    Since $x^{\prime}$ and $y$ were arbitrary, we infer $\lambda^{+}(x) \leq \tilde{\lambda}(x)$ for all $x$.

    From the definition of $\tilde{\lambda}$ and the inequality $\lambda^{+}\leq\tilde{\lambda}$, we infer $(e_{A}, e_{P})\in E(\tilde{\lambda})$.
    Finally, for all $x$, \Cref{lemma:tight_operator_improvement,lemma:relaxed_operator_wlog} imply $\Pi_{m}(x) \leq \hat{\Pi}(x, \lambda_{m}(x), e_{A}(x), e_{P}(x))$, and $\lambda_{m}\leq\lambda^{+}\leq\tilde{\lambda}$ implies $\hat{\Pi}(x, \lambda_{m}(x), e_{A}(x), e_{P}(x)) \leq \hat{\Pi}(x, \tilde{\lambda}(x), e_{A}(x), e_{P}(x))$.
\end{pf}

\subsubsection{Regularity and type-wise best-replies: \headercref{Theorem}{{thm:fixed_point}}.}\label{appendix:tight_mechanisms:comparative_statics_approach}
We prove a slightly stronger claim than \Cref{thm:fixed_point} that also spells out the profit and the agent's utility.
\begin{theorem}\label{cor:tight_mechs_are_fps}
    If $m$ is tight, then $\lambda_{m}\in\Lambda$, $m\in T(\lambda_{m})$, and all $x\in[\ubar{x}, \bar{x}]$ satisfy
    \begin{align*}
        \Pi_{m}(x) = &\min\lbrace x - u_{A}(e_{A}(x)), \lambda_{m}(x)\rbrace - c_{A}(e_{A}(x)) - (1 - e_{A}(x))c_{P}(e_{P}(x)),
        \\
        U_{m}(x) = &\max\lbrace u_{A}(e_{A}(x)), x-\lambda_{m}(x)\rbrace.
    \end{align*}
\end{theorem}

\begin{pf}[Proof of \Cref{cor:tight_mechs_are_fps}]
    Invoking \Cref{lemma:lambda_regularization}, find $\tilde{\lambda} \in\Lambda$ such that $(e_{A}, e_{P})\in E(\tilde{\lambda})$, $\lambda_{m}\leq\tilde{\lambda}$, and $\Pi_{m}(x) \leq \hat{\Pi}(x, \tilde{\lambda}(x), e_{A}(x), e_{P}(x))$ for all $x$.
    Note, $\tilde{\lambda}\in \Lambda$ implies $\tilde{\lambda}\geq\ubar{x}\geq 0$.
    Since also $(e_{A}, e_{P}) \in E(\tilde{\lambda})$, we may invoke \Cref{lemma:relaxed_operator_wlog} to find a mechanism $m^{\ast} \in T(\tilde{\lambda})$ such that $\hat{\Pi}(x, \tilde{\lambda}(x), e_{A}(x), e_{P}(x)) \leq \Pi_{m^{\ast}}(x)$ for all $x$.
    Since $m^{\ast}\in T(\tilde{\lambda})$, \Cref{lemma:tight_operator_improvement} also implies $\tilde{\lambda} \leq \lambda_{m^{\ast}}$.
    Summarizing, we find $\Pi_{m} \leq \Pi_{m^{\ast}}$ and $\lambda_{m}\leq \tilde{\lambda} \leq \lambda_{m^{\ast}}$.
    Since $m$ is tight, we conclude $\Pi_{m} = \Pi_{m^{\ast}}$ and $\lambda_{m}= \tilde{\lambda} = \lambda_{m^{\ast}}$.
    Thus, $\lambda_{m}\in\Lambda$.
    Further, $\Pi_{m} = \Pi_{m^{\ast}}$ and $m^{\ast}\in  T(\tilde{\lambda})$ and $\lambda_{m} = \tilde{\lambda}$ together imply $m\in T(\lambda_{m})$.
    The formulae for $\Pi_{m}$ and $U_{m}$ follow from \Cref{lemma:relaxed_operator_wlog}.
\end{pf}

\subsubsection{Minimal principal effort.}\label{appendix:tight_mechanisms:minimal_ep}

To study the comparative statics, we further simplify the maximization problem by eliminating the principal's effort: it is set as small as possible, following the virtual-cost heuristic from \Cref{sec:how_to_tighten:virtual_cost_comparison}. 
We obtain an objective that depends only on the agent's effort, as summarized in \Cref{prop:minimal_principal_effort} below.

For all $y\in [\ubar{x}, \bar{x}]$ and $\lambda\in\Lambda$ and $\tilde{e}_{A}\in [0, 1]$, the smallest feasible principal effort $\tilde{e}_{P}$ such that $(\tilde{e}_{A}, \tilde{e}_{P}) \in E(y, \lambda)$ will be $\alpha(y)$ or $\beta(y, \tilde{e}_{A})$, defined as follows.
\begin{definition}\label{def:alpha_beta}
    Let $\lambda\in\Lambda$.
    For all $y\in [\ubar{x}, \bar{x}]$, $\tilde{e}_{A}\in [0, 1]$, and $s\in\mathbb{R}$, let
    \begin{align*}
        \alpha(y) &= 
        \begin{cases}
           \max\left\lbrace0, \sup_{x\in (y, \bar{x}]} \frac{\lambda(x) - y}{x - y}\right\rbrace, &\text{if $y < \bar{x}$},\\
           0, &\text{if $y = \bar{x}$};
        \end{cases}
        \\
        \beta(y, \tilde{e}_{A}) &= \max\left\lbrace0, \max_{x\in [\ubar{x}, \bar{x}]} \frac{\lambda(x) - \lambda(y) - u_{A}(\tilde{e}_{A})}{x + \tau}\right\rbrace;
        \\
        \hat{Z}(s) &= \argmax_{x\in [\ubar{x}, \bar{x}]}\frac{\lambda(x) - s}{x + \tau}.
    \end{align*}
    Let $\hat{z}$ be an increasing selection from $\hat{Z}$.\footnote{For example, for all $s$ let $\hat{z}(s) = \max \hat{Z}(s)$. The objective defining $\hat{Z}$ is supermodular in $(x, s)$.}
\end{definition}
Since $\lambda$ is continuous, $\beta$ is well-defined and continuous, and $\hat{Z}$ has non-empty compact values.
Further, $\lambda\leq\id$ and $0 < \tau$ imply $\beta(y, \tilde{e}_{A}) < 1$.

For tight mechanisms, the efforts $\alpha(y)$ and $\beta(y, e_{A}(y))$ correspond to the minimal principal-effort described in \cref{eq:minimal_eP}.
Namely, the minimal effort is $\alpha(y)$ when $r_{\emptyset}(y) = 0$; it is $\beta(y, \tilde{e}_{A})$ when $r_{P}(y) = y + \tau$ and $r_{\emptyset}(y)$ is chosen to provide type $y$'s missing information rent.
These cases are exhaustive when the refunds are chosen optimally.

We will distinguish multiple regimes depending on whether the principal's effort equals $\alpha(y)$ or $\beta(y, \tilde{e}_{A})$, and on whether the minimum in the objective \eqref{eq:relaxed_operator:objective} is given by $y - u_{A}(\tilde{e}_{A})$ or $\lambda(y)$.
It turns out that $y - u_{A}(\tilde{e}_{A}) < \lambda(y)$ holds only if $\alpha(y) \geq \beta(y, \tilde{e}_{A})$ holds.\footnote{
Specifically, $\alpha(y)$ is the smallest $\tilde{e}_{P}\in [0, 1]$ such that $\lambda(x) \leq \tilde{e}_{P} x + (1 - \tilde{e}_{P})y$ for all $x\in [y, \bar{x}]$, while $\beta(y, \tilde{e}_{A})$ is the smallest $\tilde{e}_{P}\in [0, 1]$ such that $\lambda(x) \leq \tilde{e}_{P} x + \lambda(y) + u_{A}(\tilde{e}_{A}) + \tilde{e}_{P}\tau$ for all $x\in [y, \bar{x}]$.
Thus, $\alpha(y) < \beta(y)$ only if $(1 - \alpha(y))y > \lambda(y) + u_{A}(\tilde{e}_{A}) + \alpha(y)\tau$, which implies $y \geq \lambda(y) + u_{A}(\tilde{e}_{A})$.}
Intuitively, the minimum equals $\lambda(y)$ if type $y$ has a binding incentive to be truthful. The principal's effort $e_{P}(y)$ equals $\alpha(y)$ whenever the no-evidence refund $r_{\emptyset}(y)$ equals $0$. If $y - u_{A}(\tilde{e}_{A}) < \lambda(y)$, then type $y$ has a strict incentive, meaning the principal should decrease $r_{\emptyset}(y)$ to $0$ since this tightens incentives and raises profits. 

Thus, we distinguish three intersecting regimes: first, $y - u_{A}(\tilde{e}_{A}) \leq \lambda(y)$; second, both $y - u_{A}(\tilde{e}_{A}) \geq  \lambda(y)$ and $\alpha(y) \geq \beta(y, \tilde{e}_{A})$; third, $\alpha(y) \leq \beta(y, \tilde{e}_{A})$.
The respective profits are:
\begin{definition}\label{def:auxiliary_profit functions}
Let $\lambda\in\Lambda$, $y\in [\ubar{x}, \bar{x}]$, and $\tilde{e}_{A}\in [0, 1]$.
Let
\begin{align*}
    \pi_{1}(y, \tilde{e}_{A}) 
    &= y - u_{A}(\tilde{e}_{A}) - c_{A}(\tilde{e}_{A}) - (1 - \tilde{e}_{A}) c_{P}(\alpha(y));
    \\
    \pi_{2}(y, \tilde{e}_{A}) 
    &= \lambda(y) - c_{A}(\tilde{e}_{A}) - (1 - \tilde{e}_{A}) c_{P}(\alpha(y));
    \\
    \pi_{3}(y, \tilde{e}_{A}) 
    &= \lambda(y) - c_{A}(\tilde{e}_{A}) - (1 - \tilde{e}_{A}) c_{P}(\beta(y, \tilde{e}_{A}));
    \\
    \ubar{\pi}(y, \tilde{e}_{A}) 
    &= \min_{i\in\lbrace 1, 2, 3\rbrace} \pi_{i}(y, \tilde{e}_{A}).
\end{align*}
\end{definition}

For $i\in\lbrace 1, 2 \rbrace$, let $\partial_{2}\pi_{i}$ denote the partial derivative of $\pi_{i}$ with respect to its second argument (agent effort).
For fixed $y$, the function $\tilde{e}_{A}\mapsto \pi_{3}(y, \tilde{e}_{A})$ is absolutely continuous\footnote{We use Theorem 2 of \citet{milgrom2002envelope}, which applies since the assumption $0 < \ubar{x} + \tau$ ensures that the objective in the definition of $\beta$ has a bounded derivative with respect to $\tilde{e}_{A}$.} and $\partial_{2} \pi_{3}(y,  \tilde{e}_{A})$ defined as follows is a derivative:
\begin{equation*}
    \partial_{2} \pi_{3}(y, \tilde{e}_{A}) = c_{P}(\beta(y, \tilde{e}_{A})) - c_{A}^{\prime}(\tilde{e}_{A})  
    + (1 - \tilde{e}_{A}) \frac{c_{P}^{\prime}(\beta(y, \tilde{e}_{A})) u_{A}^{\prime}(\tilde{e}_{A})}{\hat{z}(\lambda(y) + u_{A}(\tilde{e}_{A}))) + \tau}\bm{1}(\beta(y, \tilde{e}_{A})) > 0)
    .
\end{equation*}
The next proposition essentially obtains by choosing $e_{P}$ as small as possible given $e_{A}$.
\begin{proposition}\label{prop:minimal_principal_effort}
    Let $\lambda\in\Lambda$.
    If $m \in T(\lambda)$, then all $y\in [\ubar{x}, \bar{x}]$ satisfy $e_{A}(y) < 1$ and $\Pi_{m}(y) = \ubar{\pi}(y, e_{A}(y)) = \max_{\tilde{e}_{A}\in [0, 1]} \ubar{\pi}(y, \tilde{e}_{A})$ and $e_{P}(y) = \max\lbrace\alpha(y), \beta(y, e_{A}(y))\rbrace$.
\end{proposition}

\begin{pf}[Proof of \Cref{prop:minimal_principal_effort}]
    Let $y\in [\ubar{x}, \bar{x}]$.
    \Cref{lemma:relaxed_operator_wlog} implies $(e_{A}(y), e_{P}(y)) \in \hat{T}(y, \lambda)$, that is $(e_{A}(y), e_{P}(y))$ maximizes \eqref{eq:relaxed_operator:objective} subject to \eqref{eq:relaxed_operator:IC}.
    The objective \eqref{eq:relaxed_operator:objective} is decreasing in $\tilde{e}_{P}$, strictly so if $\tilde{e}_{A} < 1$.
    Fixing $\tilde{e}_{A}$, the minimal value of $\tilde{e}_{P}$ satisfying \eqref{eq:relaxed_operator:IC} is $\max\lbrace\alpha(y), \beta(y, \tilde{e}_{A})\rbrace$; this step uses that $y - u_{A}(\tilde{e}_{A}) < \lambda(y)$ holds only if $\alpha(y) \geq \beta(y, \tilde{e}_{A})$ holds.
    The resulting profit is $\ubar{\pi}(y, \tilde{e}_{A})$.
    Thus, it suffices to show that $e_{A}(y) = 1$ cannot be optimal for maximizing \eqref{eq:relaxed_operator:objective} subject to \eqref{eq:relaxed_operator:IC}.
    By inspection, for all $i\in\lbrace 1, 2, 3\rbrace$, the derivative of $\pi_{i}(y, \tilde{e}_{A})$ with respect to $\tilde{e}_{A}$ at $\tilde{e}_{A}=1$ is bounded above by $-u_{A}^{\prime}(1)\bm{1}_{i=1}-c_{A}^{\prime}(1) + c_{P}(1)$.
    Since $c_{A}^{\prime}(1) > c_{P}(1)$ (\Cref{assumption:budget}) and $u_{A}$ is increasing, the derivative is strictly negative.
    Thus, $e_{A}(y) = 1$ cannot be optimal.
\end{pf}

\subsubsection{Monotone comparative statics.}\label{appendix:tight_mechanisms:SCPs}

Fixing $\tilde{e}_{A}$, the profit $\pi_{i}(y, \tilde{e}_{A})$ depends on the regime $i$ and the type $y$ because the two roles of agent-effort, i.e. providing information rent and reducing costs, are more or less important depending on $y$ and which other constraints are binding (the regime $i$).
We next derive crucial single-crossing properties for the profit $\pi_{1}$, $\pi_{2}$, and $\pi_{3}$, namely \Cref{prop:pifunctions_characterization,prop:supermodularity_across_regimes} below. These properties permit a monotone comparative statics analysis (\citet{milgrom1994monotone}).

The next lemma uses \Cref{assumption:regularity}.
\begin{lemma}\label{lemma:K3_auxiliary_order}
Let $\lambda\in\Lambda$, $y, y^{\prime} \in [\ubar{x}, \bar{x}]$, $e_{A}, e_{A}^{\prime} \in [0, 1]$
and $\lambda(y) + u_{A}(e_{A}) \leq \lambda(y^{\prime}) + u_{A}(e_{A}^{\prime})$.
If $e_{A} \leq e_{A}^{\prime}$, then $\partial_{2} \pi_{3}(y, e_{A}) \geq \partial_{2} \pi_{3}(y^{\prime}, e_{A}^{\prime})$. 
If $e_{A} < e_{A}^{\prime}$, then $\partial_{2} \pi_{3}(y, e_{A}) > \partial_{2} \pi_{3}(y^{\prime}, e_{A}^{\prime})$.  
\end{lemma}

\begin{pf}[Proof of \Cref{lemma:K3_auxiliary_order}]
    Let $q = \beta(y, e_{A})$ and $q^{\prime} = \beta(y^{\prime}, e_{A}^{\prime})$.
    Inspecting the definitions of $\beta$ and $\hat{z}$,  it holds $q \geq q^{\prime}$ and $\hat{z}(\lambda(y) + u_{A}(e_{A})) \leq \hat{z}(\lambda(y^{\prime}) + u_{A}(e_{A}^{\prime}))$.
    Thus,
    \begin{align*}
        &\partial_{2} \pi_{3}(y^{\prime}, e_{A}^{\prime}) = c_{P}(q^{\prime}) - c_{A}^{\prime}(e_{A}^{\prime}) + (1 - e_{A}^{\prime}) \frac{c_{P}^{\prime}(q^{\prime}) u_{A}^{\prime}(e_{A}^{\prime})}{\hat{z}(\lambda(y^{\prime}) + u_{A}(e_{A}^{\prime})) + \tau} \bm{1}(q^{\prime} > 0),
        \\
        & \partial_{2} \pi_{3}(y, e_{A}) 
        \geq 
        c_{P}(q^{\prime}) - c_{A}^{\prime}(e_{A}) + (1 - e_{A}) \frac{c_{P}^{\prime}(q^{\prime}) u_{A}^{\prime}(e_{A})}{\hat{z}(\lambda(y^{\prime}) + u_{A}(e_{A}^{\prime})) + \tau} \bm{1}(q^{\prime} > 0)
        .
    \end{align*}
    If $e_{A} = e_{A}^{\prime}$, then clearly $\partial_{2} \pi_{3}(y, e_{A}) \geq \partial_{2} \pi_{3}(y^{\prime}, e_{A}^{\prime})$.
    It remains to show that if $e_{A} < e_{A}^{\prime}$, then $\partial_{2} \pi_{3}(y, e_{A}) > \partial_{2} \pi_{3}(y^{\prime}, e_{A}^{\prime})$.
    To that end, it suffices to show that
        $\tilde{e}_{A}\mapsto - c_{A}^{\prime}(\tilde{e}_{A}) + (1 - \tilde{e}_{A}) \frac{c_{P}^{\prime}(q^{\prime}) u_{A}^{\prime}(\tilde{e}_{A})}{\hat{z}(\lambda(y^{\prime}) + u_{A}(e_{A}^{\prime})) + \tau} \bm{1}(q^{\prime} > 0) $
    is strictly decreasing.
    If $q^{\prime} = 0$, we are done since $c_{A}$ is strictly convex.
    Let $q^{\prime} > 0$.
    We differentiate to obtain:
    \begin{equation*}
        - c_{A}^{\prime\prime}(\tilde{e}_{A}) + \frac{c_{P}^{\prime}(q^{\prime})}{\hat{z}(\lambda(y^{\prime}) + u_{A}(e_{A}^{\prime})) + \tau}\left((1 - \tilde{e}_{A})u_{A}^{\prime\prime}(\tilde{e}_{A}) - u_{A}^{\prime}(\tilde{e}_{A})\right)
        .
    \end{equation*}
    If $(1 - \tilde{e}_{A})u_{A}^{\prime\prime}(\tilde{e}_{A}) \leq u_{A}^{\prime}(\tilde{e}_{A})$, we are done. If $(1 - \tilde{e}_{A})u_{A}^{\prime\prime}(\tilde{e}_{A}) \geq u_{A}^{\prime}(\tilde{e}_{A})$, it suffices to check $
        c_{A}^{\prime\prime}(\tilde{e}_{A}) > \frac{c_{P}^{\prime}(1)}{\ubar{x} + \tau}\left((1 - \tilde{e}_{A})u_{A}^{\prime\prime}(\tilde{e}_{A}) - u_{A}^{\prime}(\tilde{e}_{A})\right)$. This inequality holds by \Cref{assumption:regularity}.
\end{pf}

\begin{proposition}\label{prop:pifunctions_characterization}
    Let $\lambda\in\Lambda$.
    For all $i\in\lbrace 1, 2, 3\rbrace$, the function $\pi_{i}$ is submodular, and, for all $y\in[\ubar{x}, \bar{x}]$, the functions $\pi_{i}(y, \cdot)$ and $\ubar{\pi}(y, \cdot)$ are strictly quasiconcave. 
\end{proposition}

\begin{pf}[Proof of \Cref{prop:pifunctions_characterization}]
    \Cref{assumption:rent_regularity} implies that $u_{A} + c_{A}$ is strictly quasiconvex, implying that $\pi_{1}(y, \cdot)$ is strictly quasiconcave for all $y$.
    Clearly, $\pi_{2}(y, \cdot)$ is strictly concave for all $y$.
    Finally, $\pi_{1}$ and $\pi_{2}$ are submodular since $\alpha$ decreases.
    
    The second claim in \Cref{lemma:K3_auxiliary_order} implies that $\tilde{e}_{A}\mapsto \partial_{2} \pi_{3}(y, \tilde{e}_{A})$ is strictly decreasing for all $y$; thus, $\pi_{3}(y, \cdot)$ is strictly quasiconcave.
    Next, recall that $\lambda$ is increasing.
    Hence, for all $\tilde{e}_{A}\in [0, 1]$ and $y, y^{\prime}\in [\ubar{x}, \bar{x}]$, if $y < y^{\prime}$, then $\lambda(y) + u_{A}(e_{A}) \leq \lambda(y^{\prime}) + u_{A}(e_{A})$, and hence \Cref{lemma:K3_auxiliary_order} implies $\partial_{2} \pi_{3}(y, e_{A}) \geq \partial_{2} \pi_{3}(y^{\prime}, e_{A})$.
    Hence, $\pi_{3}$ is submodular.

    Finally, the function $\ubar{\pi}(y, \cdot)$ is strictly quasiconcave for each $y$ since it is the pointwise minimum of strictly quasiconcave functions.
\end{pf}

The next lemma essentially says that the benefit from incentivizing agent effort is highest when profit is given by $\pi_{3}$ and lowest when profit is given by $\pi_{1}$.
\begin{proposition}\label{prop:supermodularity_across_regimes}
    Let $\lambda\in\Lambda$.
    Let $y\in [\ubar{x}, \bar{x}]$ and $0\leq e_{A} < e_{A}^{\prime} \leq 1$.
    \begin{enumerate}
        \item If $\pi_{1}(y, e_{A}^{\prime}) \geq (>)\, \pi_{1}(y, e_{A})$, then $\pi_{2}(y, e_{A}^{\prime}) \geq (>)\, \pi_{2}(y, e_{A})$.
        \item If $\alpha(y) \leq \beta(y, e_{A}^{\prime})$ and $\pi_{2}(y, e_{A}^{\prime}) \geq (>)\, \pi_{2}(y, e_{A})$, then $\pi_{3}(y, e_{A}^{\prime}) \geq (>)\, \pi_{3}(y, e_{A})$.
    \end{enumerate}
\end{proposition}

\begin{pf}[Proof of \Cref{prop:supermodularity_across_regimes}]
    For all $\tilde{e}_{A}\in [0, 1]$, it holds $\partial_{2} \pi_{1}(y, \tilde{e}_{A}) = c_{P}(\alpha(y)) - c_{A}^{\prime}(\tilde{e}_{A}) - u_{A}^{\prime}(\tilde{e}_{A})
    \leq
    c_{P}(\alpha(y)) - c_{A}^{\prime}(\tilde{e}_{A}) = \partial_{2} \pi_{2}(y, \tilde{e}_{A})$.
    This proves claim (1).

    Next, let $\alpha(y) \leq \beta(y, e_{A}^{\prime})$. 
    Thus also $\alpha(y) \leq \beta(y, \tilde{e}_{A})$ for all $\tilde{e}_{A} \in [e_{A}, e_{A}^{\prime}]$ since $\beta$ is decreasing in its second argument.
    Thus, $\partial_{2} \pi_{3}(y, \tilde{e}_{A}) \geq c_{P}(\beta(y, \tilde{e}_{A})) - c_{A}^{\prime}(\tilde{e}_{A}) \geq c_{P}(\alpha(y)) - c_{A}^{\prime}(\tilde{e}_{A}) = \partial_{2} \pi_{2}(y, \tilde{e}_{A})$, proving claim (2).
\end{pf}

\subsubsection{Continuity.}
Given $\lambda\in\Lambda$, define $\ubar{y} = \max\lbrace y\in [\ubar{x}, \bar{x}]\colon \lambda(y) = y\rbrace$ and $\bar{y} = \min\lbrace y\in [\ubar{x}, \bar{x}]\colon \lambda(y) = \max\lambda\rbrace$.
The set $\lbrace y\in [\ubar{x}, \bar{x}]\colon \lambda(y) = y\rbrace$ is a closed interval containing $\ubar{x}$.
The next lemma shows that if $\alpha$ is interior at least once, then $\alpha$, $e_{A}$ and $e_{P}$ are continuous except possibly at $\ubar{y}$.
\begin{lemma}\label{lemma:basic_continuity}
    Let $\lambda\in\Lambda$ and $m\in T(\lambda)$.
    If $\lbrace y\in [\ubar{x}, \bar{x}]\colon \alpha(y)\in (0, 1)\rbrace$ is non-empty, then $\ubar{y} < \lambda(\bar{x}) \leq \bar{y}$ and for all $y\in [\ubar{x}, \bar{x}]$,
    \begin{enumerate}
        \item if $y\in [\ubar{x}, \ubar{y})$, then $\alpha(y) = 1$;
        \item if $y\in (\ubar{y}, \bar{x}]$, then $\alpha(y) < 1$, and $\alpha$, $e_{A}$ and $e_{P}$ are continuous at $y$;
        \item $\alpha$, $e_{A}$, and $e_{P}$ are right-continuous at $\ubar{y}$;
        \item $\alpha$ is strictly decreasing on $(\ubar{y}, \lambda(\bar{x}))$, and constantly $0$ on $[\lambda(\bar{x}),  \bar{x}]$.
    \end{enumerate}
\end{lemma}

\begin{pf}[Proof of \Cref{lemma:basic_continuity}]
    The properties of $\alpha$ and the inequality $\ubar{y} < \lambda(\bar{x}) \leq \bar{y}$ may be verified directly from the definitions of $\ubar{y}$, $\bar{y}$, and $\alpha$, and the fact that $\lambda$ is increasing, concave, and satisfies $\lambda\leq\id$ and $\lambda(\ubar{x}) = \ubar{x}$.
    Next, since $\alpha$ is continuous on $(\ubar{y}, \bar{x}]$, the profit $\ubar{\pi}(y, \tilde{e}_{A})$ is continuous in $(y, \tilde{e}_{A})$ for $y\in (\ubar{y}, \bar{x}]$.
    Since $e_{A}(y)$ maximizes $\ubar{\pi}(y, \cdot)$ (\Cref{prop:minimal_principal_effort}) and $\ubar{\pi}(y, \cdot)$ is strictly quasiconcave (\Cref{prop:pifunctions_characterization}), Berge's Maximum Theorem implies that $e_{A}$ is continuous.
    \Cref{prop:minimal_principal_effort} now implies that $e_{P}$ is continuous on $(\ubar{y}, \bar{x}]$.
    Right-continuity of $e_{A}$ and $e_{P}$ at $\ubar{y}$ follow from a similar argument.
\end{pf}

\subsection{Proof of \headercref{Theorem}{{thm:stochastic_tight_characterization}}}\label{appendix:tight_mechanisms:random_audits}
We will use definitions and results from \Cref{appendix:tight_mechanisms:preparations}.

We characterize mechanisms $m$ with random audits for which there exists $\lambda\in\Lambda$ such that $m\in T(\lambda)$, which will imply \Cref{thm:stochastic_tight_characterization}. The added generality will be useful to prove \Cref{thm:reform}.
For reference, recall $\Lambda$ is defined as the set of increasing concave $\lambda\colon[\ubar{x}, \bar{x}]\to\mathbb{R}$ such that $\lambda(\ubar{x}) = \ubar{x}$ and $\lambda\leq\id$.

Fix $\lambda\in\Lambda$ and $m\in T(\lambda)$.
We next define the candidates for the five sets described in \Cref{sec:tight_mechanisms}.
The sets distinguish whether the profit $\ubar{\pi}$ is given by $\pi_{1}$, $\pi_{2}$ or $\pi_{3}$.
The following functions $d_{1}$ and $d_{2}$ govern these cases.
For all  $y\in [\ubar{x}, \bar{x}]$ and $\tilde{e}_{A}\in [0, 1]$, let
\begin{align*}
    d_{1}(y, \tilde{e}_{A}) &= y - \lambda(y) - u_{A}(\tilde{e}_{A}); 
    \\
    d_{2}(y, \tilde{e}_{A}) &= y - \lambda(y) - u_{A}(\tilde{e}_{A}) - \alpha(y)(y + \tau).
\end{align*}
Both $d_{1}(y, \tilde{e}_{A})$ and $d_{2}(y, \tilde{e}_{A})$ are decreasing in $\tilde{e}_{A}$ but increasing in $y$ (since $y - \lambda(y)$ is increasing in $y$, while $\alpha$ is decreasing) and $d_{2}(y, \tilde{e}_{A}) \leq d_{1}(y, \tilde{e}_{A})$ holds.
The following are verified by straightforward manipulations: for all $y, \tilde{e}_{A}$,
\begin{align}\label{eq:constraint_SCP}
    \begin{aligned}
    d_{1}(y, \tilde{e}_{A}) \leq 0 \quad&\Leftrightarrow\quad \ubar{\pi}(y, \tilde{e}_{A}) = \pi_{1}(y, \tilde{e}_{A});
    \\
    d_{2}(y, \tilde{e}_{A}) \leq 0 \leq d_{1}(y, \tilde{e}_{A}) \quad&\Leftrightarrow\quad \ubar{\pi}(y, \tilde{e}_{A}) = \pi_{2}(y, \tilde{e}_{A});
    \\
    0\leq d_{2}(y, \tilde{e}_{A}) \quad&\Leftrightarrow\quad \ubar{\pi}(y, \tilde{e}_{A}) = \pi_{3}(y, \tilde{e}_{A}).
    \end{aligned}
\end{align}

We shall derive the five-intervallic structure using the monotonicity properties of $d_{1}$ and $d_{2}$, coupled with the single-crossing properties of $\pi_{1}$, $\pi_{2}$, and $\pi_{3}$ (\Cref{prop:pifunctions_characterization,prop:supermodularity_across_regimes}).
Specifically, the five candidate sets are as follows:
\begin{align}\label{eq:candidate_five_intervals}
\begin{aligned}
    Y^{\circ} &= \lbrace y\in [\ubar{x}, \bar{x}] \colon e_{P}(y) \in (0, 1)\rbrace,
    \\
    \iSP &= \lbrace y\in Y^{\circ}\colon d_{1}(y, e_{A}(y)) < 0\rbrace,
    \\
    \iP &= \lbrace y\in Y^{\circ}\colon d_{2}(y, e_{A}(y)) < 0 = d_{1}(y, e_{A}(y))\rbrace,
    \\
    \iM &= \lbrace y\in Y^{\circ}\colon d_{2}(y, e_{A}(y)) < 0 < d_{1}(y, e_{A}(y))\rbrace,
    \\
    \iR &= \lbrace y\in Y^{\circ}\colon d_{2}(y, e_{A}(y)) = 0 < d_{1}(y, e_{A}(y)) \rbrace,
    \\
    \iSR &= \lbrace y\in Y^{\circ}\colon 0 < d_{2}(y, e_{A}(y))\rbrace.
\end{aligned}
\end{align}
The sets $\iSP, \ldots, \iSR$ partition $Y^{\circ}$ since $d_{2} \leq d_{1}$.
At this point, the names (``SuperLow,'' etc.) are purely suggestive.

For $y\in Y^{\circ}$ and using $e_{P}(y) = \max\lbrace \alpha(y), \beta(y, e_{A}(y))\rbrace$, straightforward manipulations show
\begin{equation*}
    \alpha(y) \geq \beta(y, e_{A}(y)) \Leftrightarrow  d_{2}(y, e_{A}(y)) \leq 0.
\end{equation*}
In particular, $e_{P}(y)=\alpha(y)$ if $y \in \iSP\cup\iP\cup\iM\cup\iR$, and $e_{P}(y) = \beta(y, e_{A}(y))$ if $y\in \iR\cup\iSR$.

Finally, let  $\ubar{y} = \max\lbrace y\in [\ubar{x}, \bar{x}]\colon \lambda(y) = y\rbrace$ and $\bar{y} = \min\lbrace y\in[\ubar{x}, \bar{x}]\colon \lambda(y) = \max\lambda\rbrace$.

For (possibly empty) $A, B \subseteq \mathbb{R}$, we write $A < B$ to mean $a < b$ for all $a\in A$ and $b\in B$.
We write $A< B < C$ to mean $A < B$ and $B< C$ and $A<C$, and so on.

\Cref{thm:stochastic_tight_characterization} from the main text is a corollary of \Cref{cor:tight_mechs_are_fps} and the following:

\begin{theorem}\label{thm:stochastic_BR_characterization}
    Let $\lambda\in\Lambda$ and $m\in T(\lambda)$.
    Let $Y^{\circ}$ be non-empty, i.e. let $m$ have random audits.
    Then, the sets $\iSP, \ldots, \iSR$ defined in \eqref{eq:candidate_five_intervals} are consecutive intervals, each with a non-empty interior, satisfying $\iSP < \ldots < \iSR$, and such that $m$ satisfies all claims in \Cref{thm:stochastic_tight_characterization}.
    Further, it holds $\ubar{y} = \inf (Y^{\circ}) = \inf \iSP$ and $\bar{y} = \sup\iSR$ and for all $y\in [\ubar{x}, \bar{x}]$ it holds
    \begin{subequations}
    \begin{align}
        \label{eq:canonical_refunds:remptyset}
        (1 - e_{P}(y))(y - r_{\emptyset}(y)) &= \min\left\lbrace (1 - e_{P}(y))y, \lambda(y) + u_{A}(e_{A}(y)) + e_{P}(y)\tau\right\rbrace;
        \\
        \label{eq:canonical_refunds:rp}
        e_{P}(y)(y - r_{P}(y)) &= \min\left\lbrace y, \lambda(y) + u_{A}(e_{A}(y))\right\rbrace -  (1 - e_{P}(y))(y - r_{\emptyset}(y)).
    \end{align}
    \end{subequations}
\end{theorem}

\begin{pf}[Proof of \Cref{thm:stochastic_BR_characterization}]
\setcounter{step}{0}

We proceed in several steps.
\begin{step}
$\iSP < \ldots < \iSR$
\end{step}
\begin{pf}
    We show $\iSP < \iSR$, the other cases being similar.
    Towards a contradiction, let $z\in\iSP$ and $y\in\iSR$ be such that $y < z$.
    Since $z\in\iSP$ and $y\in\iSR$, it holds $d_{1}(z, e_{A}(z)) < 0 < d_{2}(y, e_{A}(y))$, meaning $\ubar{\pi}(z, e_{A}(z))=\pi_{1}(z, e_{A}(z))$ and $\ubar{\pi}(y, e_{A}(y))=\pi_{3}(y, e_{A}(y))$.
    Since both $d_{1}$ and $d_{2}$ are increasing in the first argument but decreasing in the second argument, we have $e_{A}(y) < e_{A}(z)$.
    Find $\varepsilon > 0$ sufficiently close to $0$ such that $e_{A}(y) < e_{A}(y) + \varepsilon < e_{A}(z) - \varepsilon < e_{A}(z)$ and $d_{1}(z, e_{A}(z) - \varepsilon) < 0 < d_{2}(y, e_{A}(y) + \varepsilon)$.
    Thus, $\ubar{\pi}(z, e_{A}(z) - \varepsilon) = \pi_{1}(z, e_{A}(z) - \varepsilon)$ and $\ubar{\pi}(y, e_{A}(y) + \varepsilon) = \pi_{3}(y, e_{A}(y) + \varepsilon)$.
    Since for every type $x$ the effort $e_{A}(x)$ maximizes $\ubar{\pi}(x, \cdot)$ (\Cref{prop:minimal_principal_effort}), we have $\pi_{1}(z, e_{A}(z)) \geq \pi_{1}(z, e_{A}(z) -\varepsilon)$ and $\pi_{3}(y, e_{A}(y)) \geq \pi_{3}(y, e_{A}(y) + \varepsilon)$.

    We next claim $\pi_{1}(y, e_{A}(y)) \geq \pi_{1}(y, e_{A}(y) + \varepsilon)$.
    Since $y\in \iSR$, we have $\alpha(y)\leq \beta(y, e_{A}(y))$.
    Since $\pi_{3}(y, e_{A}(y)) \geq \pi_{3}(y, e_{A}(y) + \varepsilon)$, invoking claim (2) from \Cref{prop:supermodularity_across_regimes} yields $\pi_{2}(y, e_{A}(y)) \geq \pi_{2}(y, e_{A}(y) + \varepsilon)$, and then invoking claim (1) from \Cref{prop:supermodularity_across_regimes} yields $\pi_{1}(y, e_{A}(y)) \geq \pi_{1}(y, e_{A}(y) + \varepsilon)$, as claimed.
    
    Since $\pi_{1}(y, e_{A}(y)) \geq \pi_{1}(y, e_{A}(y) + \varepsilon)$ and $y < z$, submodularity of $\pi_{1}$ implies $\pi_{1}(z, e_{A}(y)) \geq \pi_{1}(z, e_{A}(y) + \varepsilon)$.
    In summary, $\pi_{1}(z, e_{A}(y)) \geq \pi_{1}(z, e_{A}(y) + \varepsilon)$ and $\pi_{1}(z, e_{A}(z)) \geq \pi_{1}(z, e_{A}(z) -\varepsilon)$ hold.
    These two inequalities contradict the strict quasiconcavity of $\pi_{1}$ (\Cref{prop:pifunctions_characterization}) since $e_{A}(y) < e_{A}(y) + \varepsilon < e_{A}(z) - \varepsilon < e_{A}(z)$.  
\end{pf}  

\begin{step}
    For all $y\in [\ubar{x}, \bar{x}]$, it holds $e_{P}(y) > 0$ if and only if $e_{A}(y) > 0$.
\end{step}
\begin{pf}
    This step uses the assumption $c_{A}^{\prime}(0) = c_{P}^{\prime}(0) = 0$.
    Namely, if $0 < e_{P}(y)$ and $e_{A}(y) = 0$ (resp. if $0 = e_{P}(y)$ and $e_{A}(y) > 0$), then for all $i\in\lbrace 1, 2, 3\rbrace$ the derivative $\partial_{2}\pi_{i}(y, \tilde{e}_{A})$ is strictly positive (resp. negative) for all $\tilde{e}_{A}$ sufficiently close to $e_{A}(y)$, contradicting that $e_{A}(y)$ maximizes $\ubar{\pi}(y, \cdot)$.
\end{pf}

\begin{step}
    The set $\lbrace y\in [\ubar{x}, \bar{x}]\colon \alpha(y) \in (0, 1)\rbrace$ is non-empty.
\end{step}
\begin{pf}
    Towards a contradiction, let $\lbrace y\in [\ubar{x}, \bar{x}]\colon \alpha(y) \in (0, 1)\rbrace$ be empty.
    Inspecting the definition of $\alpha$, one may verify that $\lambda(y) = \min\lbrace y, \ubar{y}\rbrace$ holds for all $y\in [\ubar{x}, \bar{x}]$.
    We thus have $\alpha(y) = \bm{1}(y < \ubar{y})$ for all $y$ and we have $\beta(y, \tilde{e}_{A}) = 0$ for all $y\in [\ubar{y}, \bar{x}]$ and all $\tilde{e}_{A}\in [0, 1]$.
    Since $e_{P}(y) = \max\lbrace \alpha(y), \beta(y, e_{A}(y))\rbrace$ for all $y$ (\Cref{prop:minimal_principal_effort}), we infer that $e_{P}$ maps to $\lbrace 0, 1\rbrace$, meaning $Y^{\circ}$ is empty; contradiction.    
\end{pf}
The previous step lets us use \Cref{lemma:basic_continuity} henceforth.

\begin{step}
    For all $y\in [\ubar{x}, \ubar{y})$, it holds $e_{P}(y) = 1$.
    For all $y\in [\bar{y}, \bar{x}]$, it holds $e_{P}(y) = 0$.
    Further, $[\ubar{x}, \ubar{y}) < \iSP \ldots < \iSR < [\bar{y}, \bar{x}]$ and $\ubar{y} = \inf Y^{\circ}$ and $\sup Y^{\circ} = \bar{y}$.
    Finally, $\lambda$ strictly increases on $[\ubar{x}, \bar{y}]$, and equals $\max \lambda$ on $[\bar{y}, \bar{x}]$.
\end{step}
\begin{pf}
    Recall $e_{P}(y) = \max\lbrace \alpha(y), \beta(y, e_{A}(y)) \rbrace$ for all $y\in [\ubar{x}, \bar{x}]$.

    Since $\lambda = \id$ on $[\ubar{x}, \ubar{y}]$, we find $\alpha = 1$ on $[\ubar{x}, \ubar{y})$ by inspecting the definition of $\alpha$.
    Hence, also $e_{P} = 1$ on $[\ubar{x}, \ubar{y})$.
    For all $y\in [\bar{y}, \bar{x}]$, we have $\lambda(y) = \max\lambda$. Inspecting the definitions of $\alpha$ and $\beta$ implies $\alpha(y) = \beta(y, e_{A}(y)) = 0$, and hence $e_{P}(y) = 0$.

    We next show $\ubar{y} = \inf Y^{\circ}$. Since $\alpha(y) < 1$ for all $y\in (\ubar{y}, \bar{x}]$ (\Cref{lemma:basic_continuity}) and since $\beta < 1$, we find $e_{P}(y) < 1$ for all $y\in (\ubar{y}, \bar{x}]$.
    We have just argued that $e_{P} = 1$ holds on $[\ubar{x}, \ubar{y})$.
    Since $\iSP < \ldots< \iSR$ and since these five intervals partition the non-empty set $Y^{\circ}$ where $e_{P}$ is interior, we infer $\ubar{y} = \inf \iSP$.

    We next show $\bar{y} = \sup Y^{\circ}$.
    Abbreviate $x = \sup Y^{\circ}$.
    We already know $x\leq\bar{y}$ and $e_{P}(y) = 0$ for all $y\in [\bar{y}, \bar{x}]$.
    Since $e_{P}(y) > 0$ holds definitionally for all $y\in Y^{\circ}$, we have $\bar{y} \geq x$.
    Thus, we show $\bar{y}\leq x$.
    We already know $\ubar{y} = \inf Y^{\circ} \leq \sup Y^{\circ} = x$.\footnote{Recall, $Y^{\circ}$ is non-empty by assumption, so that $\inf Y^{\circ} \leq \sup Y^{\circ}$.}
    Since $e_{P}$ is continuous on $(\ubar{y}, \bar{x}]$ (\Cref{lemma:basic_continuity}), it follows $e_{P}(x) = 0$.
    Thus, also $\beta(x, e_{A}(x)) = 0$.
    Since $e_{P}(x) = 0$, an earlier step implies $e_{A}(x) = 0$ and, hence, $\beta(x, 0) = 0$.
    Inspecting the definition of $\beta$, we infer $\lambda(x) = \max\lambda$.
    Thus, $x\geq \bar{y}$, by the definition of $\bar{y} = \min\lbrace y\colon\lambda(y) =\max\lambda\rbrace$.

    Summarizing, $[\ubar{x}, \ubar{y}) < \iSP< \ldots < \iSR < [\bar{y}, \bar{x}]$.
    Since $\lambda$ is increasing and concave, $\lambda$ strictly increases on $[\ubar{x}, \bar{y}]$, and constantly equals $\max \lambda$ on $[\bar{y}, \bar{x}]$.
\end{pf}

By the previous step and \Cref{lemma:basic_continuity}, $e_{A}$ and $e_{P}$ are continuous on $(\inf\iSP, \bar{x}]$.

\begin{step}
    Each of $\iSP, \ldots, \iSR$ is an interval with non-empty interior.
\end{step}
Since in particular $\iSP$ and $\iSR$ are non-empty, this step will also imply $\ubar{y} = \inf Y^{\circ} = \inf \iSP$ and $\bar{y} = \sup Y^{\circ} = \sup \iSR$.
\begin{pf}
    Step 4 implies $Y^{\circ} = (\ubar{y}, \bar{y})$ or $Y^{\circ} = [\ubar{y}, \bar{y})$.
    Since $[\ubar{x}, \ubar{y}) < \iSP < \ldots < \iSR < (\ubar{y}, \bar{x}]$ and since $\iSP, \ldots, \iSR$ partition $Y^{\circ}$, each of the sets $\iSP, \ldots, \iSR$ is an interval. 

    In an intermediate step, we show $d_{1}(\ubar{y}, e_{A}(\ubar{y})) < 0 < d_{2}(\lambda(\bar{x}), e_{A}(\lambda(\bar{x})))$.
    \begin{itemize}
        \item Since $\alpha(\ubar{y}) > 0$, we have $e_{P}(\ubar{y}) > 0$.
        By a previous step, $e_{P}(\ubar{y}) > 0$ implies $e_{A}(\ubar{y}) > 0$.
        Since $\lambda(\ubar{y}) = \ubar{y}$, it follows $\ubar{y}<\lambda(\ubar{y}) + u_{A}(e_{A}(\ubar{y}))$, i.e. $d_{1}(\ubar{y}, e_{A}(\ubar{y})) < 0$.
        \item Towards a contradiction, let $d_{2}(\lambda(\bar{x}), e_{A}(\lambda(\bar{x}))) \leq 0$.
        It holds $\alpha(\lambda(\bar{x})) = 0$ (by inspecting the definition of $\alpha$).
        The assumption $d_{2}(\lambda(\bar{x}), e_{A}(\lambda(\bar{x}))) \leq 0$ thus reads $\lambda(\bar{x}) - \lambda(\lambda(\bar{x})) - u_{A}(e_{A}(\lambda(\bar{x}))) \leq 0$.
        Thus, $\beta(\lambda(\bar{x}), e_{A}(\lambda(\bar{x}))) = 0$ (by inspecting the definition of $\beta$).
        Thus, $e_{P}(\lambda(\bar{x})) = 0$.
        Thus, also $e_{A}(\lambda(\bar{x})) = 0$ (by a previous step).
        Returning to $d_{2}(\lambda(\bar{x}), e_{A}(\lambda(\bar{x}))) \leq 0$, we find $\lambda(\bar{x}) - \lambda(\lambda(\bar{x})) \leq 0$, which is impossible since $\ubar{y} < \lambda(\bar{x})$ implies $\lambda(\lambda(\bar{x})) < \lambda(\bar{x})$ (by definition of $\ubar{y}$).
    \end{itemize}
    
    We also claim $\ubar{y} < \lambda(\bar{x}) \leq \bar{y}$.
    First, by definition of $\ubar{y} = \max\lbrace y\colon \lambda(y) = y\rbrace$ and since $\lambda$ strictly increases on $[\ubar{x}, \bar{y}]$ (by a previous step) and since $\ubar{y} < \bar{x}$ (\Cref{lemma:basic_continuity}), it holds $\ubar{y} = \lambda(\ubar{y}) < \lambda(\bar{x})$.
    Second, by definition of $\bar{y} = \min\lbrace y\colon \lambda(y) = \max\lambda\rbrace$ and since $\lambda\leq\id$, it holds $\bar{y} \geq \lambda(\bar{y}) = \max\lambda \geq \lambda(\bar{x})$.
    
    We now show that $\iSP, \ldots, \iSR$ are all non-empty.
    \Cref{lemma:basic_continuity} implies that $\alpha$ and $e_{A}$ are continuous on $(\ubar{y}, \bar{x}]$, and right-continuous at $\ubar{y}$.
    In particular, since $\ubar{y} < \lambda(\bar{x}) \leq \bar{y}$, both $\hat{d}_{1}(y) \overset{\mathrm{def}}{=} d_{1}(y, e_{A}(y))$ and $\hat{d}_{2}(y) \overset{\mathrm{def}}{=} d_{2}(y, e_{A}(y))$ are continuous in $y$ for all $(\ubar{y}, \lambda(\bar{x})]$, and right-continuous at $\ubar{y}$.
    Moreover, $(\ubar{y}, \lambda(\bar{x}))\subseteq Y^{\circ}$ since $\ubar{y} = \inf Y^{\circ}$ and $\lambda(\bar{x}) \leq \bar{y} = \sup Y^{\circ}$ (by a previous step).
    Finally, $d_{2}(y, e_{A}(y)) < d_{1}(y, e_{A}(y))$ holds for all $y\in (\ubar{y}, \lambda(\bar{x}))$ since for such $y$ we have $\alpha(y) > 0$ (by inspection of $\alpha$).
    Since we have shown $\hat{d}_{1}(\ubar{y}) < 0 < \hat{d}_{2}(\lambda(\bar{x}))$, it follows there exists $y_{\iSP} \in (\ubar{y}, \lambda(\bar{x}))$ such that $\hat{d}_{2}(y_{\iSP}) < \hat{d}_{1}(y_{\iSP}) < 0$; then $y_{\iSP}\in Y^{\circ}$ since $(\ubar{y}, \lambda(\bar{x}))\subseteq Y^{\circ}$, implying $y_{\iSP}\in \iSP$ by definition of $\iSP$.
    Similarly, there is $y_{\iP} \in (\ubar{y}, \lambda(\bar{x}))$ such that $\hat{d}_{2}(y_{\iP}) < \hat{d}_{1}(y_{\iP}) = 0$; then $y_{\iP}\in \iP$.
    Continuing in this manner, we conclude that $\iSP, \ldots, \iSR$ are all non-empty.

    We next argue that each of $\iSP, \ldots, \iSR$ has a non-empty interior.
    Since $e_{A}$ and $\alpha$ are continuous on $(\ubar{y}, \bar{x}]$ and right continuous at $\ubar{y} = \inf\iSP$, it follows that each of $\iSP$, $\iM$, and $\iSR$ is open in $[\ubar{x}, \bar{x}]$, and we already know that these sets are non-empty.
    
    We next show that $\iR$ has a non-empty interior; an analogous argument applies to $\iP$.
    Since $\iR$ is a non-empty interval, it suffices to show $\iR$ is not a singleton.
    Towards a contradiction, let $\iR$ be a singleton $x$.
    Thus, $x = \sup \iM = \inf \iSR$.
    Find a sequence $(z_{n})_{n}$ in $\iM$ and a sequence $(y_{n})_{n}$ in $\iSR$, both converging to $x$.
    For all $n$ the effort $e_{A}(z_{n})$ satisfies the first-order condition $0 = c_{P}(e_{P}(z_{n})) - c_{A}^{\prime}(e_{A}(z_{n}))$ for maximizing $\pi_{2}(z_{n}, \cdot)$.
    Similarly, for all $n$, since $e_{A}(y_{n})$ maximizes $\ubar{\pi}(y_{n}, \cdot)$ and $y_{n}\in\iSR$, there is a sequence $(\varepsilon_{n, k})_{k\in\mathbb{N}}$ converging to $0$ from above such that $0 \geq \partial_{2} \pi_{3}(y_{n}, e_{A}(y_{n}) + \varepsilon_{n, k})$ for all $k$.
    Bounding $\partial_{2} \pi_{3}$, we find
    \begin{align*}
        0 
        \geq & \partial_{2} \pi_{3}(y_{n}, e_{A}(y_{n}) + \varepsilon_{n, k})
        \\
        \geq &c_{P}(\beta(y_{n}, e_{A}(y_{n}) + \varepsilon_{n, k})) - c_{A}^{\prime}(e_{A}(y_{n})+ \varepsilon_{n, k})
        \\
        &+ (1 - e_{A}(y_{n})) \frac{c_{P}^{\prime}(\beta(y_{n}, e_{A}(y_{n}) + \varepsilon_{n, k})) u_{A}^{\prime}(e_{A}(y_{n}) + \varepsilon_{n, k})}{\bar{x} + \tau}.
    \end{align*}
    We now take $k\to\infty$ and recall $e_{P}(y_{n}) = \beta(y_{n}, e_{A}(y_{n}))$ (since $y_{n}\in\iSR$) to find
    \begin{equation}\label{eq:thm:open_stochastic_tight_characterization_for_triples:ineq:2}
        0 \geq c_{P}(e_{P}(y_{n})) - c_{A}^{\prime}(e_{A}(y_{n})) + (1 - e_{A}(y_{n})) \frac{c_{P}^{\prime}(e_{P}(y_{n})) u_{A}^{\prime}(e_{A}(y_{n}))}{\bar{x} + \tau},
    \end{equation}    
    Both $e_{P}$ and $e_{A}$ are continuous at $x$.
    Thus, taking limits of \eqref{eq:thm:open_stochastic_tight_characterization_for_triples:ineq:2} and the earlier equation $0 = c_{P}(e_{P}(z_{n})) - c_{A}^{\prime}(e_{A}(z_{n}))$, we find $0 \geq (1 - e_{A}(x)) c_{P}^{\prime}(e_{P}(x)) u_{A}^{\prime}(e_{A}(x))$.
    Since $c_{P}^{\prime}$ and $u_{A}^{\prime}$ are both strictly positive except at $0$, we infer $e_{A}(x) = 1$ or $e_{A}(x) = 0$ or $e_{P}(x) = 0$.
    However, we know $e_{P}(x) > 0$ (since $x\in \iR)$.
    Hence also $e_{A}(x) > 0$ (by a previous step).
    From $e_{A}(x) > 0$ we also get $u_{A}^{\prime}(e_{A}(x)) > 0$.
    Finally, \Cref{prop:minimal_principal_effort} asserts $e_{A}(x) < 1$.
    Contradiction.
\end{pf}

\begin{step}
    The principal's effort $e_{P}$ is strictly decreasing on $\iSP\cup\ldots\cup \iR$, and decreasing on $\iSR$.
\end{step}

\begin{pf}
    As noted following \eqref{eq:constraint_SCP}, it holds $e_{P}(y) = \alpha(y)$ for all $y\in \iSP\cup\ldots\cup\iR$, and $e_{P}(y) = \beta(y, e_{A}(y))$ for all $y\in \iSR$.
    Since $\ubar{y} = \inf\iSP$, \Cref{lemma:basic_continuity} implies that $e_{P}$ strictly decreases on $\iSP\cup\ldots\cup \iR$.
    To show that $e_{P}$ decreases on $\iSR$, let $z, y \in\iSR$ be such that $z < y$.
    Towards a contradiction, let $\beta(y, e_{A}(y)) > \beta(z, e_{A}(z))$.
    Note $\lambda(z) \leq \lambda(y)$ since $\lambda$ is increasing.
    Thus $\lambda(y) + u_{A}(e_{A}(y)) < \lambda(z) + u_{A}(e_{A}(z))$.
    This strict inequality requires $e_{A}(y) < e_{A}(z)$.
    Thus, $e_{A}(y) + \varepsilon < e_{A}(z) - \varepsilon$ and $\lambda(y) + u_{A}(e_{A}(y) + \varepsilon) < \lambda(z) + u_{A}(e_{A}(z) - \varepsilon)$ for all sufficiently small $\varepsilon > 0$.
    According to \Cref{lemma:K3_auxiliary_order}, all such $\varepsilon$ satisfy $\partial_{2} \pi_{3}(y, e_{A}(y) + \varepsilon) > \partial_{2} \pi_{3}(z, e_{A}(z) - \varepsilon)$.
    Hence, all sufficiently small $\varepsilon > 0$ satisfy
    $\pi_{3}(y, e_{A}(y) + \varepsilon) - \pi_{3}(y, e_{A}(y)) > \pi_{3}(z, e_{A}(z)) - \pi_{3}(z, e_{A}(z) - \varepsilon)$.
    Since $d_{2}(y, e_{A}(y)) > 0$ and $d_{2}(z, e_{A}(z)) > 0$, also $d_{2}(y, e_{A}(y) + \varepsilon) > 0$ and $d_{2}(z, e_{A}(z) - \varepsilon) > 0$ for small $\varepsilon$. 
    Then,
    \begin{align*}
        0 \geq \ubar{\pi}(y, e_{A}(y) + \varepsilon) - \ubar{\pi}(y, e_{A}(y)) 
        &=  \pi_{3}(y, e_{A}(y) + \varepsilon) - \pi_{3}(y, e_{A}(y))
        \\
        &> \pi_{3}(z, e_{A}(z)) - \pi_{3}(z, e_{A}(z) - \varepsilon) 
        \\
        &= \ubar{\pi}(z, e_{A}(z)) - \ubar{\pi}(z, e_{A}(z) - \varepsilon).
    \end{align*}
    But \Cref{prop:minimal_principal_effort} asserts that $e_{A}(z)$ maximizes $\ubar{\pi}(z, \cdot)$; contradiction.
\end{pf}

\begin{step}
    The agent's effort $e_{A}$ is strictly decreasing on each of $\iSP$, $\iM$, and $\iSR$, but strictly increasing on each of $\iP$ and $\iR$.
\end{step}
\begin{pf}
    Strict monotonicity of $e_{A}$ on $\iSP$ and $\iM$ follows from inspecting the first-order conditions for maximizing $\pi_{1}$ and $\pi_{2}$, respectively, and using that $\alpha$ is strictly decreasing on both $\iSP$ and $\iM$.

    Next, consider $\iP$.
    By definition, all $y\in \iP$ satisfy $u_{A}(e_{A}(y)) = y - \lambda(y)$.
    Recall also $\lambda(\ubar{x}) = \ubar{x}$ (since $\lambda\in\Lambda$) while $\lambda(y) < y$ holds on $\iP$ (\Cref{lemma:basic_continuity}).
    Since $y - \lambda(y)$ is increasing in $y$ and $\lambda$ is concave, it follows that $y - \lambda(y)$ is strictly increasing on $\iP$.
    Thus, also $e_{A}$ is strictly increasing on $\iP$.

    Next, consider $\iR$.
    By definition, all $y\in \iR$ satisfy $\frac{u_{A}(e_{A}(y))}{y} = 1 - \alpha(y) - \frac{\alpha(y)\tau}{y} - \frac{\lambda(y)}{y}$.
    We know $\frac{\lambda(y)}{y}$ is decreasing (\Cref{appendix:tight_mechanisms:shape_of_loss}).
    We know $1 > e_{P}(y) = \alpha(y) > 0$ for all $y\in\iR$.
    Since \Cref{lemma:basic_continuity} implies that $\alpha$ is strictly decreasing whenever interior, we conclude $\alpha$ strictly decreases on $\iR$.
    Thus, $e_{A}(y)$ must strictly increase on $\iR$.

    Finally, consider $\iSR$. 
    We first show $e_{A}$ (weakly) decreases on $\iSR$.
    For all $y\in \iSR$, we know $e_{A}(y)$ maximizes $\pi_{3}(y, \cdot)$ on a neighborhood of $e_{A}(y)$ (\Cref{prop:minimal_principal_effort} and \eqref{eq:constraint_SCP}). Since $\pi_{3}(y, \cdot)$ is strictly quasiconcave (\Cref{prop:pifunctions_characterization}), the maximizer is a global maximizer. Since $\pi_{3}$ is also submodular (\Cref{prop:pifunctions_characterization}), it follows that $e_{A}$ is decreasing on $\iSR$.

    We now show that $e_{A}$ is strictly decreasing on $\iSR$.
    The idea is again to derive a first-order condition, but some care has to be taken when attempting to differentiate $\pi_{3}$ with respect to agent effort. 
    Towards a contradiction, let $[z_{0}, z_{1}]$ be a non-degenerate subinterval in the interior of $\iSR$ on which $e_{A}$ constantly equals $e_{A}(z_{0})$.
    As in previous steps, for all $y\in [z_{0}, z_{1}]$, if $\varepsilon$ is sufficiently close to $0$, then $\ubar{\pi}(y, e_{A}(z_{0}) \pm \varepsilon) = \pi_{3}(y, e_{A}(z_{0}) \pm \varepsilon)$.
    
    In an auxiliary step, we argue that for almost all $y\in [z_{0}, z_{1}]$ the function $\tilde{e}_{A}\mapsto \pi_{3}(y, \tilde{e}_{A})$ is differentiable at $\tilde{e}_{A} = e_{A}(z_{0})$.
    To that end, it suffices to check differentiability of $\tilde{e}_{A}\mapsto \beta(y, \tilde{e}_{A})$ at $\tilde{e}_{A} = e_{A}(z_{0})$ for almost all $y\in [z_{0}, z_{1}]$.
    For $y$ such that $\hat{Z}(\lambda(y) + u_{A}(e_{A}(z_{0})))$ is a singleton, Theorem 3 of \citet{milgrom2002envelope} implies that $\beta(y, \cdot)$ is differentiable at $\tilde{e}_{A} = e_{A}(z_{0})$.\footnote{We are using $\beta(y, e_{A}(z_{0})) = e_{P}(y) > 0$. Hence, also $\beta(y, \tilde{e}_{A}) > 0$ for $\tilde{e}_{A}$ close to $e_{A}(z_{0})$.}
    Thus we show that for almost all $y\in [z_{0}, z_{1}]$ the set $\hat{Z}(\lambda(y) + u_{A}(e_{A}(z_{0})))$ is a singleton.
    For all $y\in [z_{0}, z_{1}]$, the set $\hat{Z}(\lambda(y) + u_{A}(e_{A}(z_{0})))$ is an interval since $\lambda$ is concave and lies below the affine function $x\mapsto \beta(y, e_{A}(z_{0})) (x + \tau) + \lambda(y) + u_{A}(e_{A}(z_{0}))$.
    Since $\lambda$ strictly increases on $[z_{0}, z_{1}]$ (by an earlier step and since $[z_{0}, z_{1}] \subset \iSR$), a routine argument verifies that if $y < y^{\prime}$, then $\max \hat{Z}(\lambda(y) + u_{A}(e_{A}(z_{0}))) \leq \min  \hat{Z}(\lambda(y^{\prime}) + u_{A}(e_{A}(z_{0})))$.\footnote{The objective defining $\hat{Z}$ is strictly supermodular in $(x, y)$ since $\lambda$ strictly increases on $[z_{0}, z_{1}]$.}
    Hence, if $\hat{Z}(\lambda(y) + u_{A}(e_{A}(z_{0})))$ is a non-degenerate interval, then its interior contains a rational number that is not contained in an interval of the collection $\lbrace \hat{Z}(\lambda(y^{\prime}) + u_{A}(e_{A}(z_{0})))\colon y^{\prime}\in [z_{0}, z_{1}] \setminus\lbrace y\rbrace\rbrace$.
    Hence, there are at most countably many $y\in [z_{0}, z_{1}]$ for which $\hat{Z}(\lambda(y) + u_{A}(e_{A}(z_{0})))$ is non-degenerate.

    Find $y, y^{\prime} \in [z_{0}, z_{1}]$ such that $y < y^{\prime}$ and $\tilde{e}_{A}\mapsto \pi_{3}(y, \tilde{e}_{A})$ and $\tilde{e}_{A}\mapsto \pi_{3}(y^{\prime}, \tilde{e}_{A})$ are differentiable at $e_{A}(z_{0})$.
    This is possible by the above claim and since $[z_{0}, z_{1}]$ is non-degenerate by assumption.
    Since $e_{A}(z_{0})$ maximizes $\pi_{3}(y, \cdot)$ (on a neighborhood of $e_{A}(z_{0})$),
    \begin{align*}
        0
        =
        c_{P}(\beta(y, e_{A}(z_{0}))) - c_{A}^{\prime}(e_{A}(z_{0})) + (1 - e_{A}(z_{0})) \frac{c_{P}^{\prime}(\beta(y, e_{A}(z_{0}))) u_{A}^{\prime}(e_{A}(z_{0}))}{\hat{z}(\lambda(y) + u_{A}(e_{A}(z_{0}))) + \tau}.
    \end{align*}
    An analogous first-order condition holds for $y^{\prime}$ (obtained by replacing all instances of $y$ with $y^{\prime}$).
    Since $\lambda$ is strictly increasing on $[z_{0}, z_{1}]$, we have $\hat{z}(\lambda(y) + u_{A}(e_{A}(z_{0}))) + \tau \leq \hat{z}(\lambda(y^{\prime}) + u_{A}(e_{A}(z_{0})) + \tau$ and $\beta(y, e_{A}(z_{0})) > \beta(y^{\prime}, e_{A}(z_{0}))$.
    Consequently, the first-order condition cannot hold for both $y$ and $y^{\prime}$; contradiction.
\end{pf}

We next show the claims regarding the efficient agent effort $e_{A}^{\eff}$.
For $y$ in $\iSR$ (resp. in $[\ubar{x}, \sup\iSP)$), the claim $e_{A}(y) > e_{A}^{\eff}(y)$ (resp. $e_{A}(y) < e_{A}^{\eff}(y)$) follows by noting that $e_{A}^{\eff}$ maximizes $\pi_{2}$ and that, as in \Cref{prop:supermodularity_across_regimes}, the marginal benefit of $e_{A}$ is strictly higher (resp. strictly lower) when profit is given by $\pi_{3}$ (resp. by $\pi_{1}$) than when profit is given by $\pi_{2}$.
Next, on $[\bar{y}, \bar{x}]$, both $e_{A}$ and $e_{A}^{\eff}$ equal $0$.
On the interior of $\iP$, we have $e_{A} < e_{A}^{\eff}$ since on this interval $e_{A}$ is strictly increasing (as proven earlier), $e_{A}^{\eff}$ is strictly decreasing, and the two coincide at the top of the interval (i.e. at $\inf\iM$).
By a similar argument, $e_{A} > e_{A}^{\eff}$ on the interior of $\iR$.

The principal's profit $\Pi_{m}$ is increasing since all $y$ satisfy  $\Pi_{m}(y) = \max_{\tilde{e}_{A}\in[0, 1]} \ubar{\pi}(y, \tilde{e}_{A})$, where $\ubar{\pi}(y, \tilde{e}_{A})$ is increasing in $y$ for all $\tilde{e}_{A}$.
Finally, \Cref{lemma:relaxed_operator_wlog} shows that the agent's utility is given by $U_{m}(y) = \max\lbrace u_{A}(e_{A}(y)), y - \lambda(y)\rbrace$ for all $y$.
The characterization of $e_{A}$ and $\lambda$ across the intervals now implies that $U_{m}$ is $v$-shaped with minimum at $\inf\iP$; this minimum is strictly positive since $e_{A}$ is strictly positive at this type.
From the characterization of the intervals and \eqref{eq:constraint_SCP}, we also get when the agent's incentive to advance the surplus is binding, i.e. $U_{m}(y) = u_{A}(e_{A}(y)) > y - \lambda(y)$ for $y < \sup\iSP$, and $U_{m}(y) = y - \lambda(y)$ for $y \geq \sup\iSP$.

Finally, we show the claims regarding $r_{\emptyset}$ and $r_{P}$, beginning with \eqref{eq:canonical_refunds:remptyset}.

First, let $y \in [\ubar{x}, \ubar{y}]$.
Thus, $\lambda(y) = y$ and the minimum on the right side of \eqref{eq:canonical_refunds:remptyset} equals $(1 - e_{P}(y))y$.
Since $y \in [\ubar{x}, \ubar{y}]$, we also know that profit at type $y$ equals $y - u_{A}(e_{A}(y)) - c_{A}(e_{A}(y)) - (1 - e_{A}(y))c_{P}(e_{P}(y))$, which requires $R_{NA}(y) = 0$.
Hence, $e_{P}(y) = 1$ or $r_{\emptyset}(y) = 0$.
In both cases, the left side of \eqref{eq:canonical_refunds:remptyset} equals $(1 - e_{P}(y))y$.

Let $y\in (\ubar{y}, \bar{y})$.
Thus, $\lambda(y) < y$ and $0 < e_{P}(y) < 1$.
From the proof of \Cref{lemma:relaxed_operator_wlog}, we know that $(1 - e_{P})(y - r_{\emptyset}(y))$ is at most the right side of \eqref{eq:canonical_refunds:remptyset}.
Towards a contradiction, suppose it is strictly less.
Since $e_{P}(y) < 1$, thus $0 < r_{\emptyset}(y)$.
Since $\lambda(y) < y$ and $0 < e_{P}(y)$, the suprema in the definitions of $\alpha(y)$ and $\beta(y, e_{A}(y))$ are attained.
In particular, if $e_{P}(y) = \alpha(y)$, there is $x$ such that $\lambda(x) = e_{P}(y)x + (1 - e_{P}(y))y$; if $e_{P}(y) = \beta(y, e_{A}(y))$, there is $x$ such that $\lambda(x) = e_{P}(y) x + \lambda(y) + u_{A}(e_{A}(y)) + e_{P}(y)\tau$.
In both cases, since $(1 - e_{P})(y - r_{\emptyset}(y))$ is strictly less than the right side of \eqref{eq:canonical_refunds:remptyset}, it follows $\lambda(x) > e_{P}(y)x + (1 - e_{P}(y))(y - r_{\emptyset}(y))$, contradicting $m\in T(\lambda)$.

Finally, take $y \in [\bar{y}, \bar{x}]$.
Thus, $e_{P}(y) = e_{A}(y) = 0$.
Since $y \in [\bar{y}, \bar{x}]$, type $y$'s IC is binding (i.e. $U_{m}(y) = y - \lambda_{m}(y)$).
Using $e_{P}(y) = e_{A}(y) = 0$ and spelling out $y$'s IC, we find $y - r_{\emptyset}(y) = \lambda_{m}(y)$.
This equation is equivalent to \eqref{eq:canonical_refunds:remptyset}.

From \eqref{eq:canonical_refunds:remptyset} it follows that $r_{\emptyset}$ is constantly zero below $\inf\iSR$. For $y$ above $\inf\iSR$, $r_{\emptyset}(y)$ solves $(1 - e_{P}(y))(1 - \frac{r_{\emptyset}(y)}{y}) = \frac{\lambda(y) + u_{A}(e_{A}(y)) + e_{P}(y)\tau}{y}$. 
This equation requires $r_{\emptyset}(y)$ to increase since $\frac{\lambda(y)}{y}$ is decreasing everywhere and since $e_{A}$ and $e_{P}$ are decreasing from $\inf\iSR$ onwards. 

With \eqref{eq:canonical_refunds:remptyset} established, equation \eqref{eq:canonical_refunds:rp} follows from the formula for the agent's utility, $U_{m}(y) = \max\lbrace u_{A}(e_{A}(y)), y - \lambda(y)\rbrace$.
From these equations, we also get $r_{P}(y) = 0$ for all $y\leq \inf\iM$ and $r_{P}(y) = y + \tau$ for all $y\geq \sup\iM$.
On $\iM$, $r_{P}(y)$ solves $e_{P}(y) r_{P}(y) = y - \lambda(y) - u_{A}(e_{A}(y))$. Since $y - \lambda(y)$ is increasing in $y$ and $e_{A}$ is decreasing on $\iM$, this equation requires $r_{P}$ to increase on $\iM$.
\end{pf}

\addcontentsline{toc}{section}{References}
\newrefcontext[sorting=nyt]
\printbibliography

\newpage
\clearpage

\begin{center}
    {\Large Online Appendix: The Division of Surplus and the Burden of Proof}  
\end{center}


\appendix
\setcounter{section}{0}
\renewcommand{\thesection}{O.\arabic{section}}
\renewcommand{\theHsection}{O.\arabic{section}}

\section{Tightness: Proof of \headercref{Lemma}{{lemma:tight_mechanisms_wlog}}}\label{appendix:tight_mechanisms:abstract_tight_existence_proof}
The proof uses definitions from \Cref{appendix:tight_mechanisms:preparations}.
Let $m_{0}$ be a mechanism.
We show there is a tight mechanism $m^{\ast}$ such that $(\Pi_{m_{0}}, \lambda_{m_{0}}) \leq (\Pi_{m^{\ast}}, \lambda_{m^{\ast}})$.

For all $\lambda\in\Lambda_{0}$, define $\Pi^{\ast}(\lambda)\colon[\ubar{x}, \bar{x}]\to\mathbb{R}$ by $\Pi^{\ast}(y, \lambda) = \max_{m(y)\in M(y, \lambda)} \Pi(y, m(y))$ for all $y$.

We shall use Zorn's Lemma. To that end, we define a partially ordered set.\footnote{We do not work on the space of mechanisms since tightness is not obviously a partial order; two distinct mechanisms may induce the same profit and loss function.}
Let $\mathcal{L}$ be the set of $\lambda\in\Lambda_{0}$ such that $(\Pi_{m_{0}}, \lambda_{m_{0}}) \leq (\Pi^{\ast}(\lambda), \lambda)$.
Given $\lambda, \lambda^{\prime}\in\mathcal{L}$, write $\lambda\preceq\lambda^{\prime}$ if $(\Pi^{\ast}(\lambda), \lambda) \leq (\Pi^{\ast}(\lambda^{\prime}), \lambda^{\prime})$.
The set $\mathcal{L}$ is partially ordered by $\preceq$ and is non-empty (\Cref{lemma:tight_operator_improvement} implies $\Pi_{m_{0}}\leq \Pi^{\ast}(\lambda_{m_{0}})$, i.e. $\lambda_{m_{0}}\in\mathcal{L}$).

We later prove that $\mathcal{L}$ has a maximal element.
Momentarily taking existence as given, we find our candidate for $m^{\ast}$.
Let $\lambda^{\ast}$ be maximal in $\mathcal{L}$.
We pick our candidate mechanism as $m^{\ast}\in T(\lambda^{\ast})$, so that $\Pi^{\ast}(\lambda^{\ast}) = \Pi_{m^{\ast}}$.
We first show $m^{\ast}$ is tight.
Take $m$ such that $(\Pi_{m^{\ast}}, \lambda_{m^{\ast}}) \leq (\Pi_{m}, \lambda_{m})$.
\Cref{lemma:tight_operator_improvement} implies $\lambda^{\ast}\leq\lambda_{m^{\ast}}$ and $\Pi_{m} \leq \Pi^{\ast}(\lambda_{m})$.
Using also $\Pi^{\ast}(\lambda^{\ast})=\Pi_{m^{\ast}}$ (by definition) and $(\Pi_{m_{0}}, \lambda_{m_{0}})\leq (\Pi^{\ast}(\lambda^{\ast}), \lambda^{\ast})$ (since $\lambda^{\ast}\in\mathcal{L}$), we obtain the chain $(\Pi_{m_{0}}, \lambda_{m_{0}})\leq (\Pi^{\ast}(\lambda^{\ast}), \lambda^{\ast}) \leq (\Pi_{m^{\ast}}, \lambda_{m^{\ast}}) \leq (\Pi_{m}, \lambda_{m}) \leq (\Pi^{\ast}(\lambda_{m}), \lambda_{m})$.
This inequality implies $\lambda_{m}\in\mathcal{L}$.
Since $\lambda^{\ast}$ is maximal in $\mathcal{L}$, it follows $(\Pi_{m_{0}}, \lambda_{m_{0}})\leq (\Pi^{\ast}(\lambda^{\ast}), \lambda^{\ast}) = (\Pi_{m^{\ast}}, \lambda_{m^{\ast}}) = (\Pi_{m}, \lambda_{m}) = (\Pi^{\ast}(\lambda_{m}), \lambda_{m})$.
In particular, $(\Pi_{m^{\ast}}, \lambda_{m^{\ast}}) = (\Pi_{m}, \lambda_{m})$, proving that $m^{\ast}$ is tight.
To show that $m^{\ast}$ is tighter than $m_{0}$, repeat the above argument with $m = m^{\ast}$; the argument implies $(\Pi_{m_{0}}, \lambda_{m_{0}}) \leq (\Pi_{m^{\ast}}, \lambda_{m^{\ast}})$.

It remains to show that $\mathcal{L}$ has a maximal element.
We shall use Zorn's Lemma.
Let $C$ be a chain. 
We show $C$ admits an upper bound in $\mathcal{L}$.
For all $y\in [\ubar{x}, \bar{x}]$, let $\bar{\lambda}(y) = \sup_{\lambda\in C} \lambda(y)$.
The supremum exists since all functions in $\Lambda_{0}$ are bounded above by $\bar{x}$.
Thus, also $\bar{\lambda}\in \Lambda_{0}$.
Thus, to show that $\bar{\lambda}$ is in $\mathcal{L}$ and is an upper bound of $C$, we show $\sup_{\lambda\in C} \Pi^{\ast}(y, \lambda) \leq \Pi^{\ast}(y, \bar{\lambda})$ for all $y\in [\ubar{x}, \bar{x}]$.

For the remainder of the proof, fix arbitrary $y$.
Towards a contradiction, let there be $\hat{\lambda}\in C$ such that $\Pi^{\ast}(y, \bar{\lambda}) < \Pi^{\ast}(y, \hat{\lambda})$.
Let $\hat{C} = \lbrace \lambda\in C\colon \hat{\lambda}\preceq \lambda\rbrace$.
Viewed as a net with direction $\preceq$, $C$ converges pointwise to $\bar{\lambda}$.
Since $C$ is a chain, also $\hat{C}$ converges pointwise to $\bar{\lambda}$.
The profit $\Pi^{\ast}(y, \cdot)$ is definitionally increasing with respect to $\preceq$.
Thus, $\Pi^{\ast}(y, \bar{\lambda}) < \Pi^{\ast}(y, \hat{\lambda}) \leq \Pi^{\ast}(y, \lambda)$ for all $\lambda \in \hat{C}$.

For all $\lambda\in \hat{C}$, find $m(y, \lambda) \in T(y, \lambda)$, so that $\lbrace m(y, \lambda)\rbrace_{\lambda\in \hat{C}}$ is a net in $[0, 1]^{2}\times [0, y+\tau]^{3}$ with direction $\preceq$.
Since $[0, 1]^{2}\times [0, y+\tau]^{3}$ is compact, we may find a subnet of $\lbrace m(y, \lambda)\rbrace_{\lambda\in \hat{C}}$ that converges; call the limit $\hat{m}(y)$.
That is, there is a directed set $(D, \leq)$ and a function $\phi\colon D\to \hat{C}$ such that:
(i) for all $d\in D$, it holds $\phi_{d} \in \hat{C}$;
(ii) for all $\lambda\in \hat{C}$, there is $d\in D$ such that $\phi_{d^{\prime}} \succeq \lambda$ for all $d^{\prime} \geq d$;
(iii) for all $\varepsilon > 0$, there is $d\in D$ such that $\Vert m(y, \phi_{d^{\prime}}) - \hat{m}(y)\Vert_{\infty} \leq \varepsilon$ for all $d^{\prime}\geq d$.
Since $\lbrace m(y, \phi_{d})\rbrace_{d\in D}$ converges to $\hat{m}(y)$, the associated net of profits $\lbrace \Pi(y, m(y, \phi_{d}))\rbrace_{d\in D}$ converges to $\Pi(y, \hat{m}(y))$.
Since $\phi$ maps to $\hat{C}$, all $d\in D$ satisfy $\Pi(y, m(y, \phi_{d})) = \Pi^{\ast}(y, \phi_{d}) \geq \Pi^{\ast}(y, \hat{\lambda}) > \Pi^{\ast}(y, \bar{\lambda})$.
Thus, $\Pi(y, \hat{m}(y)) >  \Pi^{\ast}(y, \bar{\lambda})$.
We derive a contradiction to $\Pi(y, \hat{m}(y)) >  \Pi^{\ast}(y, \bar{\lambda})$. Indeed, we will show $\hat{m}(y)\in M(y, \bar{\lambda})$, and hence $\Pi(y, \hat{m}(y))\leq \Pi^{\ast}(y, \bar{\lambda})$ by definition of $\Pi^{\ast}$; contradiction.
Thus, we show $\hat{m}(y)\in M(y, \bar{\lambda})$.
We show \eqref{eq:def:virtual_problem:relaxing_lambda}; the arguments for \eqref{eq:def:virtual_problem:on_path_utility} and \eqref{eq:def:virtual_problem:feasible_eA} are similar; essentially, since each constraint depends only on a finite number of types, pointwise convergence of $C$ to $\bar{\lambda}$ suffices to carry properties to the limit. 
To show \eqref{eq:def:virtual_problem:relaxing_lambda}, fix arbitrary $x\in [y, \bar{x}]$.
We have to show $\bar{\lambda}(x) \leq \hat{e}_{P}(y) x + (1 - \hat{e}_{P}(y)) (y - \hat{r}_{\emptyset}(y))$.
For all $d\in D$, we know $m(y, \phi_{d})\in M(y, \phi_{d})$, implying $\phi_{d}(x) \leq e_{P, \phi_{d}}(y) x + (1 - e_{P, \phi_{d}}(y)) (y - r_{\emptyset, \phi_{d}}(y))$, where $e_{P, \phi_{d}}$ and $r_{\emptyset, \phi_{d}}$ are the principal effort and the no-evidence refund in mechanism $m(y, \phi_{d})$.
Thus, it suffices to argue that for all $\varepsilon > 0$ there is $d\in D$ such that $(\phi_{d}(x), e_{P, \phi_{d}}(y), r_{\emptyset, \phi_{d}}(y))$ is within $\varepsilon$ of $(\bar{\lambda}(x), \hat{e}_{P}(y), \hat{r}_{\emptyset}(y))$ in the supremum norm on $\mathbb{R}^{3}$.
Since $\lbrace m(y, \phi_{d})\rbrace_{d\in D}$ converges to $\hat{m}(y)$, there is $d_{0}\in D$ such that $(e_{P, \phi_{d}}(y), r_{\emptyset, \phi_{d}}(y))$ is within $\varepsilon$ of $(\hat{e}_{P}(y), r_{\emptyset}(y))$ for all $d\geq d_{0}$.
Next, since $\hat{C}$ converges pointwise to $\bar{\lambda}$, there is $\lambda_{1}\in \hat{C}$ such that $\vert \lambda(x) - \bar{\lambda}(x)\vert \leq \varepsilon$ for all $\lambda \in \hat{C}$ such that $\lambda \succeq \lambda_{1}$.
Find $d_{1}\in D$ such that $\phi_{d} \succeq \lambda_{1}$ for all $d\geq d_{1}$, which is possible by the definition of a subnet.
Let $d_{2}$ be an upper bound of $\lbrace d_{0}, d_{1}\rbrace$ in $(D, \leq)$, which exists since $D$ is directed.
For $d\geq d_{2}$, the triple $(\phi_{d}(x), e_{P, \phi_{d}}(y), r_{\emptyset, \phi_{d}}(y))$ is within $\varepsilon$ of $(\bar{\lambda}(x), \hat{e}_{P}(y), \hat{r}_{\emptyset}(y))$.
\qed

\section{Tight mechanisms with non-random audits}\label{appendix:tight_mechanisms:nonrandom_audits}
\Cref{thm:deterministic_tight_characterization} follows from \Cref{cor:tight_mechs_are_fps} and the following:
\begin{theorem}\label{thm:deterministic_BR_characterization}
    Let $\lambda\in\Lambda$ and $m\in T(\lambda)$.
    If $m$ has non-random audits, then there is $y_{0}\in [\ubar{x}, \bar{x}]$ such that $m$ is a debt-with-relief mechanism with face value $y_{0}$, relief $r_{A}^{\debtwrelief} = c_{A}^{\prime}(e_{A}^{\debtwrelief})$, and agent effort $e_{A}$ given by $e_{A}(x) = e_{A}^{\debtwrelief} \bm{1}(x \in [\ubar{x}, y_{0}))$ for all $x\in[\ubar{x}, \bar{x}]$.
\end{theorem}
In this statement, recall the definition $e_{A}^{\debtwrelief} = \argmin_{\tilde{e}_{A}\in [0, 1]} u_{A}(\tilde{e}_{A}) + c_{A}(\tilde{e}_{A}) + (1 - \tilde{e}_{A}) c_{P}(1)$.

\begin{pf}[Proof of \Cref{thm:deterministic_BR_characterization}]
    Recall $e_{P}(y) = \max\lbrace \alpha(y), \beta(y, e_{A}(y))\rbrace$ for all $y\in[\ubar{x}, \bar{x}]$ and $\beta < 1$.
    Since $e_{P}$ maps to $\lbrace 0, 1\rbrace$, we find $e_{P} = \alpha$.
    Since $\alpha$ is decreasing, there is $y_{0}\in\mathbb{R}$ such that $\alpha(y) = \bm{1}(y < y_{0})$.
    Using this formula for $\alpha$, one may verify $\lambda(y) = \min\lbrace y, y_{0}\rbrace$ for all $y$.
    Hence also $\alpha(y_{0}) = 0$ and $e_{P}(y) = \bm{1}(y < y_{0})$.
    
    We next characterize $e_{A}(y)$, for arbitrary $y$.
    Recall that $e_{A}(y)$ maximizes $\ubar{\pi}(y, \cdot)$.
    If $y \geq y_{0}$, using $e_{P}(y) = 0$, it is easy to check that $e_{A}(y) = 0$ uniquely maximizes $\ubar{\pi}(y, \cdot)$.
    For $y < y_{0}$, using $e_{P}(y) = 1$ we obtain $\ubar{\pi}(y, \tilde{e}_{A}) = y - u_{A}(\tilde{e}_{A}) - c_{A}(\tilde{e}_{A}) - (1 - \tilde{e}_{A}) c_{P}(1)$ for all $\tilde{e}_{A}\in [0, 1]$, and hence $e_{A}(y) = e_{A}^{\debtwrelief}$.
    
    Finally, consider the refund $r_{A}$.
    Fix $y$.
    Since $m\in T(\lambda)$, we know $\Pi_{m}(y) = \ubar{\pi}(y, e_{A}(y))$.
    Spelling out this equation shows $e_{P}(y) r_{P}(y) + (1 - e_{P}(y))r_{\emptyset}(y) = \min\left\lbrace y - u_{A}(e_{A}(y)), \lambda(y)\right\rbrace$.
    Using the formula for $e_{P}$, we thus find $r_{P}(y) = 0$ for $y \in [\ubar{x}, y_{0})$, and $r_{\emptyset}(y) = y - y_{0}$ for $y\in [y_{0}, \bar{x}]$.
    Thus, for all $y \in [\ubar{x}, y_{0})$ the refund $r_{A}(y)$ is given by $r_{A}^{\debtwrelief} = c_{A}^{\prime}(e_{A}^{\debtwrelief})$ to incentivize $e_{A}^{\debtwrelief}$; for $y\in [y_{0}, \bar{x}]$, the refund $r_{A}(y)$ is $0$ to incentivize zero effort.
\end{pf}

\section{Proof of \headercref{Theorem}{{thm:optimal_mech_fully_stochastic}}}\label{appendix:thm:fully_stochastic}
\setcounter{step}{0}
\begin{step}\label{lemma:thm:optimal_mech_fully_stochastic:stochastic_at_top}
    If $m$ is optimal and tight, then $e_{P}$ is strictly positive except at $\bar{x}$.
\end{step}

\begin{pf}
    Let $m$ be tight and optimal. 
    Denote $\lambda = \lambda_{m}$, so that $m\in T(\lambda)$.
    Let $\bar{y} = \inf\lbrace y\in [\ubar{x}, \bar{x}]\colon e_{P}(y) = 0\rbrace$ (recall $e_{P}(\bar{x}) = 0$).
    We show $\bar{y} = \bar{x}$.
    Since $e_{P}$ is decreasing, it follows $e_{P}$ is strictly positive except at $\bar{x}$.

Theorems \ref{thm:stochastic_BR_characterization} and \ref{thm:deterministic_BR_characterization} imply $e_{P}(\bar{x}) = 0$ and $\lambda$ is constantly $\lambda(\bar{y})$ on $[\bar{y}, \bar{x}]$.
    Since $m\in T(\lambda)$, \Cref{lemma:relaxed_operator_wlog} implies $\Pi(x, m(x)) = \hat{\Pi}(x, \lambda(x), e_{A}(x), e_{P}(x))$ and $(e_{A}(x), e_{P}(x))\in E(x, \lambda)$ for all $x$.
    
    We perturb the mechanism.
    Fix $\eta \in (0, 1)$.
    Let $y_{\eta} = \inf\lbrace y\in [\ubar{x}, \bar{x}]\colon e_{P}(y) < \eta\rbrace$, which is well-defined since $e_{P}(\bar{x}) = 0$.
    We now define $\tilde{e}_{P}(x) = \max\lbrace e_{P}(x), \eta\rbrace$ for all $x$.
    Further, we define $\tilde{\lambda}(x) = \lambda(\bar{y}) + \eta(x - \bar{y})$ for all $x\in [\bar{y}, \bar{x}]$; for all other $x$, let $\tilde{\lambda}(x) = \lambda(x)$. 
    Since $e_{P}$ is right-continuous and decreases (Theorems \ref{thm:stochastic_BR_characterization} and \ref{thm:deterministic_BR_characterization}), we have $e_{P}(y_{\eta}) \leq \eta$ and $y_{\eta}\leq\bar{y}$; further, $e_{P}$ and $\tilde{e}_{P}$ coincide before $y_{\eta}$.
    Note $\tilde{\lambda}\leq \id$ since $\lambda\leq \id$ and $\eta < 1$.
    Further, $\tilde{\lambda}\geq \lambda$ since $\lambda$ is constant on $[\bar{y}, \bar{x}]$.

    We next show that, for all $x$ and $y$ such that $y\leq x$, it holds
    \begin{equation}\label{eq:lemma:eliminate_deterministic:feasibility:1}
        \tilde{\lambda}(x) \leq \tilde{e}_{P}(y) x + \min\left\lbrace(1 - \tilde{e}_{P}(y))y, \tilde{\lambda}(y) + u_{A}(e_{A}(y)) + \tilde{e}_{P}(y)\tau\right\rbrace.
    \end{equation}
    For $x\leq \bar{y}$, inequality \eqref{eq:lemma:eliminate_deterministic:feasibility:1} follows since $\tilde{e}_{P}\geq e_{P}$ and $\tilde{\lambda}(x) = \lambda(x)$.
    Thus, let $x\geq \bar{y}$.
    Then \eqref{eq:lemma:eliminate_deterministic:feasibility:1} is established via (the second line follows from $(e_{A}, e_{P})\in E(\lambda)$ and $x \geq \bar{y}$):
    \begin{align*}
    \tilde{\lambda}(x) 
    &\leq \lambda(\bar{y}) + \tilde{e}_{P}(y)(x - \bar{y}) 
    \\ &\leq e_{P}(y) \bar{y} + \min\left\lbrace(1 - e_{P}(y))y, \lambda(y) + u_{A}(e_{A}(y)) + e_{P}(y)\tau\right\rbrace + \tilde{e}_{P}(y)(x - \bar{y}) 
    \\
    &\leq \tilde{e}_{P}(y) x + \min\left\lbrace (1 - \tilde{e}_{P}(y))y, \tilde{\lambda}(y) + u_{A}(e_{A}(y)) + \tilde{e}_{P}(y)\tau\right\rbrace
    .
    \end{align*}

    In view of inequality \eqref{eq:lemma:eliminate_deterministic:feasibility:1}, \Cref{lemma:relaxed_operator_wlog} implies that there is a mechanism whose profit is at least $\hat{\Pi}(x, \tilde{\lambda}(x), e_{A}(x), \tilde{e}_{P}(x))$ for every $x$.
    Before proceeding, we note that $x - u_{A}(e_{A}(x)) \geq \tilde{\lambda}(x) = \lambda(\bar{y}) + \eta(x - \bar{y})$ holds for all $x\in [\bar{y}, \bar{x}]$; indeed, Theorems \ref{thm:stochastic_BR_characterization} and \ref{thm:deterministic_BR_characterization} imply $e_{A}(x) = 0$ for all $x\in [\bar{y}, \bar{x}]$, and we already noted $\tilde{\lambda}(x) \leq x$ for all $x$.
    
    We obtain the following lower bound (the first inequality holds since $m$ is optimal, the second inequality uses $x - u_{A}(e_{A}(x)) \geq \tilde{\lambda}(x) = \lambda(\bar{y}) + \eta(x - \bar{y})$ for all $x\in [\bar{y}, \bar{x}]$, the third inequality uses convexity of $c_{P}$, and the final inequality is by inspection):
    \begin{align*}
        0\geq &\int_{[\ubar{x}, \bar{x}]} \left(\hat{\Pi}(x, \tilde{\lambda}(x), e_{A}(x), \tilde{e}_{P}(x)) -  \hat{\Pi}(x, \lambda(x), e_{A}(x), e_{P}(x))\right)\de F(x)
        \\
        \geq 
        &
        \int_{[\bar{y}, \bar{x}]} \left(\min\left\lbrace x - u_{A}(e_{A}(x)), \lambda(\bar{y}) + \eta(x - \bar{y})\right\rbrace - \min\left\lbrace x - u_{A}(e_{A}(x)), \lambda(x)\right\rbrace\right) \de F(x)  \\
        &- \int_{[y_{\eta}, \bar{x}]}  (1 - e_{A}(x))\left(c_{P}(\eta) - c_{P}(e_{P}(x))\right) \de F(x) 
        \\
        \geq 
        &
        \int_{[\bar{y}, \bar{x}]} \eta (x - \bar{y})\de F(x) - \int_{[y_{\eta}, \bar{x}]} (1 - e_{A}(x))\eta c_{P}^{\prime}(\eta) \de F(x)
        \\
        \geq &\int_{[\bar{y}, \bar{x}]} \eta (x - \bar{y})\de F(x) - \eta c_{P}^{\prime}(\eta)
        .
    \end{align*}
    Now divide by $\eta > 0$ and pass to the limit $\eta\to 0$ to find $0 \geq \int_{[\bar{y}, \bar{x}]} (x - \bar{y})\de F(x) - c_{P}^{\prime}(0)$.
    By assumption, $c_{P}^{\prime}(0)= 0$.
    Since $\bar{x} = \max\supp F$, we conclude $\bar{y} = \bar{x}$. 
\end{pf}

\begin{step}\label{lemma:thm:optimal_mech_fully_stochastic:stochastic}
    If $m$ is optimal and tight, then $m$ has random audits.
\end{step}
\begin{pf}
    Let $m$ have non-random audits.
    We show $m$ is not optimal.
    According to \Cref{thm:deterministic_tight_characterization}, there is a type $y_{0}$ such that $m$ is a debt-with-relief mechanism with threshold $y_{0}$ and relief $r_{A}^{\debtwrelief}$.
    By the previous step, it suffices to show $m$ is not optimal if $y_{0} = \bar{x}$.
    Thus, let $y_{0} = \bar{x}$.
    Define $k = \min_{\tilde{e}_{A}} \tilde{e}_{A}c_{A}^{\prime}(\tilde{e}_{A}) + (1 - \tilde{e}_{A}) c_{P}(1)$.
    For all $x\in [\ubar{x}, \bar{x}]$ profit is given by $\min\lbrace x, \bar{x}\rbrace  - k\bm{1}(x<\bar{x})$.
    Let $\varepsilon > 0$.
    Consider the debt-with-relief mechanism $m_{\varepsilon}$ with face value $\bar{x} - \varepsilon$ (and the same relief as $m$).
    We have (the final line uses that $F$ has no mass point at $\bar{x}$):
    \begin{align*}
        &\int_{[\ubar{x}, \bar{x}]} (\Pi_{m}(x) - \Pi_{m_{\varepsilon}}(x))\de F(x)
        \\
        =&\int_{[\ubar{x},\bar{x}]} \left(\min\lbrace x, \bar{x}\rbrace - \min\lbrace x, \bar{x} - \varepsilon\rbrace - k\bm{1}(x\in [\bar{x}-\varepsilon, \bar{x}))\right)\de F(x)
        \\
        \leq  &\varepsilon \left(1 - F(\bar{x} - \varepsilon)\right) - k\int_{[\bar{x} - \varepsilon, \bar{x})} 1\de F(x)
        \\
        =
        &
        (\varepsilon - k)\left(1 - F(\bar{x} - \varepsilon)\right).
    \end{align*}
    Since $\bar{x} = \max\supp F$ and $k > 0$, for $\varepsilon > 0$ sufficiently close to $0$, the upper bound $(\varepsilon - k)(1 - F(\bar{x} - \varepsilon))$ is strictly negative.
    Thus, $m$ is not optimal.
\end{pf}

\begin{step}\label{lemma:optimal_eP_almostnever_one}
    If $m$ is optimal and tight, then principal effort $e_{P}$ is bounded away from $1$, and the agent's utility $U_{m}$ is bounded away from $0$.
\end{step}
\begin{pf}
    By the previous step, $m$ has random audits. In particular, \Cref{thm:stochastic_BR_characterization} applies.
    Let $z = \inf\iP$, so that $z > \ubar{x}$ and $\inf U_{m} = U_{m}(z) = u_{A}(e_{A}(z)) > 0$.

    It remains to show $e_{P}(\ubar{x}) < 1$.
    We claim $e_{P}(\ubar{x})$ satisfies $1 - \frac{U_{m}(z)}{z} \geq e_{P}(\ubar{x}) + (1 - e_{P}(\ubar{x}))\frac{\ubar{x}}{\bar{x}}$.
    This inequality implies $e_{P}(\ubar{x}) < 1$ since $U_{m}(z) > 0$.
    
    Towards a contradiction, let $1 - \frac{U_{m}(z)}{z} < e_{P}(\ubar{x}) + (1 - e_{P}(\ubar{x}))\frac{\ubar{x}}{\bar{x}}$.
    From \Cref{thm:stochastic_BR_characterization}, recall that $r_{P}$ and $r_{\emptyset}$ are constantly $0$ on $[\ubar{x}, z]$.
    Recall that $e_{P}$ is decreasing and right-continuous at the supremum of the types for which $e_{P}$ equals $1$.
    Hence, there exists $z^{\prime} \in (\ubar{x}, z)$ such that 
    \begin{equation*}
        e_{P}(y) + (1 - e_{P}(y))\frac{y}{\bar{x}}
        -
        \left(1 - \frac{U_{m}(z)}{z}\right)
    \end{equation*}
    is bounded away from $0$ across $y\in [\ubar{x}, z^{\prime}]$.
    Suppose $e_{P}$ is decreased by $\varepsilon > 0$ for all types in $[\ubar{x}, z^{\prime}]$ such that 
    \begin{equation*}
        e_{P}(y) - \varepsilon + (1 - e_{P}(y) + \varepsilon)\frac{y}{\bar{x}}
        >
        1 - \frac{U_{m}(z)}{z}
    \end{equation*}
    is still valid for all $y\in [\ubar{x}, z^{\prime}]$.
    The probability $e_{P}$ is left unchanged for all other types.
    
    Since $r_{P}$ and $r_{\emptyset}$ are constantly $0$ below $z$, for types in $[\ubar{x}, z^{\prime}]$ neither the utility from advancing the full surplus nor the incentive to exert effort changes.
    The incentives to deviate to a type outside $[\ubar{x}, z^{\prime})$ are clearly unaffected.
    Thus, consider the incentive of an arbitrary type $x$ to deviate to a type $y\in [\ubar{x}, z^{\prime})$.
    Since $r_{\emptyset}(y)=0$, the expected utility from deviating to $y$ equals $x - ((e_{P}(y) - \varepsilon) x + (1 - e_{P}(y) + \varepsilon) y)$; the on-path utility is $U_{m}(x)$.
    Thus, we have to show $U_{m}(x) \geq x - ((e_{P}(y) - \varepsilon) x + (1 - e_{P}(y) + \varepsilon) y)$.
    Equivalently,
    \begin{equation*}
        1 - \frac{U_{m}(x)}{x} \leq e_{P}(y) - \varepsilon + (1 - e_{P}(y) + \varepsilon)\frac{y}{x}.
    \end{equation*}
    We distinguish two cases.
    If $x \geq z$, then $U_{m}(x) = x - \lambda_{m}(x)$ since $z = \inf\iP$ and types above $\inf\iP$ have binding ICs.
    We know $\lambda_{m}(x) / x$ is decreasing.
    Hence,
    \begin{align*}
        1 - \frac{U_{m}(x)}{x} = \frac{\lambda_{m}(x)}{x} &\leq \frac{\lambda_{m}(z)}{z} \\
        & = 1 - \frac{U_{m}(z)}{z} \\
        &< 
         e_{P}(y) - \varepsilon + (1 - e_{P}(y) + \varepsilon)\frac{y}{\bar{x}}
        \leq  e_{P}(y) - \varepsilon + (1 - e_{P}(y) + \varepsilon)\frac{y}{x}
        .
    \end{align*}
    Second, take $x \leq z$, so that type $x$'s on-path utility equals $u_{A}(e_{A}(x))$.
    We just verified IC for type $z$, i.e. $1 - \frac{U_{m}(z)}{z} \leq  e_{P}(y) - \varepsilon + (1 - e_{P}(y) + \varepsilon) \frac{y}{z}$.
    Since $e_{A}$ is decreasing below $z$, also
    $1 - \frac{U_{m}(x)}{x} \leq  e_{P}(y) - \varepsilon + (1 - e_{P}(y) + \varepsilon) \frac{y}{x}$.
    
    Thus, the perturbed mechanism is IC.
    Since $y^{\prime} > \ubar{x} = \min\supp F$, we obtain a contradiction to the optimality of $m$.
    \end{pf}
    \qed

\section{Binding incentive constraints}\label{OA:binding_ICs}
Here, we define and derive basic properties of the binding incentive constraints in a given tight mechanism. We shall use these properties to analyze how to tighten a mechanism and to analyze optimal mechanisms.

Fix a function $\lambda\in\Lambda$ and a mechanism $m\in T(\lambda)$.
Given types $y$ and $x$ such that $y\leq x$,
we say \emph{$x$ is a binding IC type of $y$} if the best deviation of $x$ is to $y$. The set of $y$'s binding IC types is denoted $\hat{X}(y)$, i.e.
\begin{equation*}
    \hat{X}(y) = \left\lbrace x\in [y, \bar{x}]\colon \lambda(x) = e_{P}(y) x + (1 - e_{P}(y))(y - r_{\emptyset}(y))\right\rbrace
    .
\end{equation*}
We say $\hat{X}$ is the binding IC correspondence of mechanism $m$.

This definition concerns pairs $(m, \lambda)$ such that $\lambda\in\Lambda$ and $m\in T(\lambda)$. This definition nests the one given in the main text for tight mechanisms $m$ since $\lambda_{m}\in\Lambda$ and $m\in T(\lambda_{m})$ hold for such mechanisms.
As shown later in \Cref{thm:tight_operator_fp}, whenever $m\in T(\lambda)$ and $\lambda\in\Lambda$, the loss function $\lambda_{m}$ of $m$ equals $\lambda$; hence the interpretation of $\hat{X}(y)$ as the set of types whose best deviation is to $y$.

The next lemma summarizes basic properties of the binding ICs.
\begin{lemma}\label{lemma:auxiliary_binding_IC}
    Let $\lambda\in\Lambda$, $m\in T(\lambda)$, and let $\hat{X}$ be the binding IC correspondence.
    Let $\ubar{y} = \max\lbrace y\colon \lambda(y) = y\rbrace$ and $\bar{y} = \min\lbrace y\colon \lambda(y) = \max\lambda\rbrace$.
    Then:
    \begin{enumerate}
        \item The correspondence $\hat{X}$ has non-empty compact convex values. On $[\ubar{y}, \bar{x}]$, the correspondence $\hat{X}$ is upper-hemicontinuous.
        \item If $y\in (\ubar{y}, \bar{y})$, then $y < \min\hat{X}(y)$.
        \item If $z, y\in [\ubar{y}, \bar{y})$ and $z < y$, then, if $e_{P}(z) > e_{P}(y)$, then $\max\hat{X}(z)  \leq \min \hat{X}(y)$; if $e_{P}(z) = e_{P}(y)$, then $\hat{X}(z)  = \hat{X}(y)$.
        \item For all $x\in [\ubar{y}, \bar{y}]$ there exists $y\in [\ubar{y}, x]$ such that $x\in\hat{X}(y)$. Moreover, the image $\hat{X}([\ubar{y}, x])$ equals the interval $[\ubar{y}, \max\hat{X}(x)]$.
    \end{enumerate}
\end{lemma}

\begin{pf}[Proof of \Cref{lemma:auxiliary_binding_IC}]
    Let us abbreviate $\phi(y) = (1 - e_{P}(y))(y - r_{\emptyset}(y))$.
    From Theorems \ref{thm:stochastic_BR_characterization} and \ref{thm:deterministic_BR_characterization}, we know $\phi(y) = \min\lbrace (1 - e_{P}(y))y, \lambda(y) + u_{A}(e_{A}(y)) + e_{P}(y)\tau\rbrace$.

    Let $y \in [\ubar{y}, \bar{x}]$. To see that $\hat{X}(y)$ is non-empty, distinguish three cases. 
    First, if $y\geq\bar{y}$, then $e_{P}(y) = e_{A}(y) = 0$ and $\lambda$ is constant above $y$, implying that $\hat{X}(y)$ is simply the interval $[y, \bar{x}]$.
    Second, let $y \in (\ubar{y}, \bar{y})$, so that $y > \lambda(y)$.
    Using $y > \lambda(y)$, one may verify that the suprema in the definitions of $\alpha(y)$ and $\beta(y, e_{A}(y))$ are attained at points weakly above $y$. Using $e_{P}(y) = \max\lbrace \alpha(y), \beta(y, e_{A}(y))\rbrace > 0$ and rearranging, one may verify $\lambda(x) = e_{P}(y)x + \phi(y)$, meaning $x\in\hat{X}(y)$. Finally, let $y \leq \ubar{y}$.
    Hence, $\lambda(y) = y$, and from here it easily follows $\lambda(y) = y = e_{P}(y)y + \phi(y)$, meaning $y\in\hat{X}(y)$.
    
    Given non-empty values, concavity and continuity of $\lambda$ imply that $\hat{X}(y)$ is a non-empty compact interval. Upper-hemicontinuity on $[\ubar{y}, \bar{x}]$ follows since $\lambda$, $e_{A}$ and $e_{P}$ are right-continuous at $\ubar{y}$ and continuous on $(\ubar{y}, \bar{x}]$ (Theorems \ref{thm:stochastic_BR_characterization} and \ref{thm:deterministic_BR_characterization}).
    
    Next, let $y \in (\ubar{y}, \bar{y})$. Thus, $e_{P}(y) > 0$ and $\lambda(y) < y$. These inequalities imply $\lambda(y) < e_{P}(y) y + (1 - e_{P}(y))y$ and $\lambda(y) < e_{P}(y) y + \lambda(y) + u_{A}(e_{A}(y)) + e_{P}(y)\tau$. In particular, $y\notin \hat{X}(y)$. 
    Since $y \leq \min\hat{X}(y)$ holds definitionally, we get $y < \min\hat{X}(y)$.
    
    Next, let $y, z\in [\ubar{y}, \bar{y})$ and $z < y$.
    Recall that $e_{P}$ is decreasing, and so $e_{P}(z) \geq e_{P}(y)$.
    If $e_{P}(z) = e_{P}(y)$, then also $\phi(y) = \phi(z)$ (since, else, one of $\hat{X}(y)$ and $\hat{X}(z)$ would be empty) and thus $\hat{X}(z) = \hat{X}(y)$.
    Thus let $e_{P}(z) > e_{P}(y)$. 
    Let $x_{y} \in \hat{X}(y)$ and $x_{z} \in \hat{X}(z)$.
    Thus $\lambda(x_{y}) = e_{P}(y) x_{y} + \phi(y) \leq e_{P}(z) x_{y} + \phi(z)$ and $\lambda(x_{z}) = e_{P}(z) x_{z} + \phi(z) \leq e_{P}(y) x_{z} + \phi(y)$ where the inequalities come from $m\in T(\lambda)$.
    Add the two inequalities to obtain $0 \leq (x_{y} - x_{z})(e_{P}(z) -  e_{P}(y))$.
    Thus $x_{z} \leq x_{y}$.

    Finally, let $x\in [\ubar{y}, \bar{y}]$. We show that the image $\hat{X}([\ubar{y}, x])$ equals the interval $[\ubar{y}, \max\hat{X}(x)]$ and that there exists $y\in [\ubar{y}, x]$ such that $x\in\hat{X}(y)$.
    From claim 3 proven above and since $e_{P}$ is decreasing, the image $\hat{X}([\ubar{y}, x])$ is contained in $[\min\hat{X}(\ubar{y}), \max\hat{X}(x)]$.
    Clearly, $\min\hat{X}(\ubar{y})$ and $\max\hat{X}(x)$ are both in $\hat{X}([\ubar{y}, x])$.
    Lemma 2 of \citet{de2008axiomatization} implies $\hat{X}([\ubar{y}, x])=[\min\hat{X}(\ubar{y}), \max\hat{X}(x)]$.\footnote{Lemma 2 of \citet{de2008axiomatization} is an ``Intermediate Value Theorem for correspondences'' and asserts: \emph{Let $[a_{0}, a_{1}]\subset\mathbb{R}$, let $s \in \mathbb{R}$, and let $\Psi\colon [a_{0}, a_{1}]\twoheadrightarrow\mathbb{R}$ be a correspondence with non-empty convex values and a compact graph. If there are $s_{0}\in\Psi(a_{0})$ and $s_{1}\in\Psi(a_{1})$ such that $s_{0} \leq s \leq s_{1}$, then there is $a \in [a_{0}, a_{1}]$ such that $s \in \Psi(a)$.}}
    Using $\lambda(\ubar{y}) = \ubar{y}$, one may verify $\lambda(\ubar{y}) = e_{P}(\ubar{y}) \ubar{y} + \phi(\ubar{y})$.
    Thus, $\hat{X}([\ubar{y}, x])$ equals the interval $[\ubar{y}, \max\hat{X}(x)]$.
    Finally, since $x \leq \max\hat{X}(x)$ and since the image $\hat{X}([\ubar{y}, x])$ equals the interval $[\ubar{y}, \max\hat{X}(x)]$, it follows there is $y\in [\ubar{y}, x]$ such that $x\in\hat{X}(y)$.
\end{pf}

\section{Proofs for \headercref{Section}{{sec:tight_mechanisms:how_to_tighten}}: how to tighten a mechanism?}\label{OA:how_to_tighten}

In this part of the appendix, we prove \Cref{thm:reform}.
We also present an algorithm (\Cref{def:tightening_algorithm} below) that turns an arbitrary given mechanism into a tight mechanism with a type-by-type higher profit.
The steps of the algorithm also proceed type-by-type.

Before defining the algorithm, recall the following notation.
The set $\Lambda_{0}$ denotes the set of functions $\lambda\colon [\ubar{x}, \bar{x}] \to [-\tau, \bar{x}]$ such that $\lambda\leq\id$.
Given $y\in [\ubar{x}, \bar{x}]$ and $\lambda\in\Lambda_{0}$, the set $M(y, \lambda)$ is the set of tuples 
$(\tilde{e}_{A}, \tilde{e}_{P}, \tilde{r}_{A}, \tilde{r}_{P}, \tilde{r}_{\emptyset}) \in [0, 1]^{2}\times [0, y + \tau]^{3}$
such that all of the following hold:
\begin{align*}
    \lambda(y) &\geq y - \left(\tilde{e}_{P} \tilde{r}_{P} + (1 - \tilde{e}_{P}) \tilde{r}_{\emptyset}\right) - u_{A}(\tilde{e}_{A});
    \\
    \forall x\in [y, \bar{x}],\quad \lambda(x) &\leq \tilde{e}_{P} x + (1 - \tilde{e}_{P}) (y - \tilde{r}_{\emptyset});
    \\
    c_{A}^{\prime}(\tilde{e}_{A}) &= \tilde{r}_{A} - \left(\tilde{e}_{P} \tilde{r}_{P} + (1 - \tilde{e}_{P}) \tilde{r}_{\emptyset}\right)
    .
\end{align*}
The profit from such a tuple is
\begin{equation*}
    \Pi(y, \tilde{e}_{A}, \tilde{e}_{P}, \tilde{r}_{A}, \tilde{r}_{P}, \tilde{r}_{\emptyset}) =  y - \tilde{e}_{P} \tilde{r}_{P} - (1 - \tilde{e}_{P})\tilde{r}_{\emptyset} 
    - u_{A}(\tilde{e}_{A}) - c_{A}(\tilde{e}_{A}) - (1 - \tilde{e}_{A})c_{P}(\tilde{e}_{P}).
\end{equation*}
Let $T(y, \lambda) = \argmax_{m(y) \in M(y, \lambda)} \Pi(y, m(y))$.
We write $m\in T(\lambda)$ to mean $m(y) \in T(y, \lambda)$ for all $y$.

\begin{definition}\label{def:tightening_algorithm}
Let $m$ be a mechanism.
The \emph{tightening algorithm} proceeds in three steps. 
\begin{enumerate}
    \item For every type $x \in [\ubar{x}, \bar{x}]$, let $\lambda^{+}(x) = \max\lbrace \ubar{x}, \sup_{y\in [\ubar{x}, x]} \lambda_{m}(y)\rbrace$.
    \item For every type $x\in [\ubar{x}, \bar{x}]$, let
    \begin{equation*}
        \tilde{\lambda}(x) = \inf_{y\in[\ubar{x}, \bar{x}]} e_{P}(y) x + \min\left\lbrace(1 - e_{P}(y))y, \lambda^{+}(y) + u_{A}(e_{A}(y)) + e_{P}(y)\tau\right\rbrace.
    \end{equation*}
    This auxiliary function is also defined in \Cref{lemma:lambda_regularization} and can be interpreted as the loss function induced by optimal refunds when the agent can also deviate by advancing more than the surplus (note the range of the infimum).
    \item Choose $m^{\ast} \in T(\tilde{\lambda})$.
\end{enumerate}
\end{definition}
Recall that $\Lambda$ denotes the set of increasing, concave functions $\lambda\colon [\ubar{x}, \bar{x}]\to \mathbb{R}_{+}$ such that $\lambda(\ubar{x}) = \ubar{x}$ and $\lambda\leq\id$.

If the input mechanism $m$ has a loss function $\lambda_{m}$ that is in $\Lambda$, then steps (1) and (2) of the algorithm are redundant; indeed, in this case $\lambda_{m} = \tilde{\lambda}$ can be shown to hold. 

The following theorem generalizes \Cref{thm:reform} from the main text.
\begin{theorem}\label{thm:tight_operator_fp}
    If $\lambda\in \Lambda$ and $m\in T(\lambda)$, then $m$ is tight and $\lambda_{m} = \lambda$.
    Conversely, if $m$ is a tight mechanism, then $\lambda_{m}\in \Lambda$ and $m\in T(\lambda_{m})$.
    
    If $m$ is a mechanism and $m^{\ast}$ is obtained by applying the tightening algorithm to $m$, then $m^{\ast}$ is tight and tighter than $m$.
\end{theorem}

The remainder of this section proves \Cref{thm:tight_operator_fp}.
We first show that the tightening algorithm tightens the input mechanism and that $\tilde{\lambda}$ is in $\Lambda$.
\begin{lemma}\label{cor:tightening_algorithm_improves}
    Let $m$ be a mechanism.
    If $\tilde{\lambda}$ and $m^{\ast}$ are obtained by applying the tightening algorithm to $m$, then $\tilde{\lambda}\in\Lambda$ and $(\Pi_{m}, \lambda_{m}) \leq (\Pi_{m^{\ast}}, \tilde{\lambda}) \leq  (\Pi_{m^{\ast}}, \lambda_{m^{\ast}})$.
\end{lemma}

\begin{pf}[Proof of \Cref{cor:tightening_algorithm_improves}]
    The function $\tilde{\lambda}$ is as in \Cref{lemma:lambda_regularization}.
    Hence, the lemma implies $\tilde{\lambda}\in\Lambda$, $\lambda_{m}\leq\tilde{\lambda}$, and $\Pi_{m}(x)\leq \hat{\Pi}(x, \tilde{\lambda}, e_{A}(x), e_{P}(x))$ for all $x$.
    Since $m^{\ast}\in T(\tilde{\lambda})$ by construction, \Cref{lemma:relaxed_operator_wlog,lemma:tight_operator_improvement} imply $\hat{\Pi}(x, \tilde{\lambda}, e_{A}(x), e_{P}(x)) \leq \Pi_{m^{\ast}}(x)$ and $\tilde{\lambda}(x) \leq \lambda_{m^{\ast}}(x)$ for all $x$.
    In sum, $\Pi_{m}\leq \Pi_{m^{\ast}}$ and $\lambda_{m}\leq\lambda_{m^{\ast}}$. 
\end{pf}

The difficult step is that the output mechanism from the algorithm cannot be tightened.
For this step, let us record the following corollary of
\Cref{prop:pifunctions_characterization,prop:minimal_principal_effort,lemma:relaxed_operator_wlog}.
\begin{corollary}\label{cor:operator_quasiunique_output}
    Let $\lambda\in\Lambda$.
    Let $x\in[\ubar{x}, \bar{x}]$.
    If $m(x)$ and $m^{\ast}(x)$ are both in $T(x, \lambda)$, then $(e_{A}(x), e_{P}(x))=(e_{A}^{\ast}(x), e_{P}^{\ast}(x))$.
    Further, $\hat{T}(\lambda)$ is a singleton.
\end{corollary}
The following is the key step to prove \Cref{thm:tight_operator_fp}.
\begin{lemma}\label{lemma:almost_tight_operator}
    Let $\lambda, \lambda^{\ast}\in\Lambda$, $m\in T(\lambda)$, and $m^{\ast}\in T(\lambda^{\ast})$. 
    If $(\Pi_{m}, \lambda) \leq (\Pi_{m^{\ast}}, \lambda^{\ast})$, then $(\Pi_{m}, \lambda) = (\Pi_{m^{\ast}}, \lambda^{\ast})$.
\end{lemma}

\begin{pf}[Proof of \Cref{lemma:almost_tight_operator}]
\setcounter{step}{0}
    It suffices to prove $\lambda = \lambda^{\ast}$ since then $m\in T(\lambda)$ and $m^{\ast} \in T(\lambda^{\ast})$ imply $\Pi_{m} = \Pi_{m^{\ast}}$.
    
    We will use the following fact, valid for all $y\in [\ubar{x}, \bar{x}]$:
    \begin{equation}
        \mbox{if}\quad \lambda(y) = \lambda^{\ast}(y), \quad\mbox{then}\quad (e_{A}(y), e_{P}(y)) = (e_{A}^{\ast}(y), e_{P}^{\ast}(y)).
        \label{eq:almost_tight_auxiliary_claim}
    \end{equation}
    To prove this fact, note that the (in)equalities $\lambda(y) = \lambda^{\ast}(y)$ and $\lambda \leq \lambda^{\ast}$ imply $M(y, \lambda^{\ast}) \subseteq M(y, \lambda)$; i.e., to sustain the pointwise higher $\lambda^{\ast}$ the principal has fewer choices than when sustaining $\lambda$. Hence, also $\Pi_{m^{\ast}}(y) \leq \Pi_{m}(y)$. Thus $\Pi_{m^{\ast}}(y) = \Pi_{m}(y)$ and $m^{\ast}(y) \in T(y, \lambda)$. 
    Since also $m(y) \in T(y, \lambda)$, \Cref{cor:operator_quasiunique_output} implies $(e_{A}(y), e_{P}(y)) = (e_{A}^{\ast}(y), e_{P}^{\ast}(y))$.

    The idea of the proof is now as follows. Given a type $y$ for which we have established $\lambda(y) = \lambda^{\ast}(y)$, the claim \eqref{eq:almost_tight_auxiliary_claim} implies $e_{P}(y) = e_{P}^{\ast}(y)$.
    This suggests $\lambda(x) = \lambda^{\ast}(x)$ for all types $x$ who contemplated deviating to $y$; specifically, we confirm this for types $x$ in $y$'s binding IC correspondence under $(m, \lambda)$.
    By ``repeating'' these steps, we deduce $\lambda=\lambda^{\ast}$ using continuity properties of the binding IC correspondence (\Cref{lemma:auxiliary_binding_IC}).

    In the remainder of the proof, we have to take some notational care and be aware of the dependence of certain objects on $(\lambda, m)$ and $(\lambda^{\ast}, m^{\ast})$.
    Specifically, let $\hat{X}$ be the binding IC correspondences for $m$.
    Further, for all $y$, let $\phi(y) = (1 - e_{P}(y))(y - r_{\emptyset}(y))$ and $\phi^{\ast}(y) = (1 - e_{P}^{\ast}(y))(y - r_{\emptyset}^{\ast}(y))$.
    Likewise, the functions $\alpha, \ldots, \ubar{\pi}$ are all as in \Cref{appendix:tight_mechanisms:minimal_ep} for $(m, \lambda)$; $\alpha^{\ast}, \ldots, \ubar{\pi}^{\ast}$ are the counterparts for $(m^{\ast}, \lambda^{\ast})$.

    Let $\ubar{y} = \max\lbrace y\colon\lambda(y) = y\rbrace$ and $\bar{y} = \min\lbrace y\colon \lambda(y) = \max \lambda\rbrace$.
    It holds $\ubar{y} \leq \bar{y}$ since $\lambda\in\Lambda$.
    We distinguish two cases.
    
    First, let $\ubar{y} = \bar{y}$.
    Thus, immediately $\lambda(y) = \lambda^{\ast}(y)$ for all $y\in [\ubar{x}, \ubar{y}]$.
    From the definitions of $\alpha$ and $\beta$, one may verify $\alpha = \beta = 0$ on $[\ubar{y}, \bar{x}]$.
    Hence, also $e_{P} = 0$ on $[\ubar{y}, \bar{x}]$.
    Since $\lambda(\ubar{y}) = \lambda^{\ast}(\ubar{y})$, the claim \eqref{eq:almost_tight_auxiliary_claim} implies $e_{P}^{\ast}(\ubar{y}) = 0$.
    From $e_{P}^{\ast}(\ubar{y}) = 0$, it follows that also $\lambda^{\ast}$ constantly equals its maxima on $[\ubar{y}, \bar{x}]$.
    Thus, $\lambda = \lambda^{\ast}$.

    In what follows, let $\ubar{y} < \bar{y}$.
    In this case, $m$ has random audits. Indeed, $\alpha$ is interior on $(\ubar{y}, \bar{y})$ (by inspection), and $\beta$ never equals $1$.
    Hence also $e_{P}$ is interior on $(\ubar{y}, \bar{y})$.
    
    Define $\bar{z} = \max\left\lbrace x\in [\ubar{x}, \bar{x}]\colon \forall y\in [\ubar{x}, x],\, \lambda(y) = \lambda^{\ast}(y)\right\rbrace$, which is well-defined since $\lambda$ and $\lambda^{\ast}$ agree at $\ubar{x}$ and are continuous.
    We show $\bar{z} = \bar{x}$, proving $\lambda = \lambda^{\ast}$.
    
    To begin with, we note $\ubar{y} \leq \bar{z}$.
    Indeed, all $y\in [\ubar{x}, \ubar{y}]$ satisfy $\lambda(y) = y$, and hence certainly $\lambda(y) = \lambda^{\ast}(y)$ since $\lambda\leq \lambda^{\ast}\leq\id$. Thus, $\ubar{y} \leq \bar{z}$.

    \begin{step}
        There is $\varepsilon > 0$ such that $(e_{A}, e_{P})$ and $(e_{A}^{\ast}, e_{P}^{\ast})$ agree on $[\ubar{y}, \ubar{y} + \varepsilon]$.
    \end{step}
    \begin{pf}
        First, we claim that the efforts $e_{A}$ and $e_{A}^{\ast}$ are bounded away from $0$ on $[\ubar{y}, \ubar{y} + \varepsilon]$ for $\varepsilon > 0$ sufficiently small. For $e_{A}$, this is easily verified using \Cref{thm:stochastic_BR_characterization} and the definition of $\ubar{y}$. 
        We already know $\lambda^{\ast}(\ubar{y}) = \lambda(\ubar{y})$, and hence also $e_{A}^{\ast}(\ubar{y}) > 0$ (by claim \eqref{eq:almost_tight_auxiliary_claim}).
        Using Theorems \ref{thm:stochastic_BR_characterization} and \ref{thm:deterministic_BR_characterization} for $(m^{\ast}, \lambda^{\ast})$, one may verify that $e_{A}^{\ast}$ is thus bounded away from $0$ on a neighborhood of $\ubar{y}$.
        
        Thus, $e_{A}$ and $e_{A}^{\ast}$ are bounded away from $0$ on $[\ubar{y}, \ubar{y} + \varepsilon]$ for $\varepsilon > 0$ sufficiently small. 
        Since $\lambda(\ubar{y}) = \lambda^{\ast}(\ubar{y}) = \ubar{y}$, for $\varepsilon$ sufficiently small and $y\in [\ubar{y}, \ubar{y} + \varepsilon]$, it holds 
        \begin{equation}\label{eq:prop:almost_tight:perturb_at_ubary}
            y - u_{A}(e_{A}(y)) < \min\lbrace \lambda(y),  \lambda^{\ast}(y)\rbrace 
            \quad\text{and}\quad
            y - u_{A}(e^{\ast}_{A}(y)) < \min\lbrace \lambda(y),  \lambda^{\ast}(y)\rbrace
            .
        \end{equation}
        Fix such $\varepsilon > 0$.
        We show $e_{A}$ and $e_{A}^{\ast}$ agree on $[\ubar{y}, \ubar{y} + \varepsilon]$.
        Let $y\in [\ubar{y}, \ubar{y} + \varepsilon]$.
        The strict inequalities \eqref{eq:prop:almost_tight:perturb_at_ubary} imply $e_{P}(y) = \alpha(y)$ and $e_{P}^{\ast}(y) = \alpha^{\ast}(y)$, and that, for all $\tilde{e}_{A} \in \lbrace e_{A}(y), e_{A}^{\ast}(y)\rbrace$, we have $\ubar{\pi}(y, \tilde{e}_{A}) = y - u_{A}(\tilde{e}_{A}) - c_{A}(\tilde{e}_{A}) - (1 - \tilde{e}_{A}) \alpha(y)$ and $\ubar{\pi}^{\ast}(y, \tilde{e}_{A}) = y - u_{A}(\tilde{e}_{A}) - c_{A}(\tilde{e}_{A}) - (1 - \tilde{e}_{A}) \alpha^{\ast}(y)$.
        Now recall that $e_{A}(y)$ maximizes $\ubar{\pi}(y, \cdot)$ (yielding profit $\Pi_{m}(y)$), while $e_{A}^{\ast}(y)$ maximizes $\ubar{\pi}^{\ast}(y, \cdot)$ (yielding profit $\Pi_{m^{\ast}}(y)$).
        Since $\lambda\leq\lambda^{\ast}$, we deduce $\alpha(y) \leq \alpha^{\ast}(y)$.
        Inspecting the profits $\ubar{\pi}$ and $\ubar{\pi}^{\ast}$, the assumption $\Pi_{m}(y) \leq \Pi_{m^{\ast}}(y)$ thus requires $\alpha(y) \geq \alpha^{\ast}(y)$. Hence, $\alpha(y) = \alpha^{\ast}(y)$, and hence also $e_{P}(y) = e_{P}^{\ast}(y)$.\footnote{The inequality $y - u_{A}(e_{A}(y)) < \lambda(y)$ from \eqref{eq:prop:almost_tight:perturb_at_ubary} implies $\alpha(y) \geq \beta(y, e_{A}(y))$, where $\beta$ is as in \Cref{def:alpha_beta} for $\lambda$. Thus, $e_{P}(y) = \alpha(y)$. A similar argument shows $e_{P}^{\ast}(y) = \alpha^{\ast}(y)$.}
        Using now that the map $\tilde{e}_{A}\mapsto y - u_{A}(\tilde{e}_{A}) - c_{A}(\tilde{e}_{A}) - (1 - \tilde{e}_{A}) \alpha(y)$ is strictly quasiconcave (\Cref{assumption:rent_regularity}), we also deduce $e_{A}(y) = e_{A}^{\ast}(y)$.
    \end{pf}

    \begin{step}
        There is $\varepsilon > 0$ such that $\ubar{y} + \varepsilon \leq \bar{z}$.
    \end{step}
    \begin{pf}
        Let $\varepsilon > 0$ be as in the previous step.
        To show $\ubar{y} + \varepsilon \leq \bar{z}$, we show $\lambda$ and $\lambda^{\ast}$ agree on $[\ubar{y}, \ubar{y} + \varepsilon]$.
        Fix $x\in [\ubar{y}, \ubar{y} + \varepsilon]$.
        \Cref{lemma:auxiliary_binding_IC} implies that there is $y\in [\ubar{y}, x]$ such that $x\in \hat{X}(y)$.
        Hence also $y\in [\ubar{y}, \ubar{y} + \varepsilon]$, and hence $(e_{A}(y), e_{P}(y)) = (e_{A}^{\ast}(y), e_{P}^{\ast}(y))$ by the choice of $\varepsilon$.
        Therefore, $\lambda(x) = e_{P}(y) x + \phi(y) = e_{P}^{\ast}(y)x + \phi^{\ast}(y) \geq \lambda^{\ast}(x)$.
        Since $\lambda \leq \lambda^{\ast}$, also $\lambda(x) = \lambda^{\ast}(x)$.
    \end{pf}

    \begin{step}
        It holds $\bar{y} \leq \bar{z}$.
    \end{step}
    \begin{pf}
        Since there is $\varepsilon > 0$ such that $\lambda$ and $\lambda^{\ast}$ agree on $[\ubar{x}, \ubar{y} + \varepsilon]$, it suffices to show the following: \emph{for all $x\in (\ubar{y}, \bar{y})$, if $\lambda$ and $\lambda^{\ast}$ agree on $[\ubar{x}, x]$, then there exists $\delta > 0$ such that $\lambda$ and $\lambda^{\ast}$ agree on $[\ubar{x}, x + \delta]$.}
        (The previous step ensures that we can ignore $x = \ubar{y}$.)
        Fix $x\in (\ubar{y}, \bar{y})$ such that $\lambda$ and $\lambda^{\ast}$ agree on $[\ubar{x}, x]$.
        \Cref{lemma:auxiliary_binding_IC} implies that the image $\hat{X}([\ubar{y}, x])$ equals the interval $[\ubar{y}, \max \hat{X}(x)]$.
        Since $x\in (\ubar{y}, \bar{y})$, we have $e_{P}(x)\in (0, 1)$ and hence, importantly, \Cref{lemma:auxiliary_binding_IC} implies $x < \min\hat{X}(x)$.
        Now put $\delta = \max (\hat{X}(x)) - x$.
        Thus, $\delta > 0$ and $\hat{X}([\ubar{y}, x]) = [\ubar{y}, x + \delta]$.
        We prove that $\lambda$ and $\lambda^{\ast}$ agree on $(x, x + \delta]$.
        Thus, let $x^{\prime} \in (x, x + \delta]$.
        Hence, there is $y \in [\ubar{x}, x]$ such that $x^{\prime}\in\hat{X}(y)$.
        Since $y \leq x$, we have $\lambda(y) = \lambda^{\ast}(y)$ (by the assumption on $x$).
        The claim \eqref{eq:almost_tight_auxiliary_claim} implies $(e_{A}(y), e_{P}(y)) = (e_{A}^{\ast}(y), e_{P}^{\ast}(y))$.
        Thus, $\lambda(x^{\prime}) = e_{P}(y) x^{\prime} + \phi(y) = e_{P}^{\ast}(y)x^{\prime} + \phi^{\ast}(y)$. Since $x^{\prime}\geq y$, also $e_{P}^{\ast}(y)x^{\prime} + \phi^{\ast}(y)\geq \lambda^{\ast}(x^{\prime})$.
        In particular, $\lambda(x^{\prime}) \geq \lambda^{\ast}(x^{\prime})$. Since $\lambda\leq \lambda^{\ast}$, we conclude $\lambda(x^{\prime}) = \lambda^{\ast}(x^{\prime})$.
    \end{pf}

    \begin{step}
        It holds $\bar{x} = \bar{z}$.
    \end{step}
    \begin{pf}
        Since $\bar{y}\leq \bar{z}$, it suffices to show that $\lambda$ and $\lambda^{\ast}$ agree on $[\bar{y}, \bar{x}]$.
        \Cref{thm:stochastic_BR_characterization} establishes $\lambda(\bar{y}) = \max \lambda$ and $e_{P}(\bar{y}) = e_{A}(\bar{y}) = 0$. 
        Since $\bar{y} \leq \bar{z}$, we know $\lambda(\bar{y}) = \lambda^{\ast}(\bar{y})$, and hence the claim \eqref{eq:almost_tight_auxiliary_claim} implies $e_{P}^{\ast}(\bar{y}) = e_{A}^{\ast}(\bar{y}) = 0$.
        Thus also $\lambda^{\ast}(\bar{y}) = \max \lambda^{\ast}$ (by inspecting the definitions of $\alpha^{\ast}$ and $\beta^{\ast}$).
        Summarizing, $\max \lambda = \lambda(\bar{y}) = \lambda^{\ast}(\bar{y}) = \max\lambda^{\ast}$. Since $\lambda$ and $\lambda^{\ast}$ are increasing, we conclude that $\lambda$ and $\lambda^{\ast}$ agree on $[\bar{y}, \bar{x}]$.
    \end{pf}
    Since $\bar{x} = \bar{z}$, we conclude $\lambda = \lambda^{\ast}$.
\end{pf}

Equipped with \Cref{lemma:almost_tight_operator}, \Cref{thm:tight_operator_fp} follows easily.
\begin{pf}[Proof of \Cref{thm:tight_operator_fp}]
    We first show: \emph{a mechanism $m$ is tight if and only if there exists $\lambda\in\Lambda$ such that $m\in T(\lambda)$; in this case, $\lambda = \lambda_{m}$.}
    We already know that if $m$ is tight, then $\lambda_{m}\in\Lambda$ and $m\in T(\lambda_{m})$ (\Cref{thm:fixed_point}).
    Thus, it remains to show: \emph{if $\lambda\in\Lambda$ such that $m\in T(\lambda)$, then $m$ is tight and $\lambda_{m} = \lambda$.}
    To show that $m$ is tight, let $m^{\prime}$ be a mechanism such that $(\Pi_{m}, \lambda_{m}) \leq (\Pi_{m^{\prime}}, \lambda_{m^{\prime}})$. We show $(\Pi_{m}, \lambda_{m}) = (\Pi_{m^{\prime}}, \lambda_{m^{\prime}})$.
    We know $\lambda\leq\lambda_{m}$ (\Cref{lemma:tight_operator_improvement}).
    By invoking \Cref{cor:tightening_algorithm_improves}, find $\tilde{\lambda}\in\Lambda$ and $m^{\ast}\in T(\tilde{\lambda})$ such that $(\Pi_{m^{\prime}}, \lambda_{m^{\prime}}) \leq (\Pi_{m^{\ast}}, \tilde{\lambda})$ (i.e. apply the tightening algorithm to $m^{\prime}$).
    Thus, $(\Pi_{m}, \lambda)\leq (\Pi_{m^{\ast}}, \tilde{\lambda})$.
    \Cref{lemma:almost_tight_operator} now implies $(\Pi_{m}, \lambda)= (\Pi_{m^{\ast}}, \tilde{\lambda})$.
    Thus, also $(\Pi_{m}, \lambda_{m}) = (\Pi_{m^{\prime}}, \lambda_{m^{\prime}})$.
    The identity $\lambda=\lambda_{m}$ follows by simply applying the same arguments to $m^{\prime} = m$.

    Finally, let $m^{\ast}$ be obtained by applying the tightening algorithm to $m$.
    \Cref{cor:tightening_algorithm_improves} immediately implies that $m^{\ast}$ is tighter than $m$.
    The previous paragraph implies that $m^{\ast}$ is tight since $\tilde{\lambda}\in\Lambda$ (\Cref{cor:tightening_algorithm_improves}) and since $m^{\ast}\in T(\tilde{\lambda})$ (by construction).
\end{pf}

\section{Optimal mechanisms}\label{OA:trade-offs}
\subsection{Results}
This appendix presents the analysis of optimal mechanisms alluded to in \Cref{sec:optimality:trade_off}.
We first prove that an optimal tight mechanism exists, and that all optimal mechanisms are essentially tight.
We then define mechanisms with doubly unique binding ICs.
For such mechanisms we describe the audit-the-poor-or-burden-the-rich perturbation and derive necessary conditions for optimality.
Then, we provide a regularity condition such that the necessary conditions are also sufficient, even without double uniqueness.
Finally, we show that optimal tight mechanisms are well-approximated by tight mechanisms with doubly unique binding ICs.
The proofs are in \Cref{OA:trade-offs:proofs}.

\subsubsection{Existence and essential tightness.}
\begin{definition}
    A mechanism $m$ is \emph{essentially tight} if there is a tight mechanism $m^{\ast}$ such that $(\Pi_{m}, U_{m}, e_{A}, e_{P})=(\Pi_{m^{\ast}}, U_{m^{\ast}}, e_{A}^{\ast}, e_{P}^{\ast})$ holds $F$-almost everywhere.
\end{definition}
\begin{lemma}\label{lemma:optimal_existence}
    A tight optimal mechanism exists. 
    All optimal mechanisms are essentially tight.
\end{lemma}

\subsubsection{Doubly unique binding ICs.}
Here, fix a tight mechanism $m$ such that $e_{P}$ is interior on $[\ubar{x}, \bar{x})$.
We focus on such mechanisms since all optimal mechanisms are essentially tight (\Cref{lemma:optimal_existence}) and since a necessary condition for optimality is that $e_{P}$ be interior on $[\ubar{x}, \bar{x})$ (\Cref{thm:optimal_mech_fully_stochastic}).
Hence, $\iSP, \ldots, \iSR$ cover $[\ubar{x}, \bar{x})$.

Given a type $y\in [\ubar{x}, \bar{x}]$, let 
\begin{equation*}
    \hat{X}(y) = \lbrace x\in [y, \bar{x}]\colon \lambda_{m}(x) = e_{P}(y) x + (1 - e_{P}(y))(y - r_{\emptyset}(y))\rbrace
\end{equation*}
be the binding IC correspondence (which is non-empty valued, by \Cref{lemma:auxiliary_binding_IC}).

\begin{definition}
    A tight mechanism $m$ has \emph{doubly unique} binding ICs if on $(\ubar{x}, \bar{x})$ the binding IC correspondence $\hat{X}$ is singleton-valued and for all $x \in (\ubar{x}, \bar{x})$ there exists at most one $y\in (\ubar{x}, \bar{x})$ such that $x\in \hat{X}(y)$.
\end{definition}
From \Cref{lemma:auxiliary_binding_IC} and since $e_{P}$ is decreasing, we know that $\hat{X}$ is upper-hemicontinuous, and that for all $y, y^{\prime}$ such that $y < y^{\prime}$, it holds $\hat{X}(y) = \hat{X}(y^{\prime})$ or $\max\hat{X}(y) \leq \min\hat{X}(y^{\prime})$.
Hence, for doubly unique binding ICs, the unique binding IC type is strictly increasing until it reaches the highest type. Formally:
\begin{lemma}
    A tight mechanism $m$ has doubly unique binding ICs if and only if on $(\ubar{x}, \bar{x})$ the binding IC correspondence $\hat{X}$ is singleton-valued and its unique selection $\hat{x}$ is continuous, and there exists $x^{\ast} \in [\ubar{x}, \bar{x}]$ such that $\hat{x}$ is strictly increasing on $(\ubar{x}, x^{\ast}]$, and all $y\in [x^{\ast}, \bar{x}]$ satisfy $\hat{x}(y) = \bar{x}$.
\end{lemma}
We will write $x^{\ast} = \min \hat{x}^{-1}(\bar{x})$, where $\hat{x}^{-1}(\bar{x})$ means the pre-image of $\hat{x}$. Note, $\hat{x}(\bar{x}) = \bar{x}$, so the pre-image is non-empty.
In optimal tight mechanisms with doubly unique binding ICs, we will show $x^{\ast} = \bar{x}$, so that $\hat{x}$ is indeed strictly increasing on all of $(\ubar{x}, \bar{x})$.

To study the trade-off across types, consider the following perturbation.
Fix $y$.
The principal increases $e_{A}(\hat{x}(y))$ (by increasing the refund $r_{A}(\hat{x}(y))$), thereby reducing the incentive of $\hat{x}(y)$ to deviate to $y$.
Thus, the principal can decrease $e_{P}(y)$, perturbing some of $e_{A}(y)$, $r_{A}(y)$, $r_{P}(y)$, and $r_{\emptyset}(y)$ to hold type $y$'s on-path utility constant.

We decompose the overall impact of the perturbation into two parts capturing the change to the total evidence costs at $\hat{x}(y)$ and $y$.
For all $y\in (\ubar{x}, \bar{x})$, we define the \emph{direct impact $D(y;m, \hat{x})$} and the \emph{indirect impact $I(y;m, \hat{x})$} as:\footnote{In equation \eqref{eq:def:direct_impact}, we can divide by $u_{A}^{\prime}(e_{A}(y))$ since $u_{A}^{\prime}(e_{A}(y))$ is non-zero if $e_{A}(y)$ is non-zero, which is the case for $y\in (\ubar{x}, \bar{x})$. In equation \eqref{eq:def:indirect_impact}, we can divide by $\hat{x}(y) - y$ or $\hat{x}(y) + \tau$ since $y < \hat{x}(y)$ holds for all $y\in (\ubar{x}, \bar{x})$, and since $0 < \ubar{x} + \tau$ holds by assumption.}
\begin{subequations}\label{eq:def:impacts}
\begin{align}
    \label{eq:def:direct_impact}
    D(y; m, \hat{x}) &= \frac{c_{A}^{\prime}(e_{A}(y)) - c_{P}(e_{P}(y))}{u_{A}^{\prime}(e_{A}(y))},
    \\
    \label{eq:def:indirect_impact}
    I(y; m, \hat{x}) &=
    \begin{dcases}
        \frac{(1 - e_{A}(y)) c_{P}^{\prime}(e_{P}(y))}{\hat{x}(y) - y}, \qquad\hfill\mbox{if $y \in \iSP\cup \iP\cup \iM$};
        \\
        \frac{(1 - e_{A}(y)) c_{P}^{\prime}(e_{P}(y))}{\hat{x}(y) - y}
        - \frac{y + \tau}{\hat{x}(y) - y} D(y),
        \qquad\hfill\mbox{if $y \in \iR$};
        \\
        \frac{(1 - e_{A}(y)) c_{P}^{\prime}(e_{P}(y))}{\hat{x}(y) + \tau},
        \qquad\hfill\mbox{if $y \in \iSR$}.
    \end{dcases}
\end{align}
\end{subequations}

We interpret the impacts.
Increasing $e_{A}(\hat{x}(y))$ represents $\$ 1$ of surplus that falls to $\hat{x}(y)$ plus the impact $D(\hat{x}(y); m, \hat{x})$, interpreted as follows.
Increasing $e_{A}(\hat{x}(y))$ destroys some surplus through the agent's effort, $c_{A}^{\prime}(e_{A}(\hat{x}(y))$, but restores some surplus by decreasing on-path at type $\hat{x}(y)$ the principal's effort costs, $-c_{P}(e_{P}(\hat{x}(y)))$.
The normalization via $u_{A}^{\prime}(e_{A}(\hat{x}(y)))$ identifies the correct rate of change when the decrease of $e_{A}(\hat{x}(y))$ decreases the on-path utility of $\hat{x}(y)$ by $\$ 1$.

The indirect impact $I(y; m, \hat{x})$ is more delicate and depends on which of the five intervals contains $y$.
In all five intervals, the impact entails the costs $(1 - e_{A}(y)) c_{P}^{\prime}(e_{P}(y))$ from increasing $e_{P}(y)$.
The normalization, either $\hat{x}(y) - y$ or $\hat{x}(y) + \tau$, identifies the correct change of $e_{P}(y)$ when the on-path utility of $\hat{x}(y)$ increases by $\$ 1$.
To hold $y$'s on-path utility constant for $y$ in $\iSP\cup \iP\cup \iM$, the principal perturbs $r_{P}(y)$.
On $\iSR$, we have $r_{P}(y) = y + \tau$ fixed and the principal instead perturbs $r_{\emptyset}(y)$; perturbing $r_{\emptyset}(y)$, of course, impacts the incentives of others to falsely advance $y$, but this is accounted for by the perturbation of $e_{P}(y)$, which explains why the normalization on $\iSR$ is $\hat{x}(y) + \tau$.
For $y$ in $\iR$, however, the principal holds $y$'s on-path utility constant by perturbing $e_{A}(y)$ (by perturbing $r_{A}(y)$).
Perturbing $e_{A}(y)$, in turn, has the impact of $D(y; m, \hat{x})$ on the evidence production costs at $y$, interpreted just like $D(\hat{x}(y); m, \hat{x})$ but for type $y$; the factor $\frac{y + \tau}{\hat{x}(y) - y}$ identifies the correct change of $e_{A}(y)$ that holds $y$'s on-path utility constant.

\subsubsection{Necessary conditions for optimality.}
The perturbation marginally decreases profits at $\hat{x}(y)$ by $1 + D(\hat{x}(y); m, \hat{x})$, but marginally increases profits at $y$ by $I(y; m, \hat{x})$.
Optimally, these are balanced, as confirmed by equation \eqref{eq:thm:necessary_for_optimality:1} below.
\begin{theorem}\label{thm:necessary_for_optimality}
    Let the type distribution $F$ be absolutely continuous and have full support.
    Let $m$ be tight, optimal
    and have doubly unique binding ICs.
    Let $\hat{x}$ be the unique binding IC selection on $(\ubar{x}, \bar{x})$.
    Let $x^{\ast} = \min \hat{x}^{-1}(\bar{x})$.
    Then,
    \begin{subequations}
        \begin{align}
            \label{eq:thm:necessary_for_optimality:1}
            \forall [a, b]\subset(\ubar{x}, \bar{x}), 
            \quad \int_{[a, b]} I(y; m, \hat{x}) \de F(y) &= \int_{[\hat{x}(a), \hat{x}(b)]}(1 + D(y; m, \hat{x}))\de F(y);
            \\
            \label{eq:thm:necessary_for_optimality:2}
            \forall x\in\iSP,\quad D(x; m, \hat{x}) &= -1;
            \\
            \label{eq:thm:necessary_for_optimality:3}
            \forall x\in\iM,\quad D(x; m, \hat{x}) &= 0;
            \\
            \label{eq:thm:necessary_for_optimality:4}
            \forall x\in\iSR,\quad D(x; m, \hat{x}) &= I(x; m, \hat{x}).
        \end{align}
    \end{subequations}
    Further, $\inf_{y\in (\ubar{x}, \bar{x})} \hat{x}(y) = \sup\iSP$ and $x^{\ast} = \bar{x}$.
\end{theorem}

In addition to the perturbation that draws out the trade-off across types, there are perturbations affecting only the profit at a single type, yielding \eqref{eq:thm:necessary_for_optimality:2}, \eqref{eq:thm:necessary_for_optimality:3} and \eqref{eq:thm:necessary_for_optimality:4}.
These are the respective first-order conditions for maximizing $\pi_{1}$, $\pi_{2}$, and $\pi_{3}$ (as defined in \Cref{appendix:tight_mechanisms:minimal_ep}).

\subsubsection{Sufficient conditions for optimality.}
\Cref{thm:necessary_for_optimality} characterizes necessary conditions for a tight mechanism with doubly unique binding ICs to be optimal.
We now provide a regularity condition under which these conditions are also sufficient, even without double uniqueness.
The regularity condition (\Cref{assumption:global_concavity} below) is that the problem of maximizing expected profit is a concave optimization problem.

For all $y\in [\ubar{x}, \bar{x}]$, $\tilde{e}_{A}\in [0, 1]$, and $\lambda\in\Lambda$, let $\ubar{\pi}(y, \tilde{e}_{A}, \lambda)$ be the profit defined in \Cref{def:auxiliary_profit functions} (\Cref{appendix:tight_mechanisms:minimal_ep}; there we did not notate the dependence on $\lambda$).
For a tight mechanism $m$ with loss function $\lambda = \lambda_{m}$, we know $\lambda\in\Lambda$ and that for each type $y$ the profit is given by $\max_{\tilde{e}_{A}\in [0, 1]}\ubar{\pi}(y, \tilde{e}_{A}, \lambda)$.
The regularity condition is that for all $y$ the profit $\ubar{\pi}(y, \tilde{e}_{A}, \lambda)$ is concave in $(\tilde{e}_{A}, \lambda)$ for relevant values of $\tilde{e}_{A}$.
\begin{assumption}\label{assumption:global_concavity}
    There exists $\check{e}_{A}\in [0, 1]$ such that:
    \begin{enumerate}
        \item For all $y\in [\ubar{x}, \bar{x}]$ and $\lambda\in\Lambda$, if $e_{A}(y)\in \argmax_{\tilde{e}_{A}\in [0, 1]} \ubar{\pi}(y, \tilde{e}_{A}, \lambda)$, then $e_{A}(y) \leq \check{e}_{A}$.
        \item For all $y\in [\ubar{x}, \bar{x}]$, the map $(\tilde{e}_{A}, \lambda)\to \ubar{\pi}(y, \tilde{e}_{A}, \lambda)$ is concave on $[0, \check{e}_{A}]\times\Lambda$.
    \end{enumerate}
\end{assumption}
In \Cref{appendix:sufficient_condition_for_global_concavity}, we provide a sufficient condition on the cost functions $c_{A}$ and $c_{P}$ such that \Cref{assumption:global_concavity} holds.

To show that the conditions from \Cref{thm:necessary_for_optimality} are sufficient for optimality, we focus on tight mechanisms (since all optimal mechanisms are essentially tight; \Cref{lemma:optimal_existence}) such that $e_{P}$ is interior on $[\ubar{x}, \bar{x})$ (which is necessary for $m$ to be optimal; \Cref{thm:optimal_mech_fully_stochastic}).
\begin{theorem}\label{thm:sufficient_for_optimality}
    Let $F$ be absolutely continuous and have full support.
    Let \Cref{assumption:global_concavity} hold.
    Let $m$ be tight and such that $e_{P}$ is interior on $[\ubar{x}, \bar{x})$.
    Then, $m$ is optimal if there exists a continuous, strictly increasing binding IC selection $\hat{x}\colon (\ubar{x}, \bar{x}) \to (\ubar{x}, \bar{x})$ for $m$ satisfying $\inf_{y\searrow \ubar{x}} \hat{x}(y) = \sup\iSP$ and the following:
    \begin{subequations}
        \begin{align}
            \label{eq:thm:sufficient_for_optimality:1}
            \forall [a, b]\subset(\ubar{x}, \bar{x}), 
            \quad \int_{[a, b]} I(y; m, \hat{x}) \de F(y) &= \int_{[\hat{x}(a), \hat{x}(b)]}(1 + D(y; m, \hat{x}))\de F(y);
            \\
            \label{eq:thm:sufficient_for_optimality:2}
            \forall x\in\iSP,\quad D(x; m, \hat{x}) &= -1;
            \\
            \label{eq:thm:sufficient_for_optimality:3}
            \forall x\in\iM,\quad D(x; m, \hat{x}) &= 0;
            \\
            \label{eq:thm:sufficient_for_optimality:4}
            \forall x\in\iSR,\quad D(x; m, \hat{x}) &= I(x; m, \hat{x}).
        \end{align}
    \end{subequations}
\end{theorem}

\begin{remark}
    Conditions \eqref{eq:thm:sufficient_for_optimality:2} and \eqref{eq:thm:sufficient_for_optimality:3} are already implied by tightness of $m$ and do not depend on the selection $\hat{x}$. Condition \eqref{eq:thm:sufficient_for_optimality:4} does depend on $\hat{x}$.
\end{remark}

\subsubsection{Approximation.}
The next theorem shows that optimal tight mechanisms are well-approximated via tight mechanisms with doubly unique binding ICs.
\begin{theorem}\label{thm:doubly_unique_approximation}
    For all optimal tight mechanisms $m$ there exists a sequence $(m_{n})_{n\in\mathbb{N}}$ of mechanisms satisfying the following.
    \begin{enumerate}
        \item Each mechanism $m_{n}$ is tight, has doubly unique binding ICs, and $e_{P, n}$ is interior and strictly decreasing on $[\ubar{x}, \bar{x})$.
        \item All $x\in (\ubar{x}, \bar{x})$ satisfy $\lim_{n\to\infty}(e_{A, n}(x), e_{P, n}(x), \Pi_{m_{n}}(x)) = (e_{A}(x), e_{P}(x), \Pi_{m}(x))$.
    \end{enumerate}
\end{theorem}
Since the efforts determine the refunds (except when $r_{P}$ or $r_{\emptyset}$ are never paid), the entire mechanism converges pointwise almost everywhere along the sequence.

\subsection{Proofs for \headercref{Appendix}{{OA:trade-offs}}}\label{OA:trade-offs:proofs}

Given a type $x$ and $\lambda\in\Lambda$, let $\Pi^{\ast}(x, \lambda) = \max_{m(x)\in M(x, \lambda)} \Pi(x, m(x))$ be the optimal profit from applying $T$.
Further, let $\Pi^{\ast}(\lambda) = \int \Pi^{\ast}(x, \lambda)\de F(x)$.

\subsubsection{Proof of \headercref{Lemma}{{lemma:optimal_existence}}.}
    We first show that a tight optimal mechanism exists.
    For every mechanism there is a tight mechanism with a pointwise higher profit (\Cref{lemma:tight_mechanisms_wlog}). In a tight mechanism the expected profit equals $\Pi^{\ast}(\lambda)$.
    Thus, to prove that a tight optimal mechanism exists, it suffices to show that $\lambda\mapsto\Pi^{\ast}(\lambda)$ admits a maximizer across $\lambda\in\Lambda$.
    Since all functions in $\Lambda$ are Lipschitz continuous with constant $1$ (\Cref{appendix:tight_mechanisms:shape_of_loss}) and map to $[\ubar{x}, \bar{x}]$, the Arzel{\`a}--Ascoli Theorem implies that $\Lambda$ is compact in the supremum norm.
    For fixed $y$, the correspondence $\lambda\mapsto E(y, \lambda)$ is upper hemicontinuous and has non-empty compact values (by inspection), and thus $\lambda \mapsto \Pi^{\ast}(y, \lambda)$ is upper semicontinuous \citep[Lemma 17.30]{aliprantis2006infinite}.
    By Fatou's Lemma, also $\lambda\mapsto\int \Pi^{\ast}(y, \lambda) \de F(y)$ is upper semicontinuous.

    Now let $m$ be optimal.
    We show $m$ is essentially tight.
    Let $\lambda_{m}$ denote $m$'s loss function.
    Using \Cref{lemma:lambda_regularization}, find $\lambda\in\Lambda$ such that $(e_{A}, e_{P}) \in E(\lambda)$ and $\Pi_{m}(y) \leq \hat{\Pi}(y, \lambda(y), e_{A}(y), e_{P}(y))$ for all $y\in [\ubar{x}, \bar{x}]$. 
    Find $m^{\ast}\in T(\lambda)$.
    \Cref{thm:tight_operator_fp} implies $m^{\ast}$ is tight.
    \Cref{lemma:relaxed_operator_wlog} implies $(e_{A}^{\ast}, e_{P}^{\ast}) \in \hat{T}(\lambda)$ and $\Pi_{m^{\ast}}(y)  = \hat{\Pi}(y, \lambda(y), e_{A}^{\ast}(y), e_{P}^{\ast}(y)) \geq \hat{\Pi}(y, \lambda(y), e_{A}(y), e_{P}(y))$ for all $y$.
    Since $m$ is optimal, $F$-almost all $y$ satisfy
    \begin{equation*}
        \Pi_{m^{\ast}}(y) = \hat{\Pi}(y, \lambda(y), e_{A}^{\ast}(y), e_{P}^{\ast}(y)) = \hat{\Pi}(y, \lambda(y), e_{A}(y), e_{P}(y)) = \Pi_{m}(y).
    \end{equation*}
    For $y$ such that $\hat{\Pi}(y, \lambda(y), e_{A}(y), e_{P}(y)) = \hat{\Pi}(y, \lambda(y), e_{A}^{\ast}(y), e_{P}^{\ast}(y))$, the inclusions $(e_{A}^{\ast}, e_{P}^{\ast}) \in \hat{T}(\lambda)$ and $(e_{A}, e_{P}) \in E(\lambda)$ imply $(e_{A}, e_{P})\in \hat{T}(\lambda)$.
    But \Cref{cor:operator_quasiunique_output} asserts that $\hat{T}$ is a singleton. Thus, $(e_{A}, e_{P}) = (e_{A}^{\ast}, e_{P}^{\ast})$ for $F$-almost all types.
\qed

\subsubsection{Auxiliary results for the perturbation.}

\begin{lemma}\label{lemma:basic_lambda_approximation}
    Let $(\lambda_{n})_{n\in\mathbb{N}}$ be a sequence in $\Lambda$ converging to $\lambda\in\Lambda$.
    Let $m\in T(\lambda)$ and $m_{n}\in T(\lambda_{n})$ for all $n$.
    If all $y\in (\ubar{x}, \bar{x}]$ satisfy $\lambda(y) < y$, then for all $y\in (\ubar{x}, \bar{x}]$ the efforts $(e_{A, n}(y), e_{P, n}(y))_{n\in\mathbb{N}}$ converge to $(e_{A}(y), e_{P}(y))$. 
\end{lemma}

\begin{pf}[Proof of \Cref{lemma:basic_lambda_approximation}]
    Fixing $y\in (\ubar{x}, \bar{x})$, we first argue that $(e_{A, n}(y))_{n\in\mathbb{N}}$ converges to $e_{A}(y)$.
    It suffices to show that for all subsequences of $(e_{A, n}(y))_{n\in\mathbb{N}}$ there is a subsubsequence that converges to $e_{A}(y)$.
    Fix an initial subsequence. Pass to a subsubsequence along which $(e_{A, n}(y))_{n\in\mathbb{N}}$ converges.
    By relabeling, let $(e_{A, n}(y))_{n\in\mathbb{N}}$ be the convergent subsubsequence.
    For all $n$, the agent effort $e_{A, n}(y)$ is in $\argmax_{\tilde{e}_{A}\in [0, 1]} \ubar{\pi}(y, \tilde{e}_{A}, \lambda_{n})$.
    Likewise, $e_{A}(y) \in \argmax_{\tilde{e}_{A}\in [0, 1]} \ubar{\pi}(y, \tilde{e}_{A}, \lambda)$.
    Since $y > \ubar{x}$, it holds $\lambda(y) < y$ (by assumption) and hence also $\lambda_{n}(y) < y$ for sufficiently large $n$.
    One may then verify that the supremum in the definition of $\alpha(y, \lambda)$ is attained over points $[y + \delta, \bar{x}]$, for some $\delta > 0$.
    A routine application of Berge's Maximum Theorem now shows that $\alpha(y, \lambda_{n})\to\alpha(y, \lambda)$ as $n\to\infty$.
    Berge's Maximum Theorem also implies that $\beta(y, \tilde{e}_{A}, \tilde{\lambda})$ is continuous in $(\tilde{e}_{A}, \tilde{\lambda}) \in [0, 1]\times\Lambda$.
    A final application of Berge's Maximum Theorem now shows that $\tilde{\lambda}\mapsto\argmax_{\tilde{e}_{A}\in [0, 1]} \ubar{\pi}(y, \tilde{e}_{A}, \tilde{\lambda})$ is upper-hemicontinuous at $\lambda$.
    Consequently, the limit of $(e_{A, n}(y))_{n\in\mathbb{N}}$ is in $\argmax_{\tilde{e}_{A}\in [0, 1]} \ubar{\pi}(y, \tilde{e}_{A}, \lambda)$.
    Since $\ubar{\pi}$ is strictly quasiconcave in agent effort (\Cref{prop:pifunctions_characterization}), and since $e_{A}(y)$ is in $\argmax_{\tilde{e}_{A}\in [0, 1]} \ubar{\pi}(y, \tilde{e}_{A}, \lambda)$, we conclude that $(e_{A, n}(y))_{n\in\mathbb{N}}$ converges to $e_{A}(y)$.

    Since $m_{n}\in T(\lambda_{n})$ for all $n$, we have $e_{P, n}(y) = \max\lbrace\alpha(y, \lambda_{n}), \beta(y, e_{A, n}(y), \lambda_{n})\rbrace$ for all $n$.
    Repeating the arguments for continuity of $\alpha$ and $\beta$, we find that for all $y\in (\ubar{x}, \bar{x}]$ the sequence $(e_{P, n}(y))_{n\in\mathbb{N}}$ converges to $e_{P}(y)$.
\end{pf}

\begin{lemma}\label{lemma:lambda_perturbation}
    Let $\lambda\in\Lambda$ and $m\in T(\lambda)$.
    Suppose $m$ has doubly unique binding ICs and let $\lambda(y) < y$ hold for all $y\in (\ubar{x}, \bar{x}]$.
    Let $\hat{x}$ be the unique binding IC selection on $(\ubar{x}, \bar{x})$.
    Let $\eta\colon[\ubar{x}, \bar{x}]\to \mathbb{R}$ be continuous. 
    Let $\lambda + \varepsilon\eta \in \Lambda$ hold for all sufficiently small $\varepsilon > 0$.
    If at least one of the following holds:
    \begin{enumerate}
        \item $\inf_{y\searrow \ubar{x}} \hat{x}(y) \geq \sup\iSP$;\footnote{Here, $\iSP$ means the interval from \Cref{thm:stochastic_BR_characterization}. The proof verifies that \Cref{thm:stochastic_BR_characterization} applies.}
        \item $\eta$ is constantly $0$ on a neighborhood of $\ubar{x}$,
    \end{enumerate}
    then
    \begin{equation}\label{eq:lemma:lambda_perturbation:derivative}
        \lim_{\varepsilon\to 0}\frac{\Pi^{\ast}(\lambda + \varepsilon\eta) - \Pi^{\ast}(\lambda)}{\varepsilon}
        =
        \int (1 + D(y; m, \hat{x}))\eta(y) - I(y; m, \hat{x})\eta(\hat{x}(y)) \de F(y).
    \end{equation}
\end{lemma}
\begin{pf}[Proof of \Cref{lemma:lambda_perturbation}]
    Abbreviate $I(\cdot) = I(\cdot; m, \hat{x})$ and $D(\cdot) = D(\cdot; m, \hat{x})$.

    For all $\varepsilon > 0$, let $m_{\varepsilon}\in T(\lambda + \varepsilon\eta)$.
    Since $\lambda(y) < y$ for all $y\in (\ubar{x}, \bar{x}]$, the effort $e_{P}$ is interior on $(\ubar{y}, \bar{x})$ (\Cref{thm:stochastic_BR_characterization}) and \Cref{lemma:auxiliary_binding_IC} implies $y < \hat{x}(y)$ for all $y\in (\ubar{x}, \bar{x})$.
    Further, invoking \Cref{lemma:basic_lambda_approximation}, the efforts $(e_{A, \varepsilon}(y), e_{P, \varepsilon}(y))_{\varepsilon}$ converge to $(e_{A}(y), e_{P}(y))$.
    It follows that, for $\varepsilon$ sufficiently small, the mechanism $m_{\varepsilon}$ has random audits.
    Let $\iSP, \ldots, \iSR$ and $\iSP_{\varepsilon}, \ldots, \iSR_{\varepsilon}$, respectively, denote the five intervals from \Cref{thm:stochastic_BR_characterization} for $m$ and $m_{\varepsilon}$.
    Since $e_{P}$ is interior on $[\ubar{x}, \bar{x})$, the intervals $\iSP, \ldots, \iSR$ cover $[\ubar{x}, \bar{x})$.
    Since for each fixed $y$ the efforts $(e_{A, \varepsilon}(y), e_{P, \varepsilon}(y))_{\varepsilon}$ converge to $(e_{A}(y), e_{P}(y))$ and $\lambda + \varepsilon \eta$ uniformly converges to $\lambda$, also the boundaries of the intervals $\iSP_{\varepsilon}, \ldots, \iSR_{\varepsilon}$ converge to $\iSP, \ldots, \iSR$ (inspect the definitions of the intervals in \Cref{appendix:tight_mechanisms:random_audits}).
    
    We prove \eqref{eq:lemma:lambda_perturbation:derivative} via Dominated Convergence.
    Specifically, we show
    \begin{equation}\label{eq:lemma:lambda_approximation:derivative}
        \lim_{\varepsilon\to 0}\frac{\Pi^{\ast}(y, \lambda + \varepsilon\eta) - \Pi^{\ast}(y, \lambda)}{\varepsilon}
        =
        (1 + D(y))\eta(y) - I(y)\eta(\hat{x}(y))
    \end{equation}
    that holds for all $y$ in the interior of one of the intervals $\iSP, \ldots, \iSR$, and that \eqref{eq:lemma:lambda_approximation:derivative} is bounded across $y\in (\ubar{x}, \bar{x})$.
    We derive \eqref{eq:lemma:lambda_approximation:derivative} via a case distinction.
    
    Let $y \in \iSP$. Thus, $e_{A}(y)$ maximizes $\pi_{1}(y, \tilde{e}_{A}, \lambda)$ across all $\tilde{e}_{A}$ in a neighborhood $O$ of $e_{A}(y)$. For $\varepsilon$ sufficiently small, type $y$ is also in the interval $\iSP_{\varepsilon}$ of the mechanism $m_{\varepsilon}$. Using $e_{A, \varepsilon}(y) \to e_{A}(y)$, thus also $e_{A, \varepsilon}(y)$ maximizes $\pi_{1}(y, \tilde{e}_{A}, \lambda + \varepsilon \eta)$ across $\tilde{e}_{A}\in O$.
        Hence,
        \begin{align}
            \nonumber
            \Pi^{\ast}(y, \lambda + \varepsilon\eta)
            &=
            \max_{\tilde{e}_{A}\in O} y - u_{A}(\tilde{e}_{A}) - c_{A}(\tilde{e}_{A}) - (1 - \tilde{e}_{A})c_{P}(\alpha(y, \lambda + \varepsilon\eta))
            ,
            \\
            \Pi^{\ast}(y, \lambda)
            &=
            \max_{\tilde{e}_{A}\in O} y - u_{A}(\tilde{e}_{A}) - c_{A}(\tilde{e}_{A}) - (1 - \tilde{e}_{A})c_{P}(\alpha(y, \lambda))
            \label{eq:lemma:lambda_perturbation:SL_profit}
            .
        \end{align}
        Since $\lambda(y) < y$, it holds $\alpha(y, \lambda) = \max_{x\in[y+\delta, \bar{x}]} (\lambda(x) - y) / (x - y)$ for some $\delta > 0$.
        This maximizer is unique since $m$ has doubly unique binding ICs and since $e_{P}(y) = \alpha(y, \lambda)$ (for $y\in \iSP$) holds.
        Using Theorem 3 of \citet{milgrom2002envelope}, one can now show that $\alpha(y, \lambda + \varepsilon\eta)$ is differentiable in $\varepsilon$ at $\varepsilon = 0$, with derivative $\eta(\hat{x}(y)) / (\hat{x}(y) - y)$.
        Since also the maximizer in \eqref{eq:lemma:lambda_perturbation:SL_profit} is unique (since $u_{A} + c_{A}$ is strictly convex, \Cref{assumption:rent_regularity}), another application of Theorem 3 of \citet{milgrom2002envelope} implies that $\Pi^{\ast}(y, \lambda + \varepsilon\eta)$ is differentiable in $\varepsilon$ at $\varepsilon = 0$.
        Specifically, 
        \begin{equation*}
            \lim_{\varepsilon\to 0}\frac{\Pi^{\ast}(y, \lambda + \varepsilon\eta) - \Pi^{\ast}(y, \lambda)}{\varepsilon}
            =
            - (1 - e_{A}(y)) c_{P}^{\prime}(e_{P}(y))\frac{\eta(\hat{x}(y))}{\hat{x}(y) - y}
            =
            -I(y)\eta(\hat{x}(y)),
        \end{equation*}
        where we also used $e_{P}(y) = \alpha(y)$ and the definition of $I(y)$.
        Finally, since $1 + D(y) = 0$ for $y\in \iSP$, we obtain \eqref{eq:lemma:lambda_approximation:derivative}.
    
    For $y\in \iM$, the limit can be evaluated using an argument similar to the one for $\iSP$. Specifically:
        \begin{equation*}
            \lim_{\varepsilon\to 0}\frac{\Pi^{\ast}(y, \lambda + \varepsilon\eta) - \Pi^{\ast}(y, \lambda)}{\varepsilon}
            =
            \eta(y)- (1 - e_{A}(y))c_{P}^{\prime}(e_{P}(y))\frac{\eta(\hat{x}(y))}{\hat{x}(y) - y}
        \end{equation*}
        The effort $e_{A}(y)$ maximizes $\pi_{2}(y, \cdot, \lambda)$ on a neighborhood of $e_{A}(y)$.
        The first-order condition of this maximization reads $D(y) = 0$.
        Hence, \eqref{eq:lemma:lambda_approximation:derivative} holds.

        Similarly, for $y\in \iSR$,
        \begin{equation}\label{eq:lemma:lambda_approximation:SH_derivative}
            \lim_{\varepsilon\to 0}\frac{\Pi^{\ast}(y, \lambda + \varepsilon\eta) - \Pi^{\ast}(y, \lambda)}{\varepsilon}
            =
            \eta(y)- (1 - e_{A}(y))c_{P}^{\prime}(e_{P}(y))\frac{\eta(\hat{x}(y)) - \eta(y)}{\hat{x}(y) + \tau}.
        \end{equation}
        Since $m$ has doubly unique binding ICs, one can verify using Theorem 3 of \citet{milgrom2002envelope} that $\pi_{3}(y, \tilde{e}_{A}, \lambda)$ is differentiable at $\tilde{e}_{A} = e_{A}(y)$.
        Since $e_{A}(y)$ maximizes $\pi_{3}(y, \cdot, \lambda)$ on a neighborhood of $e_{A}(y)$, we obtain the first-order condition  $D(y) = I(y)$.
        Hence, \eqref{eq:lemma:lambda_approximation:SH_derivative} simplifies to \eqref{eq:lemma:lambda_approximation:derivative}.
     
     Consider $y$ in the interior of $\iP$. Hence also $y\in \iP_{\varepsilon}$ for sufficiently small $\varepsilon$ since the boundaries of the intervals converge. Since $y\in \iP\cap \iP_{\varepsilon}$, we know that $u_{A}(e_{A, \varepsilon}(y)) = y - (\lambda(y) + \varepsilon\eta(y))$ holds. Further,
        \begin{align*}
            \Pi^{\ast}(y, \lambda + \varepsilon\eta)
            &=
            \lambda(y) + \varepsilon \eta(y)- c_{A}(e_{A, \varepsilon}(y)) - (1 - e_{A, \varepsilon}(y)) c_{P}(\alpha(y, \lambda + \varepsilon\eta))
            ,
            \\
            \Pi^{\ast}(y, \lambda)
            &=
            \lambda(y) - c_{A}(e_{A}(y)) - (1 - e_{A}(y)) c_{P}(\alpha(y, \lambda))
            .
        \end{align*}
        The derivative of $\alpha(y, \lambda + \varepsilon\eta)$ is computed as for $\iSP$, and the derivative of $e_{A, \varepsilon}(y)$ via implicit differentiation. One then obtains \eqref{eq:lemma:lambda_approximation:derivative} by direct computation.
    
    Consider $y$ in the interior of $\iR$. Again, $y\in \iR_{\varepsilon}$ for sufficiently small $\varepsilon$. Thus, $u_{A}(e_{A, \varepsilon}(y)) = y - \alpha(y, \lambda + \varepsilon\eta)(y + \tau) - (\lambda(y) + \varepsilon\eta(y))$ holds. Further,
        \begin{align*}
            \Pi^{\ast}(y, \lambda + \varepsilon\eta)
            &=
           \lambda(y) + \varepsilon\eta(y) - c_{A}(e_{A, \varepsilon}(y)) - (1 - e_{A, \varepsilon}(y)) c_{P}(\alpha(y, \lambda + \varepsilon\eta))
            ,
            \\
            \Pi^{\ast}(y, \lambda)
            &=
            \lambda(y) - c_{A}(e_{A}(y)) - (1 - e_{A}(y)) c_{P}(\alpha(y, \lambda))
            .
        \end{align*}
        The derivative of $\alpha(y, \lambda + \varepsilon\eta)$ is computed as for $\iSP$, and the derivative of $e_{A, \varepsilon}(y)$ via implicit differentiation. One then obtains \eqref{eq:lemma:lambda_approximation:derivative} by direct computation.
        
    In summary, \cref{eq:lemma:lambda_approximation:derivative} holds for all $y$ in the interior of one $\iSP, \ldots, \iSR$.
    Since $\iSP, \ldots, \iSR$ cover $[\ubar{x}, \bar{x})$ and $F$ is absolutely continuous, \cref{eq:lemma:lambda_approximation:derivative} holds $F$-almost everywhere.

    It remains to show $\sup_{y\in (\ubar{x}, \bar{x})} (1 + D(y))\eta(y) - I(y)\eta(\hat{x}(y)) < \infty$.
    On $(\ubar{x}, \bar{x})$, the term $(1 + D(y))\eta(y) - I(y)\eta(\hat{x}(y))$ is continuous in $y$ and real-valued.
    Next, consider the limits as $y\to \ubar{x}$.
    Recall that $\iSP$ has a non-empty interior, so that $\sup\iSP > \ubar{x}$.
    Hence, by assumption it holds $\liminf_{y\to\ubar{x}} \hat{x}(y) - y > 0$ or it holds that $\eta$ is constantly zero on a neighborhood of $\ubar{x}$.
    In either case, we find that $(1 + D(y))\eta(y) - I(y)\eta(\hat{x}(y))$ is bounded as $y\to \ubar{x}$ by inspecting the definitions of $D$ and $I$ and recalling that $e_{A}$ is continuous and strictly positive on $[\ubar{x}, \bar{x})$. 
    Finally, consider $\bar{x}$.
    For all $y$ sufficiently close to $\bar{x}$, we have $y\in\iSR$, and hence $D(y) = I(y)$. Using $e_{P}(y) \to 0$ as $y\to\bar{x}$, one may verify $I(y)\to 0$. Thus, $(1 + D(y))\eta(y) - I(y)\eta(\hat{x}(y))$ is bounded as $y\to\bar{x}$.
\end{pf}

\begin{lemma}\label{lemma:doubly_unique_construction}
    Let $\lambda \in \Lambda$ be strictly concave and differentiable.
    If $m\in T(\lambda)$, then $m$ has doubly unique binding ICs and $e_{P}$ is strictly decreasing on $[\ubar{x}, \bar{x}]$.
\end{lemma}
\begin{pf}[Proof of \Cref{lemma:doubly_unique_construction}]
    For all $x, y$ such that $y \leq x$, we have $\lambda(x) \leq e_{P}(y) x + (1 - e_{P}(y))(y - r_{\emptyset}(y))$ since $m\in T(\lambda)$.
    That is, fixing $x$, the strictly concave function $x\mapsto\lambda(x)$ lies everywhere below the affine function $x\mapsto e_{P}(y) x + (1 - e_{P}(y))(y - r_{\emptyset}(y))$.
    Hence, the two functions intersect at most once.
    Hence, the binding IC correspondence is singleton valued.
    Let $\hat{x}$ be the unique selection, which is continuous since the binding IC correspondence is upper hemicontinuous (\Cref{lemma:auxiliary_binding_IC}).

    We next show that $e_{P}$ is strictly decreasing.
    Since $\lambda$ is strictly concave and $\lambda\leq\id$, we have $\lambda(y) < y$ on $(\ubar{x}, \bar{x}]$.
    In particular, $e_{P}$ is interior on $[\ubar{x}, \bar{x})$ and \Cref{thm:stochastic_BR_characterization} applies and the intervals $\iSP, \ldots, \iSR$ cover $[\ubar{x}, \bar{x})$.
    \Cref{thm:stochastic_BR_characterization} already asserts that $e_{P}$ strictly decreases on $\iSP, \ldots, \iR$.
    Thus, consider $y, z\in \iSR$ such that $z < y$. We have to show $e_{P}(z) > e_{P}(y)$.
    Towards a contradiction, let $e_{P}(z) = e_{P}(y)$ (we know that $e_{P}$ is decreasing), so that also $\hat{x}(z) = \hat{x}(y)$ (\Cref{lemma:auxiliary_binding_IC}).
    Abbreviate $q = e_{P}(y)$ and $x = \hat{x}(y)$.
    Since $m$ has a unique binding IC selection, one can verify using Theorem 3 of \citet{milgrom2002envelope} that $\pi_{3}(y, \tilde{e}_{A}, \lambda)$ is differentiable at $\tilde{e}_{A} = e_{A}(y)$.
    Since $y\in\iSR$, the effort $e_{A}(y)$ maximizes $\pi_{3}(y, \cdot, \lambda)$ on a neighborhood of $e_{A}(y)$ and is interior. A similar argument applies to $e_{A}(z)$.
    Spelling out the first-order conditions, we find that both $e_{A}(y)$ and $e_{A}(z)$ are solutions, in $\tilde{e}_{A}$, to the equation $0 = c_{A}^{\prime}(\tilde{e}_{A}) - c_{P}(q) - (1 - \tilde{e}_{A})\frac{c_{P}^{\prime}(q)}{x + \tau}u_{A}^{\prime}(\tilde{e}_{A})$.
    The proof of \Cref{lemma:K3_auxiliary_order} shows that the right side is strictly single-crossing as a function of $\tilde{e}_{A}$.
    Thus, $e_{A}(y) = e_{A}(z)$.
    But \Cref{thm:stochastic_BR_characterization} asserts that $e_{A}$ is strictly decreasing; contradiction.

    Let $x^{\ast} = \min \hat{x}^{-1}(\bar{x})$.
    Finally, we show that $\hat{x}$ is strictly increasing on $(\ubar{x}, x^{\ast})$.
    Let $y, z \in (\ubar{x}, x^{\ast})$ be such that $\hat{x}(y) = \hat{x}(z)$.
    We show $y = z$.
    Without loss, let $y\geq z$.
    Put $x_{0} = \hat{x}(y) = \hat{x}(z)$, so that $x_{0} > y\geq z$ (\Cref{lemma:auxiliary_binding_IC}).
    Importantly, also $x_{0} < \bar{x}$ since $y$ and $z$ are both strictly below $x^{\ast}$.
    By definition, $\lambda(x_{0}) = e_{P}(y) x_{0} + (1 - e_{P}(y))(y - r_{\emptyset}(y))$.
    For all $x\geq y$, also $\lambda(x) \leq e_{P}(y) x + (1 - e_{P}(y))(y - r_{\emptyset}(y))$ since $m\in T(\lambda)$.
    Therefore, $\lambda(x) - \lambda(x_{0}) \leq e_{P}(y)(x - x_{0})$ for all $x\geq y$.
    That is, $e_{P}(y)$ is a superdifferential of the restriction of $\lambda$ to $[y, \bar{x}]$ at $x_{0}$.
    By the same argument, also $e_{P}(z)$ is a superdifferential of the restriction of $\lambda$ to $[y, \bar{x}]$ (we are using $y\geq z$).
    Since $\lambda$ is differentiable and $\bar{x} > x_{0} > y$, the superdifferential at $x_{0}$ is a singleton.
    In particular, $e_{P}(y) = e_{P}(z)$.
    Since $e_{P}$ is strictly decreasing on $(\ubar{x}, \bar{x})$, we conclude $y = z$.
\end{pf}

\subsubsection{Proof of \headercref{Theorem}{{thm:necessary_for_optimality}}.}
        Since $m$ is tight and optimal, we know from \Cref{thm:optimal_mech_fully_stochastic} that $e_{P}$ is interior on $[\ubar{x}, \bar{x})$.
        Let $\lambda = \lambda_{m}\in \Lambda$ be $m$'s loss function.
        Since $e_{P}$ is interior on $[\ubar{x}, \bar{x})$, we have $\lambda(y) < y$ for all $y\in (\ubar{x}, \bar{x})$.
        
        Let $x_{\ast} = \inf_{y\in (\ubar{x}, \bar{x})} \hat{x}(y)$.
        We claim $\lambda$ is strictly concave on $(x_{\ast}, \bar{x})$.
        Note $\bar{x} = \sup_{y\in (\ubar{x}, \bar{x})} \hat{x}(y)$ since, by definition of the binding ICs, it holds $\hat{x} \geq \id$.
        Let $\hat{x}^{-1}$ be the continuous, strictly increasing inverse of $\hat{x}$ on $(x_{\ast}, \bar{x})$, which exists since $\hat{x}$ is continuous and strictly increasing until it reaches $\bar{x}$, i.e. on $(\ubar{x}, x^{\ast})$.
        By the definition of the loss function, for all $x\in (x_{\ast}, \bar{x})$ we have $\lambda(x) = \min_{y\in [\ubar{x}, x]} e_{P}(y) x + (1 - e_{P}(y))(y - r_{\emptyset}(y))$, and the minimum is attained at $\hat{x}^{-1}(x)$.
        Hence, for all $x, x^{\prime} \in (x_{\ast}, \bar{x})$ such that $x^{\prime} < x$, the Envelope Theorem implies $\lambda(x) - \lambda(x^{\prime}) = \int_{(x^{\prime}, x)} e_{P}(\hat{x}^{-1}(r))\de r$.
        Since $m$ has doubly unique binding ICs, \Cref{lemma:doubly_unique_construction} implies that $e_{P}$ is strictly decreasing on $(\ubar{x}, \bar{x})$.
        Hence, $\lambda$ is strictly concave on $(x_{\ast}, \bar{x})$.

        We first prove \eqref{eq:thm:necessary_for_optimality:2}, \eqref{eq:thm:necessary_for_optimality:3}, and \eqref{eq:thm:necessary_for_optimality:4}.

        Let $y\in\iSR$.
        Since $m$ has a unique binding IC selection, one can verify using Theorem 3 of \citet{milgrom2002envelope} that $\pi_{3}(y, \tilde{e}_{A}, \lambda)$ is differentiable at $\tilde{e}_{A} = e_{A}(y)$.
        Since $y\in\iSR$, the effort $e_{A}(y)$ maximizes $\pi_{3}(y, \cdot, \lambda)$ on a neighborhood of $e_{A}(y)$ and is interior. The first-order condition is $I(y; m, \hat{x}) = D(y; m, \hat{x})$, establishing \eqref{eq:thm:necessary_for_optimality:4}.

        To establish \eqref{eq:thm:necessary_for_optimality:2} and \eqref{eq:thm:necessary_for_optimality:3}, respectively, we analogously consider the first-order conditions for $\pi_{1}(y, \cdot, \lambda)$ and $\pi_{2}(y, \cdot, \lambda)$, respectively.
    
        We now prove \eqref{eq:thm:necessary_for_optimality:1}.
        Abbreviate $I(\cdot) = I(\cdot; m, \hat{x})$ and $D(\cdot) = D(\cdot; m, \hat{x})$.

        We first show $0 \leq \int ((1 + D(z)) \bm{1}(x\leq z) - I(z)\bm{1}(x\leq \hat{x}(z))) \de F(z)$ for all $x \in (x_{\ast}, \bar{x})$ such that $\lambda$ is differentiable at $x$.
        Take $y \in (\ubar{x}, \bar{x})$ such that $y < x$.
        Let $\delta = \lambda(x) - \lambda(y) - \lambda^{\prime}(x)(x - y)$.
        Since $\lambda$ is strictly concave on $(x_{\ast}, \bar{x})$, we have $\delta > 0$.
        Now define $\eta$ for all $z$ by
        \begin{equation*}
            \eta(z) = 
            \begin{cases}
                0,\quad& z<y\\
                \lambda(y) - \lambda(z) + \lambda^{\prime}(x)(z - y),\quad& z\in [y, x]\\
                -\delta,\quad&\mbox{else.}
            \end{cases}
        \end{equation*}
        Since $\lambda$ is strictly concave on $(x_{\ast}, \bar{x})$, we have $0 \geq \eta(z) \geq -\delta$ for all $z$.
        Further, $\lambda + \varepsilon\eta$ is necessarily strictly increasing and concave, and equals $\ubar{x}$ at $\ubar{x}$.
        Since $y \in (\ubar{x}, \bar{x})$, the function $\lambda$ is bounded away from the identity on $[y, \bar{x}]$, and hence $\lambda + \varepsilon\eta$ is below the identity for $\varepsilon > 0$ sufficiently close to $0$.
        In particular, $\lambda + \varepsilon\eta\in\Lambda$ for $\varepsilon > 0$ sufficiently small.
        Since $m$ is optimal, $\lambda$ maximizes $\Pi^{\ast}$ across $\Lambda$.
        Thus, $0 \geq ( \Pi^{\ast}(\lambda + \varepsilon\eta) - \Pi^{\ast}(\lambda)) / \varepsilon$.
        Clearly $\eta$ equals $0$ on a neighborhood of $\ubar{x}$.
        Thus, \Cref{lemma:lambda_perturbation} implies $0 \geq  \lim_{\varepsilon\to 0} ( \Pi^{\ast}(\lambda + \varepsilon\eta) - \Pi^{\ast}(\lambda)) / \varepsilon=\int (1 + D(z)) \eta(z) -  I(z)\eta(\hat{x}(z))\de F(z)$.
        Now, fixing $x$, as $y\to x$, the value $\delta$ vanishes. 
        Thus, $\eta / \delta$ converges pointwise to $z\mapsto -\bm{1}(x \leq z)$.
        Here, we are using that $\lambda$ is strictly concave on $(x_{\ast}, \bar{x})$, so that $\delta> 0$ for all $y$ strictly less than $x$.
        Therefore, $0 \leq \int ((1 + D(z)) \bm{1}(x\leq z) - I(z)\bm{1}(x\leq \hat{x}(z))) \de F(z)$.

        A similar argument establishes the reverse inequality $0 \geq \int ((1 + D(z)) \bm{1}(x\leq z) - I(z)\bm{1}(x\leq \hat{x}(z))) \de F(z)$ for all $x \in (x_{\ast}, \bar{x})$ such that $\lambda$ is differentiable at $x$.

        Thus, $0 = \int ((1 + D(z)) \bm{1}(x\leq z) - I(z)\bm{1}(x\leq \hat{x}(z))) \de F(z)$ for all $x\in (x_{\ast}, \bar{x})$ such that $\lambda$ is differentiable at $x$.
        Since $\lambda$ is concave, $\lambda$ is differentiable almost everywhere, and this equality holds for all $x\in (x_{\ast}, \bar{x})$.

        We next argue $x^{\ast} = \bar{x}$, before completing the proof of \eqref{eq:thm:necessary_for_optimality:1}.
        Let $(x_{n})_{n\in\mathbb{N}}$ be a sequence in $(x_{\ast}, \bar{x})$ converging to $\bar{x}$.
        Since all $n$ satisfy $0 = \int ((1 + D(z)) \bm{1}(x_{n}\leq z) - I(z)\bm{1}(x_{n}\leq \hat{x}(z))) \de F(z)$ and $F$ is absolutely continuous, we infer
        $0 = \int I(z)\bm{1}(\bar{x}\leq \hat{x}(z))) \de F(z)$. By the definition of $x^{\ast}$, therefore $0 = \int I(z)\bm{1}(x^{\ast} \leq z) \de F(z)$.
        Since $e_{P}$ is interior on $(\ubar{x}, \bar{x})$, one may verify that $I$ is strictly positive on $(\ubar{x}, \bar{x})$. Since also $\bar{x}=\max\supp F$, the last equation requires $x^{\ast} = \bar{x}$.

        We now complete the proof of \eqref{eq:thm:necessary_for_optimality:1}.
        We have shown $0 = \int ((1 + D(z)) \bm{1}(x\leq z) - I(z)\bm{1}(x\leq \hat{x}(z))) \de F(z)$ for all $x\in (x_{\ast}, \bar{x})$.
        Since $(x_{\ast}, \bar{x})$ is the image of $(\ubar{x}, x^{\ast})$ under $\hat{x}$, and since $x^{\ast} = \bar{x}$, we also get $0 = \int ((1 + D(z)) \bm{1}(\hat{x}(y)\leq z) - I(z)\bm{1}(y\leq z)) \de F(z)$ for all $y\in (\ubar{x}, \bar{x})$.
        This equation yields \eqref{eq:thm:necessary_for_optimality:1}.

        Finally, we show $x_{\ast} = \inf_{y\in (\ubar{x}, \bar{x})} \hat{x}(y) = \sup\iSP$.
        To show $x_{\ast} \geq \sup\iSP$, recall $1 + D = 0$ on $\iSP$.
        Using that $e_{P}$ is interior on $[\ubar{x}, \bar{x})$, one may verify $I > 0$ on $\iSP$.
        Hence, \eqref{eq:thm:necessary_for_optimality:1} and the fact that $F$ has full support implies $x_{\ast} \geq \sup\iSP$.
        To show $x_{\ast} \leq \sup\iSP$, take arbitrary $x\in\iP$.
        By definition, $\lambda(x) = \inf_{y\in [\ubar{x}, x]} e_{P}(y) x + (1 - e_{P}(y))(y - r_{\emptyset}(y))$. Using \Cref{thm:stochastic_tight_characterization}, one can check that the function under the infimum is continuous in $y$. Hence, the infimum is attained at some point $y\in [\ubar{x}, x]$, which simply means $x \in \hat{X}(y)$.
        Thus, $x \geq x_{\ast}$.
        Since $x\in \iP$ was arbitrary and $\inf\iP = \sup\iSP$, also $\sup\iSP \geq x_{\ast}$.
        \qed

\subsubsection{Proof of \headercref{Theorem}{{thm:sufficient_for_optimality}}.}

Recall the following: if $\tilde{\lambda}\in\Lambda$ and $\tilde{m}\in T(\tilde{\lambda})$, then $\tilde{m}$ is tight with loss function $\tilde{\lambda}$; moreover, for all $y\in [\ubar{x}, \bar{x}]$ we have $e_{A}(y) \in \argmax_{\tilde{e}_{A}\in [0, 1]} \ubar{\pi}(y, \tilde{e}_{A}, \lambda)$, and the profit at $y$ is given by $ \ubar{\pi}(y, e_{A}(y), \lambda)$.
Thus, \Cref{assumption:global_concavity} implies $e_{A}(y) \in [0, \check{e}_{A}]$.
Given a pair $\tilde{e}_{A}\colon [\ubar{x}, \bar{x}]\to [0, \check{e}_{A}]$ and $\tilde{\lambda}\in\Lambda$, let $\ubar{\Pi}(\tilde{e}_{A}, \tilde{\lambda}) = \int\ubar{\pi}(y, \tilde{e}_{A}(y), \tilde{\lambda})\de F(y)$ be the expected profit.

Towards a contradiction, suppose $m$ is not optimal.
Let $\tilde{m}$ be a tight optimal mechanism (\Cref{lemma:optimal_existence}), and let $\tilde{\lambda}$ be its loss function.
Let $\lambda$ be the loss function of $m$.
Thus, both $e_{A}$ and $\tilde{e}_{A}$ map to $[0, \check{e}_{A}]$, and the expected profits of $m$ and $\tilde{m}$, respectively, are $\ubar{\Pi}(e_{A}, \lambda)$ and $\ubar{\Pi}(\tilde{e}_{A}, \tilde{\lambda})$, respectively.
Thus, $\ubar{\Pi}(e_{A}, \lambda)<\ubar{\Pi}(\tilde{e}_{A}, \tilde{\lambda})$.

Find a sequence $(\lambda_{n})_{n}$ of differentiable, strictly concave functions in $\Lambda$ converging uniformly to $\lambda$.
For all $n$, let $m_{n}\in T(\lambda_{n})$.
Again, $e_{A, n}$ maps to $[0, \check{e}_{A}]$.
According to \Cref{lemma:doubly_unique_construction}, mechanism $m_{n}$ has doubly unique binding ICs.
Let $\hat{x}_{n}$ be the unique selection on $(\ubar{x}, \bar{x})$.
By possibly passing to a subsequence and using Helly's Selection Theorem, let $(\hat{x}_{n})_{n\in\mathbb{N}}$ converge pointwise to a function $\hat{z}$.

Since $e_{P}$ is interior on $[\ubar{x}, \bar{x})$, we have $\lambda(y) < y$ for all $y\in (\ubar{x}, \bar{x}]$. In particular, \Cref{lemma:basic_lambda_approximation} implies that for fixed $y\in (\ubar{x}, \bar{x})$ the sequence $(e_{A, n}(y), e_{P, n}(y))_{n\in\mathbb{N}}$ converges to $(e_{A}(y), e_{P}(y))$.
It follows readily that $\lim_{n\to\infty}\ubar{\Pi}(e_{A, n}, \lambda_{n}) = \ubar{\Pi}(e_{A}, \lambda)$ holds, and that $\hat{z}$ is a binding IC selection for $m$ on $(\ubar{x}, \bar{x})$.

We next argue that the limit binding IC selection $\hat{z}$ equals the binding IC selection $\hat{x}$ from the assumption of the claim we are trying to prove.
We distinguish two cases.

First, for all $y$ in the interior of $\iSP \cup \ldots \cup \iR$, we show that $m$ has a unique binding IC, so that necessarily $\hat{x}(y) = \hat{z}(y)$.
    Towards a contradiction, let $\hat{X}(y)$ not be the singleton $\lbrace\hat{x}(y)\rbrace$.
    We derive a contradiction to \eqref{eq:thm:sufficient_for_optimality:1}.
    Since $e_{P}$ strictly decreases on $\iSP \cup \ldots \cup \iR$, \Cref{lemma:auxiliary_binding_IC} implies $\max\hat{X}(y)\leq \hat{x}(y + \varepsilon)$ and $\min\hat{X}(y)\geq \hat{x}(y - \varepsilon)$ for all $\varepsilon > 0$ sufficiently close to $0$.
    In particular, $\hat{x}(y + \varepsilon) - \hat{x}(y - \varepsilon)$ is bounded away from $0$ as $\varepsilon\to 0$.
    From the definition of $D$ and the characterization of the intervals, one may verify that $1 + D(x; m, \hat{x}) > 0$ is strictly positive and continuous in $x$ on the interior of $\iP \cup \ldots \cup \iR$.
    Since also $\inf_{z\searrow\ubar{x}}\hat{x}(z) = \sup\iSP$ and $\hat{x}$ is strictly increasing, we find that $\int_{[\hat{x}(y - \varepsilon)], \hat{x}(y+\varepsilon)]} (1 + D(x; m, \hat{x})) \de F(x)$ is strictly positive and bounded away from $0$ as $\varepsilon\to 0$.
    Clearly, $\int_{[y - \varepsilon, y + \varepsilon]} I(x; m, \hat{x})\de F(x)$ vanishes as $\varepsilon\to 0$ since $I$ is bounded.
    Thus, \cref{eq:thm:sufficient_for_optimality:1} is contradicted.

Second, take $y\in \iSR$. By the assumption \eqref{eq:thm:sufficient_for_optimality:4}, it holds $I(y; m, \hat{x}) = D(y; m, \hat{x})$. We also know $I(y; m_{n}, \hat{x}_{n}) = D(y; m_{n}, \hat{x}_{n})$ for sufficiently large $n$ since $m_{n}\in T(\lambda_{n})$ and $y\in \iSR_{n}$ for sufficiently large $n$. 
    Since 
    \begin{equation*}
        \lim_{n\to\infty}(I(y; m_{n}, \hat{x}_{n}), D(y; m_{n}, \hat{x}_{n})) = (I(y; m, \hat{z}), D(y; m, \hat{z})),
    \end{equation*}
    also $I(y; m, \hat{z}) = D(y; m, \hat{z})$.
    Inspecting the definitions, it holds $D(y; m, \hat{x}) = D(y; m, \hat{z})$ and $I(y; m, \hat{x}) = I(y; m, \hat{z}) (\hat{z}(y) + \tau) / (\hat{x}(y) + \tau)$.
    Thus, $I(y; m, \hat{x}) = I(y; m, \hat{x}) (\hat{z}(y) + \tau) / (\hat{x}(y) + \tau)$
    For $y\in \iSR$, we have $e_{A}(y) < 1$ and $e_{P}(y) > 0$, implying $I(y; m, \hat{x}) \neq 0$.
    Thus, also $\hat{x}(y) = \hat{z}(y)$. 

Thus, $\hat{x} = \hat{z}$ on $(\ubar{x}, \bar{x})$.
In particular, $(\hat{x}_{n})_{n\in\mathbb{N}}$ converges pointwise to $\hat{x}$ on $(\ubar{x}, \bar{x})$.
Further, $\lim_{n\to\infty}(I(y; m_{n}, \hat{x}_{n}), D(y; m_{n}, \hat{x}_{n})) = (I(y; m, \hat{x}), D(y; m, \hat{x}))$ for all $y\in (\ubar{x}, \bar{x})$.

Abbreviate $\mu = (e_{A}, \lambda)$, $\tilde{\mu} = (\tilde{e}_{A}, \tilde{\lambda})$, and $\mu_{n} = (e_{A, n}, \lambda_{n})$ for all $n$.

For all $n$, since $e_{A}$, $\tilde{e}_{A}$, and $e_{A, n}$ map to $[0, \check{e}_{A}]$, \Cref{assumption:global_concavity} implies that $\ubar{\Pi}$ is concave on the convex hull of $\mu$, $\tilde{\mu}$, and $\mu_{n}$. 

Let $\Delta = \ubar{\Pi}(\tilde{\mu}) - \ubar{\Pi}(\mu)$, so that $\Delta > 0$.
Since $\lim_{n\to\infty}\ubar{\Pi}(\mu_{n}) = \ubar{\Pi}(\mu)$, there is $n_{0}\in\mathbb{N}$ such that $n\geq n_{0}$ implies $\ubar{\Pi}(\tilde{\mu}) - \ubar{\Pi}(\mu_{n}) \geq \Delta/2$.
By concavity of $\ubar{\Pi}$, for all $n$ the ratio $(\ubar{\Pi}(\mu_{n} + \varepsilon (\tilde{\mu} - \mu_{n})) - \ubar{\Pi}(\mu_{n})) / \varepsilon$ is decreasing in $\varepsilon$.
Thus, for all $n\geq n_{0}$ and $\varepsilon \in (0, 1]$, it holds $(\ubar{\Pi}(\mu_{n} + \varepsilon (\tilde{\mu} - \mu_{n})) - \ubar{\Pi}(\mu_{n})) / \varepsilon \geq \Delta / 2$.
Since $m_{n}\in T(\lambda_{n})$ for all $n$, we have $\ubar{\Pi}(\mu_{n}) = \Pi^{\ast}(\lambda_{n})$.
It necessarily holds $\ubar{\Pi}(\mu_{n} + \varepsilon (\tilde{\mu} - \mu_{n})) \leq \Pi^{\ast}(\lambda_{n} + \varepsilon(\tilde{\lambda} - \lambda_{n}))$ for all $\varepsilon$ and $n$.
Thus, all $n\geq n_{0}$ and $\varepsilon \in (0, 1]$ satisfy
\begin{equation}\label{eq:thm:sufficient_for_optimality:derivative_bound}
    \frac{\Pi^{\ast}(\lambda_{n} + \varepsilon(\tilde{\lambda} - \lambda_{n})) - \Pi^{\ast}(\lambda_{n})}{\varepsilon} \geq \frac{\Delta}{2}
    .
\end{equation}

Fixing $n$, abbreviate $\eta_{n} = \tilde{\lambda} - \lambda_{n}$ and $\eta = \tilde{\lambda} - \lambda$.
Note $\lambda_{n} + \varepsilon \eta_{n}$ is a convex combination of functions in $\Lambda$ and, hence, also in $\Lambda$ for all $\varepsilon$.
Since $\lambda_{n}\in\Lambda$ was chosen to be strictly concave, we have $\lambda_{n}(y) < y$ for all $y\in (\ubar{x}, \bar{x}]$.
We also recall that $m_{n}$ has doubly unique binding ICs.
Thus, \Cref{lemma:lambda_perturbation} and \eqref{eq:thm:sufficient_for_optimality:derivative_bound} imply
\begin{equation*}
    \int (1 +D(y; m_{n}, \hat{x}_{n}))\eta_{n}(y) - I(y; m_{n}, \hat{x}_{n})\eta_{n}(\hat{x}_{n}(y)) \de F(y) \geq \frac{\Delta}{2}.
\end{equation*}
We now consider the limit as $n\to\infty$.
Clearly, $\eta_{n} \to \eta$ uniformly as $n\to\infty$.
From earlier, recall $\lim_{n\to\infty}(I(y; m_{n}, \hat{x}_{n}), D(y; m_{n}, \hat{x}_{n})) = (I(y; m, \hat{x}), D(y; m, \hat{x}))$.
Using the assumption $\liminf_{y\to\ubar{x}} \hat{x}(y) = \sup\iSP$, one may verify that $\sup_{y\in (\ubar{x}, \bar{x})} (1 + D(y; m, \hat{x}))\eta(y) - I(y; m, \hat{x})\eta(\hat{x}(y))$ exists, as in the proof of \Cref{lemma:lambda_perturbation}.
Therefore, since $F$ is absolutely continuous, Dominated Convergence implies
\begin{equation*}
    \int (1 + D(y; m, \hat{x}))\eta(y) - I(y; m, \hat{x})\eta(\hat{x}(y)) \de F(y) \geq \frac{\Delta}{2}
    .
\end{equation*}
Since $\eta$ is continuous while $\hat{x}$ is continuous and strictly increasing, there is a finite collection of intervals $\lbrace [a_{k}, b_{k}]\rbrace_{k}$ and weights $\lbrace w_{k}\rbrace_{k}$ such that
\begin{equation*}
    \sum_{k} w_{k} \int_{[a_{k}, b_{k}]} (1 + D(y; m, \hat{x})) \de F(y) - w_{k}\int_{[\hat{x}^{-1}(a_{k}), \hat{x}^{-1}(b_{k})]} I(y; m, \hat{x}) \de F(y) 
    \geq \frac{\Delta}{4}
    .
\end{equation*}
However, for all $k$, the sum of the integrals is zero, by the assumption on the mechanism (consider \eqref{eq:thm:sufficient_for_optimality:1} for $[a^{\prime}_{k}, b^{\prime}_{k}] = [\hat{x}^{-1}(a_{k}), \hat{x}^{-1}(b_{k})]$); contradiction.
\qed

\subsubsection{Proof of \headercref{Theorem}{{thm:doubly_unique_approximation}}.}

Find a sequence $(\lambda_{n})_{n}$ of differentiable, strictly concave functions in $\Lambda$ converging uniformly to $\lambda$.
For all $n$, let $m_{n}\in T(\lambda_{n})$, so that $m_{n}$ is tight (\Cref{thm:tight_operator_fp}).
For all $n$, \Cref{lemma:doubly_unique_construction} implies that $m_{n}$ has doubly unique binding ICs and that $e_{P, n}$ is strictly decreasing.
Since $m$ is optimal, $e_{P}$ is interior on $[\ubar{x}, \bar{x})$.
Hence, we have $\lambda(y) < y$ for all $y\in (\ubar{x}, \bar{x}]$. In particular, \Cref{lemma:basic_lambda_approximation} implies that for fixed $y\in (\ubar{x}, \bar{x})$ the sequence $(e_{A, n}(y), e_{P, n}(y))_{n\in\mathbb{N}}$ converges to $(e_{A}(y), e_{P}(y))$.
It follows readily that all $y\in (\ubar{x}, \bar{x})$ satisfy $\lim_{n\to\infty}\Pi_{m_{n}}(y) = \lim_{n\to\infty}\ubar{\pi}(y, e_{A, n}, \lambda_{n}) = \ubar{\pi}(y, e_{A}, \lambda) = \Pi_{m}(y)$.
\qed

\subsection{Sufficient conditions for \headercref{Assumption}{{assumption:global_concavity}}}\label{appendix:sufficient_condition_for_global_concavity}
Here, let $c_{P}$ be given by $c_{P}(e_{P}) = \kappa \tilde{c}_{P}(e_{P})$ for all $e_{P}\in [0, 1]$, where $\kappa > 0$.
\begin{lemma}\label{lemma:assumptions_for_global_concavity}
    Let $c_{A}^{\prime\prime}(e_{A}) > 0$ and $u_{A}^{\prime\prime}(e_{A}) + c_{A}^{\prime\prime}(e_{A}) > 0$ for all $e_{A}\in [0, 1]$.
    Let $c_{P}^{\prime} / c_{P}^{\prime\prime}$ be bounded across $[0, 1]$.
    Then, \Cref{assumption:global_concavity} holds if $\kappa$ is sufficiently small.
\end{lemma}
\begin{pf}[Proof of \Cref{lemma:assumptions_for_global_concavity}]
    Fix arbitrary $\check{e}_{A}\in [0, 1]$.
    For all $i\in\lbrace 1, 2, 3\rbrace$, one may verify $\partial_{2}\pi_{i}(y, \check{e}_{A}, \lambda) \leq - c_{A}^{\prime}(\check{e}_{A}) + \kappa \tilde{c}_{P}(1) + (1 - \check{e}_{A})\kappa \tilde{c}_{P}^{\prime}(1) u_{A}^{\prime}(\check{e}_{A}) / (\ubar{x} + \tau)$. This upper bound converges to $- c_{A}^{\prime}(\check{e}_{A})$ as $\kappa\to 0$. Thus, if $\kappa$ is below some value $\kappa_{1}$, then $\partial_{2}\pi_{i}(y, \check{e}_{A}, \lambda) < 0$ for all $y$ and $\lambda$. Since $\pi_{i}$ is quasiconcave in agent effort, also $\partial_{2}\pi_{i}(y, e_{A}, \lambda) < 0$ for all $e_{A} \in [\check{e}_{A}, 1]$ and all $y$ and $\lambda$. It follows that for $\kappa$ below $\kappa_{1}$ and all $y$ and $\lambda$, no maximizer of $\pi_{1}(y, \cdot, \lambda)$, $\pi_{2}(y, \cdot, \lambda)$, or $\pi_{3}(y, \cdot, \lambda)$ is above $\check{e}_{A}$.

    In what follows, fix $\kappa \leq \kappa_{1}$.
    Fixing $x$ and $y$ such that $y < x$, consider the following as functions of $(e_{A}, \ell) \in [0, \check{e}_{A}]\times [\ubar{x}, \bar{x}]$ or $(e_{A}, d) \in [0, \check{e}_{A}]\times [0, \bar{x} - \ubar{x}]$:
    \begin{align}
        \label{eq:lemma:global_concavity:obj1}
        & u_{A}(e_{A}) + c_{A}(e_{A}) + (1 - e_{A})\kappa \tilde{c}_{P}\left(\frac{\ell - y}{x - y}\right)
        ,
        \\
        \label{eq:lemma:global_concavity:obj2}
        &c_{A}(e_{A}) + (1 - e_{A})\kappa \tilde{c}_{P}\left(\frac{\ell - y}{x - y}\right)
        ,
        \\
        \label{eq:lemma:global_concavity:obj3}
        & c_{A}(e_{A}) + (1 - e_{A}) \kappa \tilde{c}_{P}\left(\frac{d - u_{A}(e_{A})}{x + \tau}\right)
        .
    \end{align}
    Here, $\tilde{c}_{P}$ is smoothly extended to negative numbers by setting $\tilde{c}_{P}^{\prime\prime}(e_{P}) = \tilde{c}^{\prime\prime}_{P}(0)$ for all $e_{P} \leq 0$.
    By inspecting the traces and determinants of the Hessians of \eqref{eq:lemma:global_concavity:obj1}, \eqref{eq:lemma:global_concavity:obj2}, and \eqref{eq:lemma:global_concavity:obj3}, one can check that if $\kappa$ is sufficiently small then all three of \eqref{eq:lemma:global_concavity:obj1}, \eqref{eq:lemma:global_concavity:obj2}, and \eqref{eq:lemma:global_concavity:obj3} are convex in their respective arguments.
    This step uses that $e_{A}$ is bounded above by $\check{e}_{A} < 1$, in addition to the other assumptions on $c_{A}$ and $c_{P}$.

    Fixing sufficiently small $\kappa$, it follows that for all $x$ and $y$ such that $y < x$, also
    \begin{align*}
        & y - u_{A}(e_{A}) - c_{A}(e_{A}) - (1 - e_{A})\kappa \tilde{c}_{P}\left(\frac{\lambda(x) - y}{x - y}\right)
        ,
        \\
        & \lambda(y) - c_{A}(e_{A}) - (1 - e_{A})\kappa \tilde{c}_{P}\left(\frac{\lambda(x) - y}{x - y}\right)
        ,
        \\
        & \lambda(y) - c_{A}(e_{A}) - (1 - e_{A}) \kappa \tilde{c}_{P}\left(\frac{\lambda(x) - \lambda(y) - u_{A}(e_{A})}{x + \tau}\right)
    \end{align*}
    are all concave in $(e_{A}, \lambda) \in [0, \check{e}_{A}]\times\Lambda$. 
    Now, by construction, $\tilde{c}_{P}(t) \geq c_{P}([t]_{+})$, for all $t \in \mathbb{R}$, and $c_{P}([t]_{+}) = 0$ if $t \leq 0$.
    Clearly, $y - u_{A}(e_{A}) - c_{A}(e_{A})$ and $\ell - c_{A}(e_{A})$ are both concave in $(e_{A}, \ell)$.
    It follows that also
    \begin{align}
        \label{eq:lemma:global_concavity:obj1_restated}
        & y - u_{A}(e_{A}) - c_{A}(e_{A}) - (1 - e_{A})\kappa c_{P}\left(\left[\frac{\lambda(x) - y}{x - y}\right]_{+}\right)
        ,
        \\
        \label{eq:lemma:global_concavity:obj2_restated}
        & \lambda(y) - c_{A}(e_{A}) - (1 - e_{A})\kappa c_{P}\left(\left[\frac{\lambda(x) - y}{x - y}\right]_{+}\right)
        ,
        \\
        \label{eq:lemma:global_concavity:obj3_restated}
        & \lambda(y) - c_{A}(e_{A}) - (1 - e_{A}) \kappa c_{P}\left(\left[\frac{\lambda(x) - \lambda(y) - u_{A}(e_{A})}{x + \tau}\right]_{+}\right)
    \end{align}
    are all concave in $(e_{A}, \lambda) \in [0, \check{e}_{A}]\times\Lambda$. 
    Fixing $y \in [\ubar{x}, \bar{x})$, the profit $\ubar{\pi}(y, e_{A}, \lambda)$ equals the infimum across \eqref{eq:lemma:global_concavity:obj1_restated}, \eqref{eq:lemma:global_concavity:obj2_restated}, and \eqref{eq:lemma:global_concavity:obj3_restated}, and across all $x$ such that $y < x$.
    The pointwise infimum of concave functions is concave. Thus, $\ubar{\pi}(y, e_{A}, \lambda)$ is concave in $(e_{A}, \lambda) \in [0, \check{e}_{A}]\times\Lambda$.    

    For $y = \bar{x}$, it is readily seen that $\ubar{\pi}(y, e_{A}, \lambda) = \min\lbrace y - u_{A}(e_{A}), \lambda(y)\rbrace - c_{A}(e_{A})$ holds, which is concave.
\end{pf}

\end{document}